\documentclass[12pt, letterpaper]{article}

\usepackage{lineno,hyperref}
\modulolinenumbers[20]
\usepackage[titlenumbered,ruled]{algorithm2e}
\usepackage{graphicx, subfigure}
\usepackage{color}
\usepackage{booktabs}
\usepackage{longtable}
\usepackage{lipsum}
\usepackage{enumitem}
\usepackage{multirow}

\usepackage[numbers]{natbib}

\begin{document}
\let\WriteBookmarks\relax
\def\floatpagepagefraction{1}
\def\textpagefraction{.001}

\title{Android Malware Clustering using Community Detection on Android Packages Similarity Network}

\author{ElMouatez Billah Karbab, Mourad Debbabi\\ Abdelouahid Derhab, Djedjiga
Mouheb}

\date{May 2020}

\maketitle

\begin{abstract}
The daily amount of Android malicious applications (apps) targeting the app repositories is increasing, and their number is overwhelming the process of fingerprinting. To address this issue, we propose an enhanced \textsf{Cypider} framework, a set of techniques and tools aiming to perform a systematic detection of mobile malware by building a scalable and obfuscation resilient similarity network infrastructure of malicious apps.  Our approach is based on our proposed concept, namely \textit{malicious community}, in which we consider malicious instances that share common features are the most likely part of the same malware family.  Using this concept, we presumably assume that multiple similar Android apps with different authors are most likely to be malicious. Specifically, \textsf{Cypider} leverages this assumption for the detection of variants of known malware families and zero-day malicious apps.  \textsf{Cypider} applies \textit{community detection algorithms} on the similarity network, which extracts sub-graphs considered as suspicious and possibly malicious communities.  Furthermore, we propose a novel fingerprinting technique, namely \textit{community fingerprint},  based on a one-class machine learning model for each malicious community.  Besides, we proposed an enhanced \textsf{Cypider} framework, which requires less memory, $\approx x650\%$, and less time to build the similarity network, $\approx x700$, compared to the original version, without affecting the fingerprinting performance of the framework.  We introduce a systematic approach to locate the best threshold on different feature content vectors, which simplifies the overall detection process. \textsf{Cypider} shows excellent results by detecting $60\%-80\%$ coverage of the malware dataset in one detection iteration with higher precision $85\%-99\%$ in the detected malicious communities. On the other hand, the \textit{community fingerprints} are promising as we achieved $86\%$, $93\%$, and $94\%$ in the detection of the malware family, general malware, and benign apps respectively.
\end{abstract}

\section{Introduction}

Mobile devices and their applications have become an integrated part of our everyday life. People nowadays are using mobile devices to store and send their personal and sensitive data, which makes them attractive targets to cyber criminals. Android \cite{android_os} is the most popular and dominant operating system among mobile users.  In September 2019,  Android had 76\% of the market share worldwide \cite{statcounter}, as it is deployed on different devices such as smartphones and tablets. It also succeeded in reaching other smart devices such as TVs \cite{android_tv}, watches \cite{android_wear}, and cars \cite{android_auto}. Moreover, Android OS is increasingly integrated into IoT systems, especially with the appearance of Google's Android Things \cite{IoT2016}, an embedded operating system that is designed for low-power and resource-constrained IoT devices.

Android OS implements some mechanisms such as \textit{sandboxing}, to provide security for smart devices. However, the Android security mechanisms failed to defend against the different threats. Due to its popularity,  Android is increasingly targeted by malware developers.  According to G DATA \cite{GDATA_2019},  there is an average of more than $10k$ new malware that appeared per day in the first half of 2019, which corresponds to a new malicious app every 8 seconds.

Malware could be divided into two main classes: i)  \textit{malware variant}, which is a new version of known malware, and ii) \textit{unseen malware} or \textit{zero-day malware}, which is a new malicious app, which is unknown and not discovered by security investigators and anti-malware vendors. Due to a large number of released Android apps along with the daily emergence of malicious apps, the manual investigation, and analysis of new apps have become a cumbersome task. The conventional approaches rely on matching \textit{signature-based} patterns with known malware families. However, these approaches are not practical and not effective, as they rely on cryptographic hashes. As a result, malware developers could easily evade detection by employing minor modifications in the original malware app. Furthermore, signature-based approaches are not capable of detecting new malware families, which raises serious concerns about the systematic detection of new malware without or with minimum human intervention. Other approaches are based \textit{heuristic-based} or \textit{machine-learning} techniques to produce \textit{learning-based} patterns, which aim to detect known and new malware.  They are more effective than the \textit{signature-based} ones at detecting zero-day malware. Still, their detection accuracy essentially depends on the training set and the used features to generate the detection model.

In this context, we aim at dealing with large-scale Android malware by reducing the length of the \textit{analysis window} of newly detected malware. This time window ranges from malware detection to signature generation by security vendors. When the analysis window is long, the malicious apps are given more time to infect the users' devices. The existing techniques incur a large window due to the overwhelming number of Android apps that appear daily . Besides, they employ a manual investigation to analyze malware. Thus, there is a need to reduce reliance on manual analysis to reduce the length of the \textit{analysis window}. To achieve this objective, we propose systematic tools, techniques, and approaches for the detection of both known and unseen malware (i.e., malware variants and zero-day malware). We assume that two Android apps, each of which has a different author and certificate, are most likely to be malicious if they are highly similar. This assumption relies on the fact that the malware developer usually injects the same malicious payload into multiple repackaged applications to hide it.  Thus, it is unlikely to find such a malicious payload in a benign Android app. Consequently, two Android apps should not be highly similar with respect to their components, except for known libraries. This observation could be used as a basis to design and develop a security framework for Android malware detection.  

In this paper, we are motivated by the earlier-mentioned to propose a cyber-security framework, called \textsf{Cypider} (\textit{Cyber-Spider For Android Malware Detection}), to identify and cluster Android malware without the need of any prior knowledge about the \textit{signature-based} or \textit{learning-based} patterns of Android malware apps. \textsf{Cypider} is a novel framework, which combines techniques and methods to deal with the issue of Android malware clustering and fingerprinting. \textsf{Cypider}, and automatic framework that adopts an unsupervised approach, is designed (1) to detect repackaged malware (or malware families), which represent most of the Android malicious apps  \cite{zhou2012dissecting}, and (2) to detect new malware apps.  The principal idea of \textsf{Cypider} is based on building a \textit{similarity network} that connects the input apps. After that, sub-graphs with high connectivity, called \textit{communities}, are extracted from the \textit{similarity network}. These communities are most likely composed of malicious apps. This step is followed by generating a novel fingerprint for each extracted community,  called \textit{community fingerprinting}. Instead of employing a hash or fuzzy hash-based signature of the app, the  \textit{One-Class Support Vector Machine} learning model (OC-SVM) \cite{Scholkopf:2001} is used to compute the \textit{community fingerprint} of the Android malware family or sub-family.  OC-SVM is a machine learning technique, which is trained by using the features of  \textit{one class} to generate the learning model, i.e., community fingerprint, to decide whether a given unseen Android app belongs to the community or not. After constructing the communities of malicious apps, \textsf{Cypider} framework triggers a periodic process, which aims to build new communities from the remaining apps (i.e., apps that do not belong to any community)  and the new arrival apps to \textsf{Cypider}, which together form what we call the \textit{active dataset}.

The scalability issue of \textsf{Cypider} is addressed through three main techniques: (1) the performance of \textsf{Cypider} is negatively impacted by the high dimensionality of the statistical features that are extracted from the Apk file. For this purpose, \textit{feature hashing} \cite{qinfeng09hashk} technique is employed to reduce the high dimensional vector to a fixed-size vector without affecting the detection rate. (2) Dimensionality reduction is applied to reduce to compress the feature hashing vectors. (3) FAISS \footnote{https://github.com/facebookresearch/faiss}, which is a similarity computation technique, is used to compute the similarity among the apps' vectors,  (4) a scalable \textit{community detection} algorithm \cite{fast08blondel} is used to extract the  \textit{malicious communities}. To sum up,  \textsf{Cypider} framework is a set of algorithms, mechanisms, and techniques, which are integrated into one approach to detect Android malicious app without requiring to any pre-knowledge about the malware and their families.  Besides,  \textsf{Cypider} is designed to generate the unsupervised fingerprints of the possible malware threats by leveraging the proposed community fingerprint concept.

In the Original \textsf{Cypider}, we made the following contributions: 
\begin{enumerate}
\item We designed and implemented an unsupervised-based Android malware detection and family attribution framework. The proposed framework has shown to be effective and efficient by applying a clustering approach, which leverages the community concept and graph partition techniques. 

\item We proposed a community fingerprint, a novel detection model to represent the pattern of a given community, which could be an Android malware family or subfamily.
\end{enumerate}

In this paper, the main contributions of the enhanced \textsf{Cypider} are:

\begin{enumerate}
\item We propose an enhanced design for \textsf{Cypider} by applying dimensionality reduction on feature hashing vectors to produce a short dense vector for android malware (see Section \ref{sec:feture_preprocessing}).  This enhancement increases the efficiency of the similarity computation to $\approx x600$ times (see Section \ref{sec_efficiency}) compared to the original design \cite{karbab2016cypider}.

\item We propose a systematic mechanism to determine the best threshold trade-off for the different contents of the APK file. This proposed heuristic solves the issue of manual threshold search for each content type as in the original design \cite{karbab2016cypider}.  Furthermore, it provides the security practitioner with a simple mechanism to customize \textsf{Cypider} between the recall and precision settings (see Section \ref{sec:lsh_similartiy}).

\item We conduct detailed evaluations that assess the effect of our solution hyper-parameters on the performance of the framework. We also assess the scalability and the effectiveness of \textsf{Cypider} on large-scale datasets namely: (1) Genome malware dataset \cite{zhou2012dissecting}, (2) Drebin malware dataset \cite{Drebin_Dataset, arp2014drebin}, (3) AndroZoo (66k) malware dataset, and (4) the combined previous datasets with benign (50k) dataset. The evaluation experiments show good results in the context of unsupervised malware detection (see Section \ref{sec:evaluation}, \ref{sec_hyperparametersAnalyses}, and \ref{sec_recallPrecisionSettings}).

\item We conduct extensive experiments to measure the effect of different obfuscation techniques on our framework using different obfuscated datasets, PRAGuard \cite{Davide2015Praguard} and Obfuscated Drebin (using DroidChameleon \cite{Vaibhav2013DroidChameleon})  and non-obfuscated datasets (MalGenome). \textsf{Cypider} shows high resiliency against common obfuscation and code transformation techniques (see Section \ref{sec_obfuscationResliency}).

\end{enumerate}

\section{Overview}

\subsection{Threat Model} 
\label{sec:threat_model} 

The main objective of \textsf{Cypider} is to detect Android malicious apps without the need for any prior knowledge about these malware samples and their patterns. More precisely, \textsf{Cypider} focuses mainly on the bulk detection of malware families and malware variants rather than the detection of an individual malicious app by extracting the malicious communities from the similarity network of the apps dataset. The second objective of \textsf{Cypider} is to provide a scalable yet accurate solution that can process the overwhelming number of malware that appear daily, and which could compromise the users' smart devices. \textsf{Cypider} is robust (Section \ref{sec_obfuscationResliency}) but not immune against obfuscated apps contents. \textsf{Cypider} could resist against some types of obfuscations as it analyzes different static contents of the Android package. \textsf{Cypider} could be more resilient to obfuscation as it could fingerprint malware apps with other static contents that are not obfuscated, such as app permissions. It is important to mention that static contents could be obfuscated in one app and not in another, which is related mostly to how the malware developer writes the malware code.

\textsf{Cypider} is evaluated under real malware datasets along with random Android apps chosen from Google Play, where the apps are assumed to be obfuscated through ProGuard, to test the effectiveness and efficiency of \textsf{Cypider} in real word obfuscation scenarios. \textsf{Cypider} could not detect dynamic transform attacks, e.g., when the malicious payload does not exist in APK static file and is downloaded at runtime. The aim of \textsf{Cypider} is to detect homogeneous and pure malicious communities, where each community corresponds to one malware family and hence makes the malware analysis process easier. Besides, \textsf{Cypider} aims to provide a high detection accuracy along with a low false-positive rate.  
\subsection{Usage Scenarios} 
\label{sec:usage}

\textsf{Cypider} is a generic approach that is used to investigate the existence of similarity among apps and could be applied to various usage scenarios. The first usage scenario is about the comparison of software programs, where the input is a set of software programs, and the output is the set of programs that have similar features. The possible application of this scenario is the authorship attribution, where the obtained communities represent the binary programs that are written by the same author. The second scenario is about detecting plagiarized software, a community of binary programs with different authors, but similar to a given copyrighted software. The third and fourth usage scenarios are malware detection and family attribution, respectively, i.e., performing bulk detection by identifying communities of similar malicious apps, and consequently infer their respective families. In this paper, our focus is on Android malware detection.

\textsf{Cypider} is evaluated using two types of experimental scenarios. In the first experiment, we only consider malicious Android apps. This experiment aims to speed up the analysis process by performing bulk detection and attribute the detected malware to their corresponding families. This is also achieved by performing automatic analysis with minimal human intervention.  The family could be attributed to a community by starting with a small set of apps, i.e., one app in most cases. On the other hand, unassigned apps are considered suspicious and require manual investigation. The malware analysis approaches could be divided into:

\begin{enumerate}[wide, labelwidth=!, labelindent=0pt]

\item \textit{Semi-supervised Approach:} We use a dataset of known malicious apps and their families, in addition to unknown malicious apps. The dataset is fed to \textsf{Cypider} to generate malicious communities. First, known malicious apps from the same family are grouped (i.e., strongly-connected graph), and assigned to the same community. Second, the unknown malicious apps, which have similar features as an existing community, will most likely join that community. Analogically, the communities of known malicious apps could play the role of a magnet for unknown ones, if known and unknown apps share the same features, which implies the same malware family.

\item \textit{Unsupervised Approach:} We use a dataset that does not have any information about the tested apps that are fed to \textsf{Cypider}. The aim is to find communities instead of a single malicious app. By focusing on finding communities (i.e., bulk detection), security investigators could drastically increase their productivity. In this paper, we consider the unsupervised approach.  
\end{enumerate}

In the second experimental scenario, we consider a mixed dataset of Android apps (i.e., malicious and benign). Such a dataset could be the result of a preliminary suspiciousness app filtering, which might lead to high \textit{false positives}. Let us suppose that benign apps (i.e., \textit{false positives}) represent $50\%-75\%$ of the size of suspicious apps; we could identify malicious apps by detecting the communities that have a common payload, as we assume that highly similar apps are likely to be malicious. In other words, malicious apps from the same family tend to be very similar due to the shared malicious payload. This assumption could be extended to benign apps. Benign apps from the same category such as games tend to be similar due to the shared features. The similarity between apps helps group malware into malicious communities, but also helps group benign apps into benign communities.
In addition, \textsf{Cypider} could filter benign apps (i.e., less suspicious apps) that are not similar to the suspicious apps in the active dataset. The previously described approaches, i.e., \textit{Semi-supervised} and \textit{Unsupervised}, could be applied to the mixed dataset to target Android malware detection and family attribution. More specifically, we adopt the target of the unsupervised approach, which detects malicious apps without any prior-knowledge about malware.

\section{Methodology}
\label{sec:methodology}

\textsf{Cypider} framework is applied to a dataset of unlabeled apps (malicious or mixed dataset), aiming to generate \textit{community fingerprints} for the processed apps. The overall architecture of \textsf{Cypider} is shown in Figure \ref{fig:approach_overview}, and is comprised of the following steps:

\begin{figure}
  \centering
      \includegraphics[width=0.99\textwidth]{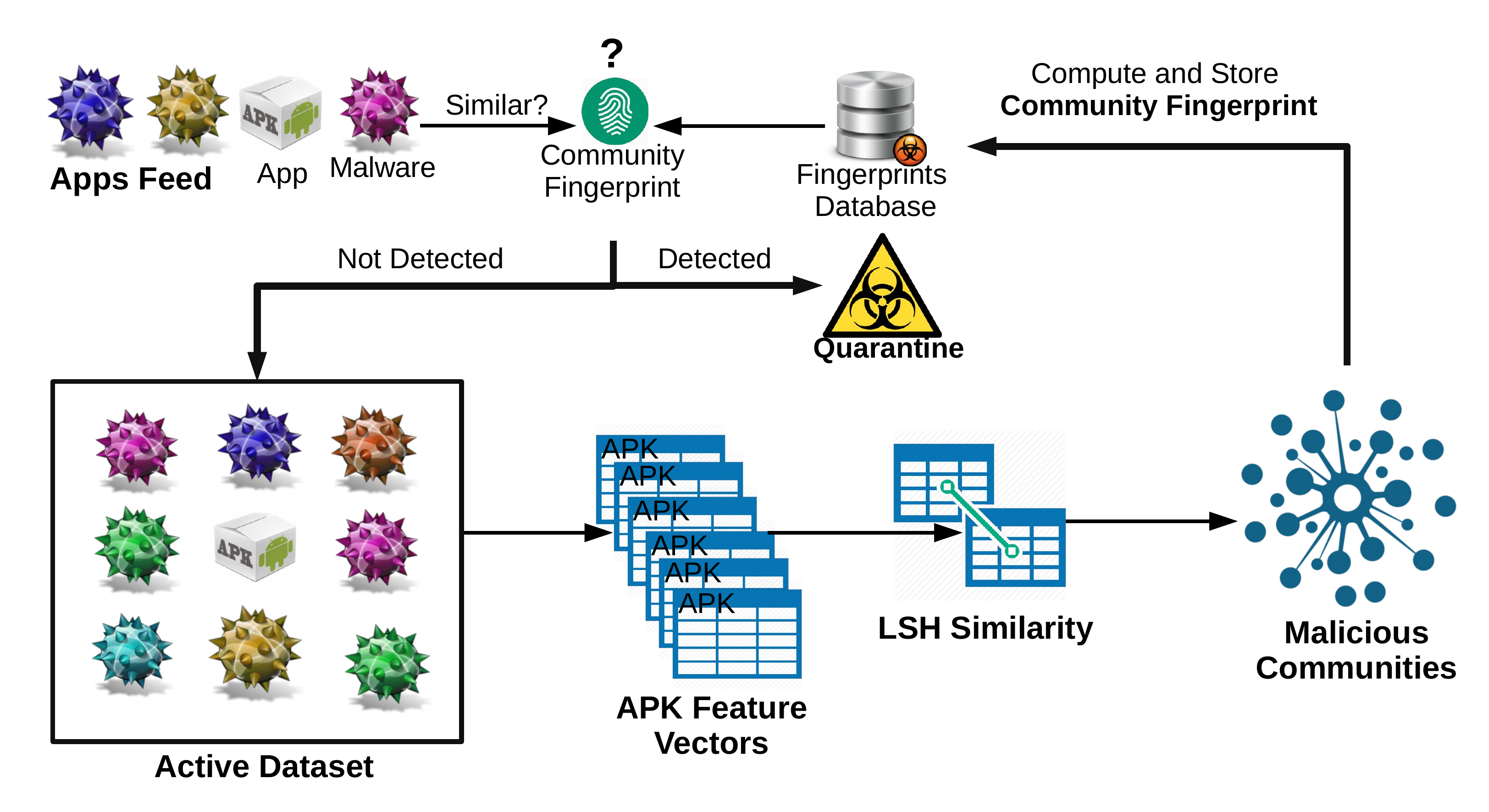}
  \caption{\textsf{Cypider} Overview}
  \label{fig:approach_overview}
\end{figure}

In the first step, the apps that are developed by the same author, also known as  \textit{sibling apps}, are filtered out. The main reason is to remove the noise that could be produced by having sibling apps in the app communities, as the authors tend to use the same components in their different apps. Sibling apps could be identified through the information that is available in the META-INF directory in the APK file, such as app version, app hash, and author signature. Thus, only apps with different authors are kept as adversaries prefer to use different fake author identities to evade detection in case one of their apps is caught. To deal with many apps having the same author, \textsf{Cypider} randomly chooses one app, and if it is found malicious after analysis, all its sibling apps are considered malicious.

After filtering out the sibling apps, we extract meaningful information (i.e., features) from the apps, to identify these apps and compute the similarity among them. To this end, \textit{Cypider} extracts statistical features from all the apps. It is worth noting that the selected features must be resilient against the attacker's evasion detection techniques. To achieve this aim, we choose features that are broad enough to cover most of the static characteristics of an Android app. The more extensive the features are, the more resilient they are. To this end, we extract from the APK file, its static contents such as dex, resources, assembly, etc., as explained in Section \ref{sec:statistical_features}.

\textsf{Cypider} uses the extracted features from each content to computes a fixed-length \textit{feature vector} for each content features.  \textsf{Cypider} leverages a machine learning pre-processing technique, named \textit{feature hashing} \cite{qinfeng09hashk}  to normalize and reduce the size of the feature vectors, as described in Section \ref{sec:feture_preprocessing}. Based on the extracted features of the previous stage, \textsf{Cypider} generates multiple small and fixed-size feature vectors. The number of the generated feature vectors depends on the number of APK contents that are considered during feature extraction, i.e., each content type corresponds to one feature vector.

Afterward, \textsf{Cypider} performs a dimensionality reduction technique, specifically by applying Principle Component Analysis (PCA), to reduce the size of the feature vector from $V$ to compressed embedding $CV$. The feature vector size and its compression are hyperparameters, which are set to $|V|=2^{16}$ and $|CV|=100$ in all the experiments. To ensure the efficient comparison between the apps, \textsf{Cypider} is empowered with a highly scalable similarity computation technique, based on FAISS \footnote{https://github.com/facebookresearch/faiss}, which computes the similarities between apps, as described in Section \ref{sec:lsh_similartiy}. 

Then, we compute the similarity between a pair of apps with respect to content \textit{feature vectors},  to decide if the two apps are connected or not in the similarity network.  The result of this similarity computation is an undirected network (or similarity network), where the nodes are Android apps, and the edges represent the high similarity between two apps for one content. For highly similar apps, multiple edges are expected to be formed between two apps. Besides, the more edges are formed, the more likely the apps malicious.

\textsf{Cypider} takes \textit{similarity network} as input in order to detect communities of malicious apps. However, this step mainly depends on the usage scenario (Section \ref{sec:usage}). In the case of a malicious dataset, \textsf{Cypider} extracts communities of highly connected apps and removes them out from the dataset. The rest of the apps, i.e., apps that do not belong to any community, will be considered in another \textsf{Cypider} malware detection iteration. We expect to get a pure or near-pure community if its apps belong to the same or almost the same Android malware family. In the case of a mixed dataset, \textsf{Cypider} first excludes all the nodes with degree $1$ from the similarity network, (i.e., the app is only self-similar), which are most likely to be benign apps. Afterward, \textsf{Cypider} extracts the communities of malicious apps.

The rest of the apps, which are not part of any community, will be processed in future \textsf{Cypider} iterations. At this point, we expect to get some communities of benign apps as \textit{false positives}. However, \textsf{Cypider}'s decision is explainable because, based on the similarity network, the security investigators can track for which contents the apps are similar.  The previous option could also be used to help in narrowing down the statistical features to prevent benign apps from being considered in malicious communities. As for community detection (Section \ref{sec:community_detection}), a highly scalable algorithm \cite{fast08blondel} is adopted to improve \textsf{Cypider}'s community detection module.
At this stage, we get a set of malicious communities, which are most likely to be a malware family or a subfamily. \textsf{Cypider} applies a one-class classifier \cite{Scholkopf:2001} on the malicious communities to produce a detection model, also called \textit{community fingerprint} (Section \ref{sec:community_fingerprint}). Instead of employing traditional crypto-hashing or fuzzy hashing of one malware instance, this fingerprint represents the pattern of a given detected community and captures the features of their apps. It is also used to decide if new malware apps are part of a family or not. The result is multiple \textit{community fingerprints}, each of which corresponds to a detected community. The generated fingerprints are stored in the \textit{signature database}, to be used later in analyzing new malware.

At this stage, \textsf{Cypider} will execute another detection iteration using a new dataset that also comprises the rest of the apps from the previous iteration, which has not been assigned to any community. The same previous steps will be followed in the new iteration. However, we need first to compare the \textit{feature vectors} of the new apps with the known \textit{malware communities fingerprint} that are stored in the database. The apps that are matched to a community fingerprint are labeled as malicious without adding them to the \textit{active dataset}. The unmatched apps are added to the \textit{active dataset} and are considered in the next iteration of the detection process. 

\textsf{Cypider} approach can be considered as an iterative process. In each iteration, \textsf{Cypider} detects and extracts communities from the \textit{active dataset} that continuously gets new apps (malware only or mixed) daily, in addition to the unassigned apps from the previous iterations. 

\section{Statistical Features} 
\label{sec:statistical_features}

In this section, we present the statistical features that are extracted from Android packaging (\textit{APK}) files, and which will be the input to the similarity computation process. As mentioned earlier, the features need to be broad enough to cover most of the static content of the APK file. The features could be classified according to the following APK content categories: (1) Binary features, which are derived from the byte-code (Dex file) of the Dalvik virtual machine. The Dex file and its hex dump file are also considered. (2) Assembly features, which are computed from the assembly of \textit{classes.dex}. (3) Manifest features, which are extracted from the Manifest file. (4) \textit{APK features}, which represent the rest of APK file contents, such as \textit{resources} and \textit{assets}. In this section, we present the statistical features, which are based on the N-gram concept. For more clarification, we first present the structure of Android packaging.

\subsection{Android APK Format} 
\label{sec:apk_format}

Android Application Package (\textit{APK}) is the official format for Android packaging, which is used to distribute and install  Android apps. It is similar to \textit{EXE} file in Windows and \textit{DEB}/\textit{RPM} file in Linux.  More specifically, \textit{APK} is a \textit{ZIP}  file that includes the required components to run the app.   \textit{APK} is structured into a set of directories, namely: \textbf{lib}, \textbf{res}, \textbf{assets} and files namely: \textbf{AndroidManifest.xml} and \textbf{classes.dex}. The purpose of these items is the following: i) \textbf{AndroidManifest.xml} includes the app meta-data, e.g., name, version, required permissions, and used libraries. ii) \textbf{classes.dex} includes the compiled classes of the Java code. iii) The \textbf{lib} directory includes C/C++ native libraries \cite{android_ndk}.  iv) The resources directory includes the non-source code files that are packaged into the \textit{APK} file during compilation. It mostly includes media files such as audio, image, and video files.

\subsection{N-grams} 
\label{sec:ngrams} 

The \textit{N-gram} is a technique found in Natural Language Processing (NLP), and is defined as the sequence of \textit{N} adjacent items that are extracted from a larger sequence. In this work, we use N-gram to extract the sequences that exist in Android malware to differentiate between the malware samples. We apply the N-gram technique on different contents of Android app package, such as \textit{classes.dex}, which captures the semantic pattern of the APK. For each APK content, we compute multiple feature vectors. Each vector \textit{V}   $\in D$  ($|D|= \Phi^N$ where $\Phi$ represents all the possible N-grams of a given  APK content). Each element from the vector \textit{V} represents the number of occurrences of a particular N-gram in APK content.

\subsection{The classes.dex Bytes N-grams}

To increase the amount of extracted information form \textit{classes.dex} file, we employ two types of N-gram: \textit{opcodes N-grams} and \textit{bytes N-grams}, which are computed from \textit{Dex} file and its assembly respectively. In addition, we compute \textit{Byted N-grams} from the \textit{hexdump} of the \textit{classes.dex} file,  by sliding a window of the hex string, where one field in that string is a byte, as depicted in Figure \ref{fig:ngrams}.

\subsection{Assembly opcodes N-grams}

\begin{figure}
  \centering
      \includegraphics[width=0.75\textwidth]{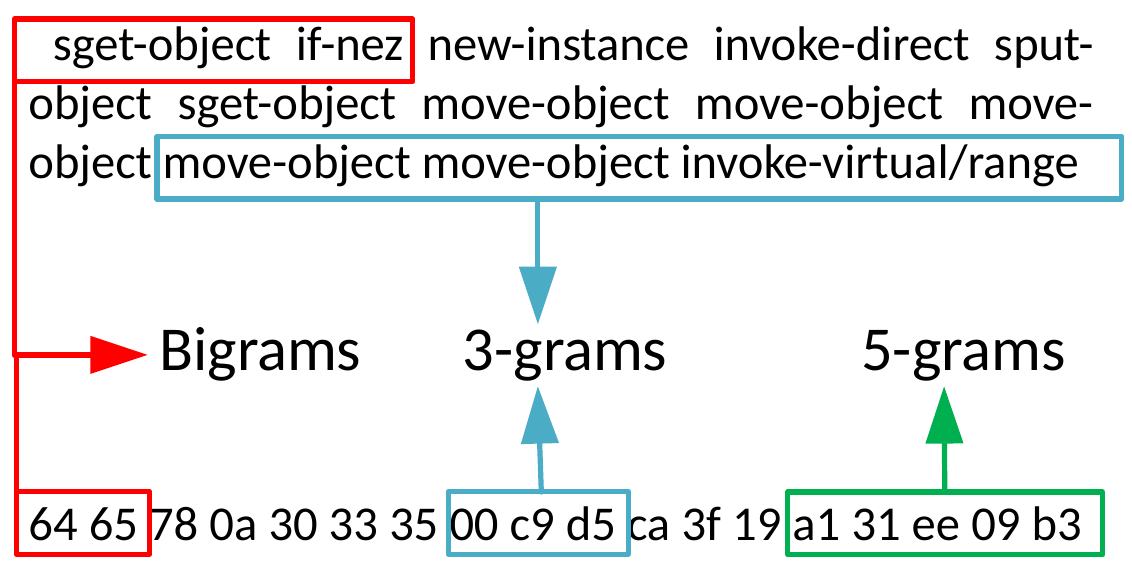}
  \caption{Opcodes \& Bytes From AnserverBot}
  \label{fig:ngrams}
\end{figure}

The opcode N-grams are the only sequences in the disassembly of \textit{classes.dex} file, where we can separate the instructions from their operands. Figure \ref{fig:ngrams} shows an example of this N-gram. We choose opcodes instead of the complete instruction for many reasons: i) The opcodes are more resilient to simple obfuscations that change some operands such as hard-coded IPs or URLs. ii) The opcodes could be more robust to changes made by malware writers who only modify or rename some operands when repackaging the Android app. iii) Also,  the extracted opcodes incur low computational cost compared to the complete instruction.

The information obtained from the opcode N-gram could be more meaningful by only considering functions that use sensitive APIs such as SMS API. We can also decrease the noise from the N-gram information by excluding the most common opcode sequence. The  N-gram extraction process will result in a list of unique N-grams along with their number of occurrence for each content category, i.e., \textit{opcode instructions}, \textit{classes.dex}. Figure \ref{fig:ngrams} depicts the extracted N-grams from the instructions and the bytes of the first portion of the \textbf{AnserverBot} malware. Besides, we consider the names of classes and methods as assembly features.

\subsection{Native Library N-grams}

The \textit{Native Library} identifies the C/C++ shared libraries \cite{android_ndk} that are used by malware.  Native library is important in some cases to differentiate  between two Android malware samples. For instance, in case of \textbf{DroidKungFu} malware family, the malware is more likely to be \textbf{DroidKungFu2} and not  \textbf{DroidKungFu1}, as  \textbf{DroidKungFu2} uses C/C++ library and \textbf{DroidKungFu1} only uses \textit{Java byte code}.

\subsection{APK N-grams}

The N-gram of the APK file allows us to get an overview of the APK file semantics. For instance, most of the repackaged apps are built by making minor modifications to original apps  \cite{deshotels2014droidlegacy}. Therefore, by applying N-gram computation on the APK file, we can detect a high similarity between the original app and the repackaged one. Besides, adversaries preserve some components in the APK file, e.g., images and GUI layout structures, primarily if they aim to build a phishing malware. In this case, both apps are visually similar, and consequently, their N-gram sequences are similar with respect to the resource directory.

\subsection{Manifest File Features}

\textit{AndroidManifest.xml} file, which contains the \textit{permissions} that are requested by apps provides important features that could help in detecting malicious apps. For example, apps that request \textit{SMS send} permission are more suspicious, as an important percentage of Android malware apps sends SMS to premium phone numbers. In addition, other features from \textit{AndroidManifest.xml} are extracted namely, \textit{services}, \textit{activities}, and \textit{receivers}.

\subsection{Android API Calls} 

The request permissions give a global view of the possible behavior of an app.  However, a more granular view could be obtained by tracking the \textit{Android API calls}, as single permission allows access to multiple \textit{API calls}. Therefore, we add the list of \textit{API calls} list, which is used by the apps, to the feature list. In addition, we focus more on the \textit{suspicious APIs}, such as \textit{sendTextMessage()} and \textit{orphan APIs}, which correspond to  undeclared permissions. We also extract the list of permissions that have no used \textit{APIs}  in the app.

\subsection{Resources}

In this category, we extract features that are related to the \textit{APK} resources, such as text string, file names, and their content. We mainly extract the files that do not include the names of standard files, e.g., \textit{String.xml}. Also, we consider the contents of files by computing \textbf{md4} hashes on each resource file, as it incurs low computation cost compared to more modern cryptographic hashing algorithms such as MD5 and SHA1.  Therefore, the scalability of the system is enhanced. Finally,  we apply text string selection on the text resources by leveraging \textit{tf-idf} (term frequency-inverse document frequency) \cite{itidf_wiki} technique for this purpose.

\subsection{APK Content Types}

Table \ref{tab:stat_features} summarizes the categories of the proposed features based on APK contents of the app. It also shows the features that are considered in the current implementation of \textsf{Cypider}. We believe that the used features give a more accurate representation of Android packages as shown in previous work \cite{karbab2016dna}. In the other hands, The features that we excluded such as {\it Text Strings} and {\it Assembly Class Names} are highly vulnerable to common obfuscation techniques. Also, excluded features such as {\it Manifest Receivers} generate very sparse features vectors which effect the overall accuracy.

\begin{table}
\centering
\begin{tabular}{|c||l|c|}
\hline
\hline
\# &   \textbf{Content Type Features} & \textbf{Implemented Feature} \\ 
\hline
\hline
0  & APK Byte N-grams             & X   \\ \hline
1  & Classes.dex Byte N-grams     & X   \\ \hline
2  & Native Library Bytes N-grams & X   \\ \hline
3  & Assembly Opcodes Ngrams      & X   \\ \hline
4  & Assembly Class Names         &     \\ \hline
5  & Assembly Method Names        &     \\ \hline
6  & Android API                  & X   \\ \hline
7  & Orphan Android API           &     \\ \hline
8  & Manifest Permissions         & X   \\ \hline
9  & Manifest Activities          & X   \\ \hline
10 & Manifest Services            & X   \\ \hline
11 & Manifest Receivers           &     \\ \hline
12 & IPs and URLs                 & X   \\ \hline
13 & APK Files names              & X   \\ \hline
14 & APK File light hashes (md4)  &     \\ \hline
15 & Text Strings                 &     \\ 
\hline
\hline
\end{tabular}
\caption{Content feature categories}
\label{tab:stat_features}
\end{table}

\section{Feature Preprocessing} \label{sec:feture_preprocessing}

The proposed framework is based mainly on two atomic operations, which are Feature extraction and similarity computation. Hence, it is of paramount importance to optimize both their design and implementation to achieve the intended scalability. The \textit{feature processing} operation is expected to output a feature vector, which can be used to compute the similarity between apps in a straightforward manner. The produced vectors will be used as input to the \textsf{Cypider} community detection system. One of the drawbacks of the N-gram technique (Section \ref{sec:ngrams}) is its very high dimensionality $D$. The computation and the memory needed by \textsf{Cypider} for Android malware detection depend dramatically on the dimension number $D$. Moreover, the complexity of computing the extracted N-grams features increases exponentially with $N$. For example, for the \textit{opcodes N-grams}, described in Section \ref{sec:ngrams}, the dimension $D$ equals to $ R^2 $ for bi-grams, where $R=200$, the number of possible opcodes in Dalvik VM.  Similarly, for \textit{3-grams}, the dimension $D= R^3$; for \textit{4-grams}, $D= R^4$. Furthermore, $N$ has to be at least  3 or 5 to capture the semantics of some Android APK content.

\begin{algorithm}
    \SetKwInOut{Input}{input}
    \SetKwInOut{Output}{output}

    \Input{\textbf{Content Features}: Set, \\
           \textbf{L}: Feature Vector Length}
    \Output{Feature Vector}

    vector = \textbf{new} vector[\textbf{L}]\;
    \For{\textbf{Item} in \textbf{Content Features}}{
        H = hash(\textbf{Item}) \;
        feature\_index = H mod \textbf{L} \;
        vector[feature\_index] = vector[feature\_index] + 1 \;
    }
 \caption{Feature Vector Computation}
 \label{algo:feature_hashing}
\end{algorithm}

To reduce the high dimensionality of an arbitrary vector, we leverage the \textit{hashing trick} technique \cite{qinfeng09hashk}, to output a fixed-size feature vector. More formally, \textit{hashing trick} reduces a vector $V$ with $D=R^N$ to a compressed version with $D=R^M$, where $M<<N$. The compressed vector allows \textsf{Cypider}'s clustering system to handle a large volume of Android apps, which boosts \textsf{Cypider} in terms of both computation and memory. As shown in previous work, \cite{Weinbergeretal09, qinfeng09hashk}, the hash kernel approximately preserves the vector distance. Moreover, the hashing technique used to reduce dimensionality affects the computational cost, which linearly increases with the number of samples and groups. Algorithm \ref{algo:feature_hashing} illustrates the overall process of computing the compacted feature vector from an N-grams set. Furthermore, it helps to control the length of the compressed vector in an associated feature space.

Despite getting a fixed-length vector using the feature hashing technique, the vector size is still significant (size=$2^{16}$) to leverage very fast approximate similarity computation \cite{NIPS2015_5893}. To overcome this issue, one could suggest decreasing the size of the feature hashing vector. However, due to hashing collision, this will drastically reduce the quality of the produced vector \cite{NIPS2015_5893},  which is not useful for malware fingerprinting \cite{Hu13MutantX}.  Therefore, we resort to dimensionality reduction techniques to compress the feature hashing vectors while keeping the maximum of information. To this end, we aim to reduce the size of the feature hashing vector $V$ from $2^{16}$ to only compressed embedding $CV$ of size $S={100}$.

To this point, we have a sizeable discrete vector obtained from one of the previous setups, mainly using feature hashing. The aim is to decrease its dimensionality by using dimensionality reduction techniques. To this end, we choose the used PCA (principal component analysis) in the current implementation of feature hashing due to its simplicity. We could also leverage other techniques, and consider the comparison between them regarding information preservation as future work. We apply PCA to choose the top $S=100$ important PCA components which render our digest. We use PCA on the matrix, where rows are the discrete vectors of the samples.

Compressing the feature hashing vector using dimensionality reduction (PCA) will economize the space ($\approx x650$ times less memory and disk space) and boost the similarity computation time ($\approx x700$ as will be presented in the next sections). However, this raises two concerns, namely portability and computation time. Regarding portability, the question is whether the produced digest for a given sample is reusable. This means that we could compute the digest using a matrix from dataset $A$ from epoch $T_i$ and use it to calculate the similarity with another digest generated using a matrix from dataset $B$ from epoch $T_j$. Our experiments confirm the portability of our digest across different datasets' epochs. 

As for the computation concern, we question the scalability of PCA (using SVD implementation) to large datasets since the asymptotic complexity exceeds cubic in the best cases. To solve this issue, we compute the principal components of PCA, using only a fraction of the dataset. Afterward, we employ those components to compress the feature hashing of the dataset. It is essential to mention that the asymptotic complexity is constant for the computation of the components in fixed-size sampled datasets ($\approx 1k$ records from the input dataset) because of the fixed size. Also, using the computed principal components on the whole input dataset has linear growth with the size of the input dataset; also, it is speedy (matrix multiplication). As such, we employ this technique in all our operations. To this end, we have a set of compressed digests for each content type, i.e., a digest for each malware sample.

\section{Similarity Computation}
\label{sec:lsh_similartiy}

The backbone of \textsf{Cypider} is the \textit{similarity network}, which is obtained by computing the pair-wise similarity between each feature vector of the apps APKs. Based on the number of these content vectors, we collect multiple similarities, which gives flexibility and modularity to \textsf{Cypider}. Any new feature vector could be added to the similarity network without disturbing the \textsf{Cypider} process. Moreover, the features could be smoothly removed, making the process of selecting the best features more convenient. \textit{This is of crucial importance to investigators. Having multiple similarities between apps static contents in the similarity network allows them to make explainable decisions since they can track contents leading to apps similarity in the final similarity network}. 

It is important to conduct similarity computation in an efficient way that is much faster than brute-force computation. To this end, in the original \textsf{Cypider} design, we leverage \textit{Locality Sensitive Hashing} (LSH) techniques, in particular \textit{LSH Forest} \cite{lshforest05bawa}. The latter is a tunable high-performance algorithm used in similarity computation of \textsf{Cypider}.  In \textit{LSH Forest}, similar items hashed using LSH are most likely to be in the same bucket (collide) and dissimilar items in different ones. This property of \textit{LSH} function could be achieved using many similarity measures. In our case, this is achieved using the well-known \textit{Euclidean} distance.

\begin{figure}
  \centering
  \includegraphics[width=0.9\textwidth]
  {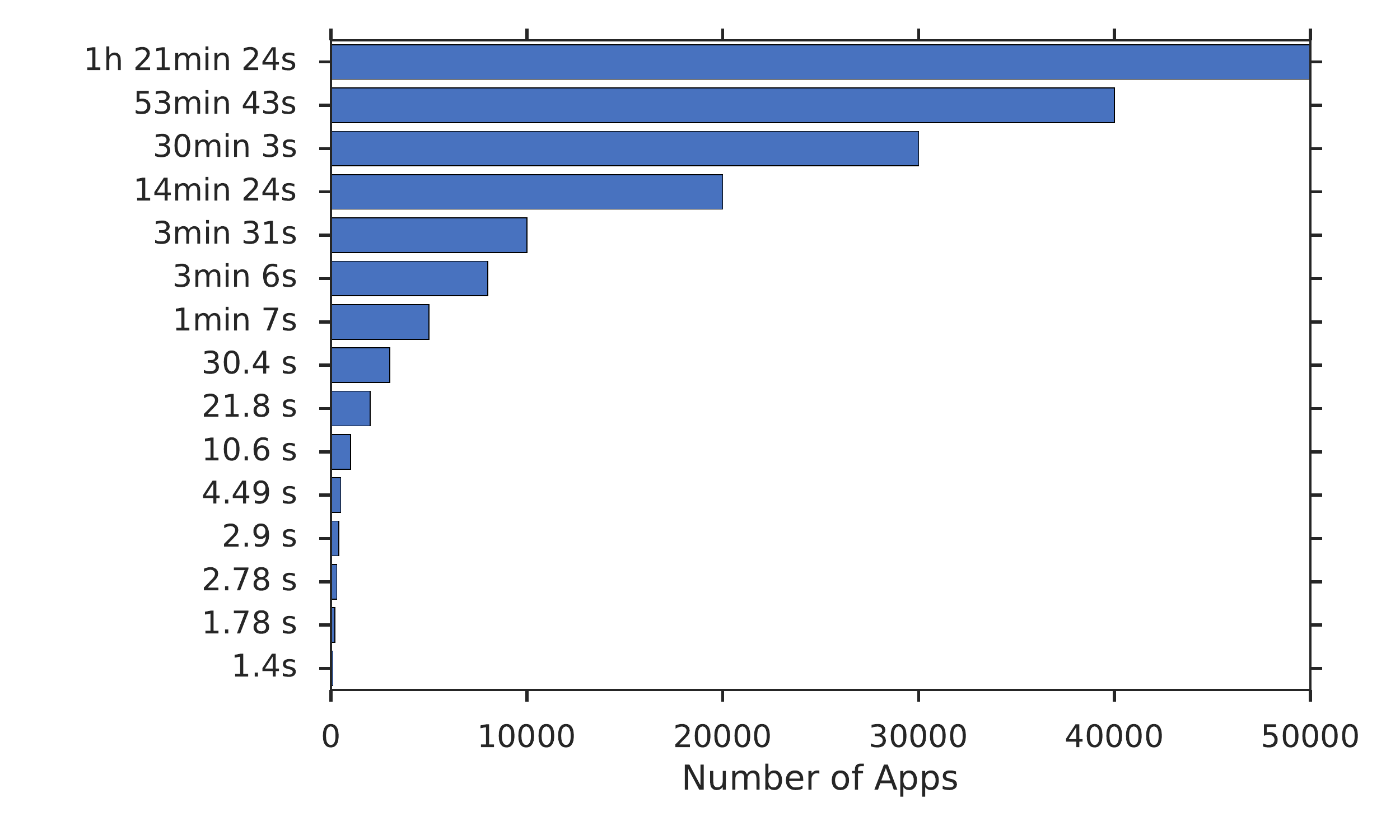}
  \caption{Sparse Vector Similarity Computation Time}
  \label{fig:lsh_compute_time}
\end{figure}

\begin{equation}
 d(m, n) = \|V_m - V_n\|  = \sqrt{\sum_{|i=1|}^{|S|} (V_m(i) - V_n(i))^2}
 \label{equ:euclidean_computation}
\end{equation}

As depicted in Formula \ref{equ:euclidean_computation}, after extracting the content feature vectors \textit{m} and \textit{n} of one APK content from a given pair of Android apps, we use the \textit{Euclidean} distance to compute the distance between the two feature vectors \textit{m} and \textit{n}. Figure \ref{fig:lsh_compute_time} depicts the LSH computational time with respect to the number of apps using \textit{one CPU core} and \textit{one thread} for the permission feature vector. It is worth noticing that the current performance using \textit{LSH Forest} is acceptable even for a large number of daily malware samples (reaching $40k$ apps per hour). In the enhanced \textsf{Cypider}, these results are drastically improved thanks to using compressed embedding (size=$100$) instead of feature hashing vectors (size=$2^{16}$). 

\begin{figure}
  \centering
  \includegraphics[width=0.9\textwidth]
  {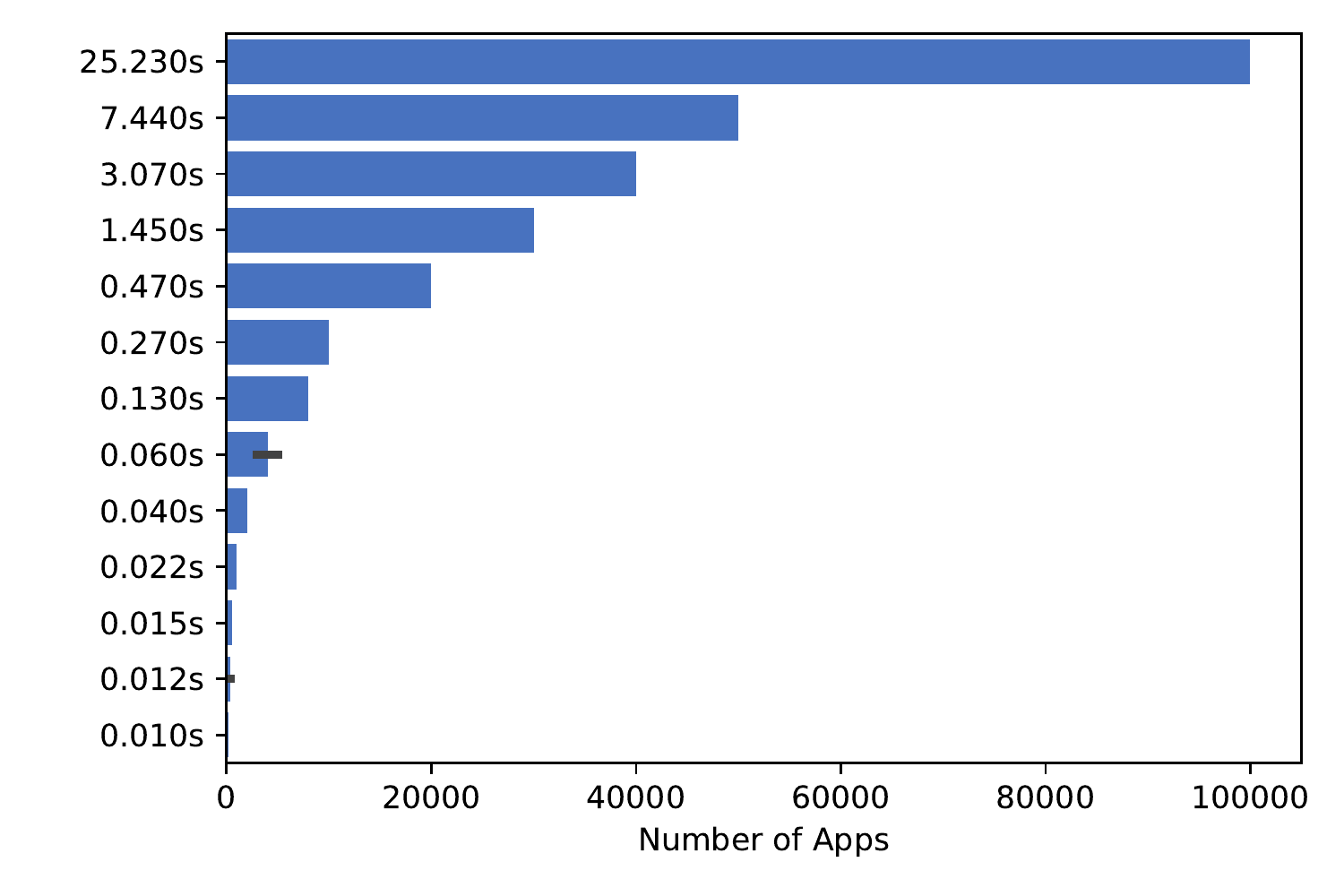}
  \caption{Embedding Similarity Computation Time}
  \label{fig:Newlsh_compute_time}
\end{figure}

Furthermore, instead of \textit{LSH Forest}, we could use fuzzy similarity techniques, such as FAISS \footnote{https://github.com/facebookresearch/faiss} implementation, which exploits all the CPU cores in addition to multi-threading. Figure \ref{fig:Newlsh_compute_time} presents the new similarity computation time (in seconds) after using the compressed embedding and FAISS library. \textsf{Cypider} is $\approx x700$ times faster compared to the previous setup on commodity hardware (Intel(R) Xeon(R) CPU E5-2630 v3 @ 2.40GHz). In the previous setup, \textsf{Cypider}  took $\approx 80$ minutes to compute the similarity between $50k$ apps on the \textbf{permission content vectors}. In contrast,  \textsf{Cypider}  takes only $\approx 8$ seconds to compute the similarity in the current setup.

Afterward, we build the similarity connections between the app feature vectors if the pairwise similarity exceeds a certain threshold, named \textbf{similarity threshold}. We propose a systematic mechanism to choose the \textbf{similarity threshold} for different content vectors (opcodes, functions, and binary bytes), which have different search spaces. The process of choosing the threshold is: (1) For a given vector content, we compute the similarity between all the dataset digests. (2) We calculate the average of the similarity values for this content vector. (3) To this point, we use a percentage from the computed average as the threshold. The threshold percentage is fixed and applied to the different averages of content types.  Therefore, exploring and searching for a suitable threshold needs only to tune the value of the threshold percentage to fulfill the security practitioner requirements toward the trade-off between purity and coverage.

In summary, the \textbf{similarity threshold} is the percentage of the \textbf{average similarity} of a given content. In other words, we compute the average value of all the pairwise similarities for each feature content.  Then, we fix a percentage from this average to be the final threshold.  Finally, we use the same percentage of \textbf{similarity threshold} for all the feature contents despite having different average values. The \textbf{similarity threshold} is manually fixed based on our evaluation, and the same threshold is used in all experiments. We investigate the effect of the \textbf{similarity threshold} on \textsf{Cypider} performance in the evaluation section (Section \ref{sec:evaluation}). The final result of the similarity computation is a \textit{heterogeneous network}, where the nodes are the apps, and the edges represent the similarity between apps if it exceeds a certain threshold.

\section{Community Detection}
\label{sec:community_detection}

A scalable community detection algorithm is essential to extract \textit{suspicious communities}. As such, we resort to the \textit{Fast unfolding community detection} algorithm \cite{fast08blondel}, which can scale to billions of network links. The algorithm yields excellent results by measuring the \textit{modularity} of communities. The latter is a scalar value $M \in [-1, 1]$ that measures the density of edges inside a given community compared to the edges between communities. The algorithm uses an approximation of the \textit{modularity} since finding the exact value is computationally hard \cite{fast08blondel}. The previous algorithm requires a homogeneous network as input to work properly.

Accordingly, we leverage a \textit{majority-voting} mechanism to homogenize the heterogeneous network generated by the similarity computation. Given the number of content similarity links $s$, the \textit{majority-voting} method decides whether a pair of apps are similar or not by computing the ratio $s/S$, where $S$ is the number of all contents used in the current \textsf{Cypider} configuration. If the ratio is above the average, the apps will only have one link in the \textit{similarity network}. Otherwise, all the links will be removed. 

It is important to mention that content similarity links could be kept for later use, such as to conduct thorough investigations about given apps.  Investigators could use them, for example, to figure out how similar the apps are, and in which content they are similar. The prior use case could be of great importance to security analysts.

\begin{figure}
  \centering
  \includegraphics[width=0.85\textwidth]
  {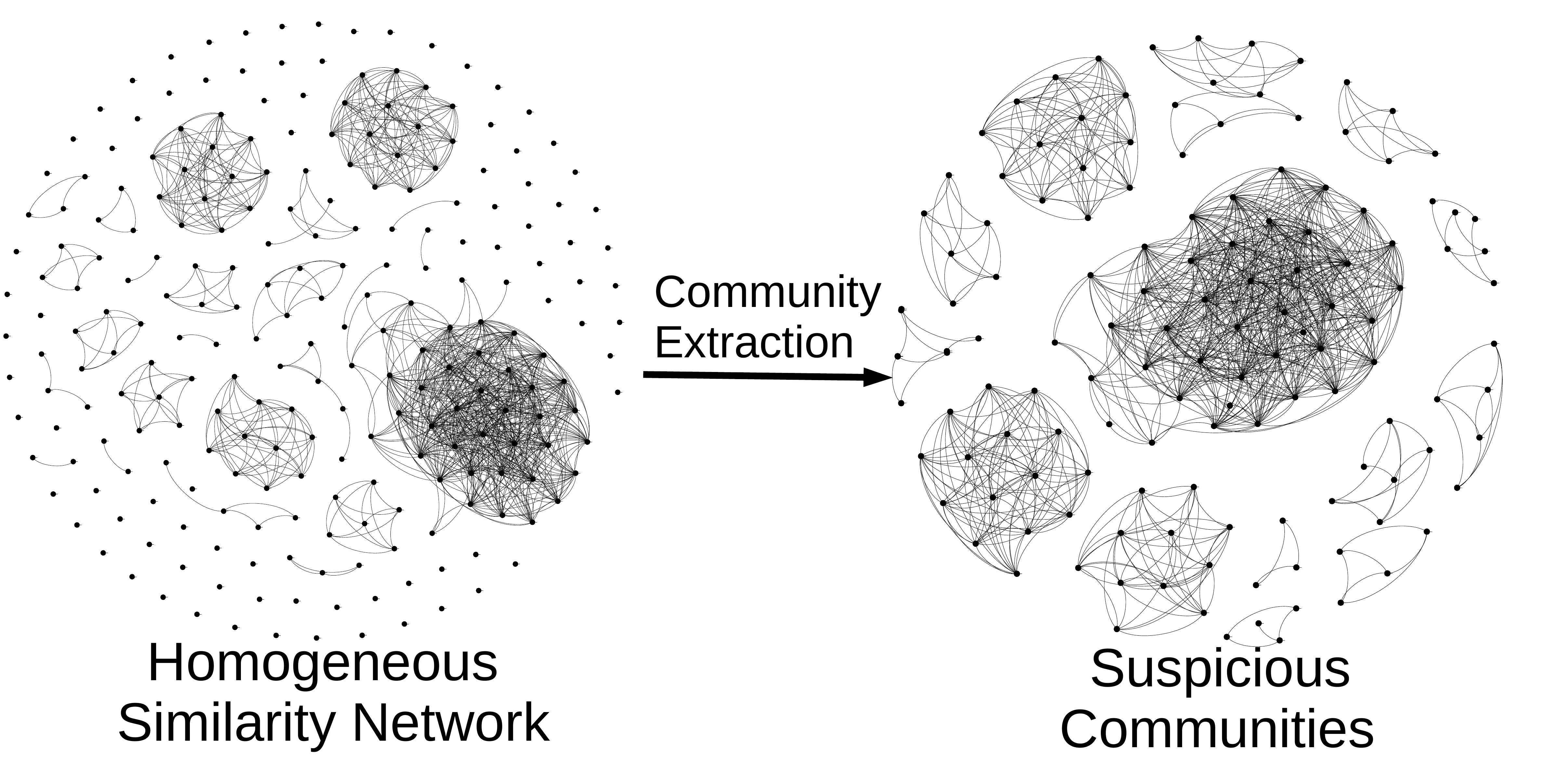}
  \caption{Applying \textsf{Cypider} On a Small Dataset}
  \label{fig:toy_network}
\end{figure}

To prevent getting inaccurate \textit{suspicious communities}, we propose to leverage the \textit{majority-voting} mechanism by filtering the links that exist between the nodes (apps). More precisely, we filter all links with a degree that is less than a \textit{degree filtering} parameter. This hyperparameter keeps only edges of a given node whose number is above the threshold. We name this hyperparameter a \textbf{content threshold} and is defined as the minimum number of similar contents needed to keep a link in the final similarity network.

As mentioned above, only nodes with high connectivity maintain their edges, which are supposedly like similar malicious apps. Notice that all the parameters have been fixed in our evaluations. In a \textit{mixed dataset} scenario, we use the degree $1$ to filter all apps having a similarity link to themselves since they are not similar to any other app in the \textit{active dataset}. \textsf{Cypider} filters these apps and consider them as benign.  \textsf{Cypider} also extracts a set of communities with different sizes using the community detection algorithm \cite{fast08blondel}. Afterward, \textsf{Cypider} filters all communities whose community cardinal is less than the \textit{minimum community size} parameter (fixed for all the evaluations). As previously mentioned, the purpose of filtration is to discard inaccurate communities. This is illustrated in Figure \ref{algo:feature_hashing} on a small Android dataset (250 malware apps), where the process of the community detection starts with a \textit{homogeneous network} and ends up with \textit{suspicious communities}. The \textbf{content threshold} and \textbf{community size} hyper parameters are thoroughly investigated in the evaluation section (Section \ref{sec:evaluation}).

\section{Community Fingerprint} 
\label{sec:community_fingerprint}

In addition to detecting malicious communities, \textsf{Cypider} also aims at generating fingerprints from the extracted communities. This could be done automatically since \textsf{Cypider} tends to be completely unsupervised.  Therefore, in the following iteration, \textsf{Cypider} filters the known apps without adding them to the \textit{active dataset}. In particular, \textsf{Cypider}  aims to generate a fuzzy fingerprint, not only for one app but also for the whole malicious community that represents malware family or subfamily. Unfortunately, traditional \textit{cryptography fingerprints} or \textit{fuzzy hashing} techniques are not suitable for this purpose. Therefore, we propose a novel fingerprinting technique based on the \textit{One-Class Support Vector Machine} learning model (OC-SVM) \cite{Scholkopf:2001}. The latter could be fuzzy enough to be able to detect a set of malicious apps, in particular, the \textit{soft boundary model}, of a given community. Thanks to the \textit{one-class model}, \textsf{Cypider} can classify new apps as belonging to a community or not.  

It is also important to mention that the signature database generated by \textsf{Cypider} is much more compressed compared to the traditional methods, where the signature is only for one malicious app. Besides, the computation of the signature is highly reduced since we check only with the community fingerprints instead of checking with each single malware hash signature. 

\textsf{Cypider} generates a \textit{community fingerprint} from a set of malicious apps as follows. First, \textsf{Cypider} extracts the features, as presented in Section \ref{sec:statistical_features}. Afterwards, it trains the \textit{one-class model} using the statistical features of the malicious community apps. In order to improve the accuracy of the community fingerprint (model), \textsf{Cypider} applies \textit{feature selection} techniques , as described in the following sub-section.


A large number of Android APK features may potentially impact the accuracy of the generated community fingerprint. To overcome this issue, \textsf{Cypider} only uses the best $N$ features with the highest information gains and discards irrelevant ones to reduce the overhead of unnecessary computational complexity. The selection of  these valuable features is based on two metrics, namely \textit{variance threshold} and \textit{inverse document frequency (IDF)}. After computing the \textit{variance} of the Android apps features, we use a fixed threshold of $T$ to filter the features with a  \textit{variance} that is lower than $T$. Notice that a threshold $T=0$ results in discarding all the features with the same value for all Android apps, which makes sense because the information gain from a fixed number for all the community apps is zero.

\begin{equation}
 idf(t)  = log{ |F| \over {1 + |\{f_{apk}: t \in f_{apk}\}|}}
 \label{equ:idf_computation}
\end{equation}

Furthermore, we compute the \textit{inverse document frequency (IDF)} on the list $F=\{ f_{apk1}, ...,f_{apkN}\}$, where $f_{apk}$ is the set of Android APK features and  $|F|$ is the number of input APKs. The \textit{idf(t)} is calculated using  Formula \ref{equ:idf_computation}, where $|\{f_{apk}: t \in f_{apk}\}|$ is the number of Android APKs having a feature value $t$. Notice that the number 1 in the formula is added to avoid zero-division. The computed $idf(t)$ is used as an alternative to frequencies, which exhibit less effectiveness in similar solutions.
\section{Experimental Results} 
\label{sec:evaluation}

In this section, we start with an overview of the implementation and testing setup, including the dataset and the performance measurement techniques.  Afterward, we present the achieved results regarding the defined metrics for both usage scenarios that are adopted in \textsf{Cypider} framework, namely \textit{malware only and mixed} datasets.  

\subsection{Implementation} 
\label{sec:implementation}

\textsf{Cypider} is implemented using \textit{Python} programming language and \textit{bash} command-line tools. The generation of binary N-grams is performed using \textit{xxd} tool to convert the content of the package to a sequence of bytes, in addition to a set of command tools such as \textit{awk} and \textit{grep} to filter the results. To apply reverse engineering on the \textit{Dex} byte-code, we use \textit{dexdump}, a tool that is provided by Android SDK. We filter the generated assembly using the standard Unix tools.  To extract the permissions from \textit{AndroidManifest.xml}, we use \textit{aapt}, an Android SDK tool, to convert the binary XML to a readable format. Then, we parse the produced XML file using the standard Python XML parsing library.  The efficiency of \textsf{Cypider} is evaluated under a commodity hardware server (Intel(R) Xeon(R) CPU E5-2630, 2.6GHz). 

\subsection{Dataset and Test Setup} 
\label{sec:dataset}

In order to evaluate \textsf{Cypider},  three well-known Android datasets are used, namely, (i) MalGenome malware dataset \cite{genome_malware} \cite{zhou2012dissecting}, (ii) Drebin malware dataset \cite{Drebin_Dataset} \cite{arp2014drebin,spreitzenbarth2013mobile}, and AndroZoo public Android app repository. As presented in Table \ref{tab:dataset_describe}, two additional datasets are built based on the previous ones by adding Android apps, which were randomly downloaded from Google Play between late 2014 and beginning of 2015. In order to build \textit{Drebin Mixed} and \textit{AndroZoo Mixed} datasets, $4,403$ benign apps are added to the original \textit{Drebin} dataset, which results is a mixed dataset (malware \& benign) with $50\%$ of apps in each category. In the same way, \textit{Genome Mixed} dataset is built with $75\%$ of benign apps (the \textit{mixed dataset} will be publicly available for the research community).

\begin{table}
\centering
\begin{tabular}{c|ccc}
    \hline
             & \textit{Drebin} & \textit{Genome} & \textit{AndroZoo} \\
    \hline
    Size     & 8733            & 4239            & 110k              \\
    Malware  & 4330            & 1168            & 66k               \\
    Benign   & 4403            & 3071            & 44k               \\
    Families & 46              & 14              &  /                \\
    \hline
\end{tabular}
\caption{Evaluation Datasets}
\label{tab:dataset_describe}
\end{table}

The three datasets mentioned above are used to evaluate \textsf{Cypider} in the unsupervised usage evaluation scenarios and by considering two cases: with and without benign apps, as shown in Section \ref{sec:usage}. First, \textsf{Cypider}  is evaluated against malware samples only. To this end, \textit{Drebin}, \textit{AndroZoo}, and  \textit{Genome} datasets were used. This use case is the most attractive one in bulk malware analysis since it decreases the cost and detection time of malware by considering only a sample from each detected community. Second, \textit{Cypider} is evaluated against mixed datasets. The second scenario is more challenging as we might not only  get the \textit{suspicious communities} as output but also benign communities (false positives) are expected along with filtered benign apps. 

To asses \textsf{Cypider} obfuscation resiliency, we conduct the evaluation on PRAGuard obfuscation dataset\footnote{http://pralab.diee.unica.it/en/AndroidPRAGuardDataset}, which contains $11k$ obfuscated malicious apps, generated using common obfuscation techniques \cite{Davide2015Praguard}. Besides, we generate $100k$ benign and malware obfuscated apps using the DroidChameleon obfuscation tool \cite{Vaibhav2013DroidChameleon}, which employs common and combinations of obfuscations techniques.

We define various metrics to measure \textsf{Cypider} performance under each dataset. The used metrics are the following: 

\paragraph{Apps Detection Metrics}
\begin{enumerate}[label=A\arabic*:]

\item \textit{True Malware}: It measures the number of malicious apps that are detected by \textsf{Cypider}. It is applied to both usage evaluation scenarios.

\item \textit{False Malware}: It measures the number of benign apps that are falsely detected as malware apps. It is applied only on the \textit{mixed dataset} as there are no benign apps in the other datasets.

\item \textit{True Benign}: It measures the number of filtered benign apps. It is only applied on \textit{mixed datasets}.

\item \textit{False Benign}:  It measures the number of malware apps that are incorrectly considered as benign in the \textit{mixed dataset}.

\item \textit{Detection Coverage}: It measures the percentage of detected malware from the overall dataset, i.e., it represents the number of Android apps, which are assigned to communities, divided by the total number of apps in the input dataset.

\end{enumerate}

\paragraph{Community Detection Metrics}
\begin{enumerate}[label=C\arabic*:]

\item \textit{Detected Communities}: It measures the number of suspicious communities that are extracted by \textsf{Cypider}.

\item \textit{Pure Detected Communities}: It measures the number of communities, which only contain instances of the same Android malware family.  To check the purity of a given community,  we refer to the labels of the used datasets. This metric is applied in both usage evaluation scenarios.

\item \textit{K-Mixed Communities}: It measures the number of communities with K-mixed malware families, where $K$ is the number of families in a detected community. This metric is applied in both usage evaluation scenarios.

\item \textit{Benign Communities}: It measures the number of benign communities that are falsely detected as suspicious. This metric is used in the \textit{mixed dataset} evaluation.

\end{enumerate}

\subsection{Mixed Dataset Results}

The evaluation results of applying \textsf{Cypider} on  \textit{Drebin Mixed} and \textit{Genome Mixed} datasets are presented in Table \ref{tab:mixed_apps_metrics}.  We can notice that \textsf{Cypider} succeeded in detecting \textit{half} of the actual malware in one single iteration and under both datasets, although the noise of benign apps is about $50\%-75\%$ of the actual dataset. Also, \textsf{Cypider} was able to filter out a large number of benign apps from the dataset. However, in both datasets, we got a \textit{false malware} of $190$ and $103$ apps under  \textit{Drebin Mixed} and \textit{Genome Mixed} datasets respectively. We also obtained a \textit{false benign} of $38$ and $10$ apps under the two above-mentioned datasets respectively. We can observe that the \textit{false positives,} and \textit{false negatives} appear, in most cases, in communities of apps with the same category (malware or benign).  Therefore, it is possible to carry out a straightforward investigation by analyzing some samples from a given suspicious community. Figure \ref{fig:net_drebin_mixed_raw} and Figure \ref{fig:net_drebin_mixed_clustering} show the similarity network and the generated communities respectively.

\begin{table}
\centering
\begin{tabular}{c|cc}
    \hline
     Community Metrics         & \textbf{Drebin Mixed} & \textbf{Genome Mixed} \\
    \hline
     \textbf{True Malware A1}  & 2413                  & 449  \\
     \textbf{False Malware A2} & 190                   & 103  \\
     \textbf{True Benign A3}   & 257                   & 171  \\
     \textbf{False Benign A4}  & 38                    & 10   \\
    \hline
\end{tabular}
\caption{Evaluation Using Apps Metrics (Mixed)}
\label{tab:mixed_apps_metrics}
\end{table}

Table \ref{tab:mixed_communtiy_metrics} shows the evaluation results of \textsf{Cypider} in terms of \textit{community metrics}. The main result is that \textsf{Cypider} succeeds in extracting $179$  (resp., $61$) \textit{pure detected communities} out of $188$ (resp., $61$) in case of \textit{Mixed Drebin} (resp., \textit{Mixed Genome}).  As a consequence, almost all the detected communities are composed of instances from the same family, whether malware or benign. Even the detected \textit{mixed communities} are only composed of two labels (2-mixed). It is worth mentioning that all the \textit{detected benign communities} are pure, i.e., there is no malware instance, which makes the investigation easier for security analysts. Also, according to our analysis, most malware labels in the \textit{2-mixed} malicious communities refer to the same malware but with different names, which is due to the naming convention that is made by each security vendor. For example, in one of the \textit{2-mixed} communities, we found instances of two types of malware: \textit{FakeInstaller} and \textit{Opfake}. These two names refer to the same malware family \cite{opFake_android_malware}, which is \textit{FakeInstaller}. Also, we found that \textit{FakeInstaller} and \textit{TrojanSMS.Boxer.AQ} refer to the same malware \cite{boxer_android_malware}, although they have different vendor names.

\begin{table}
\centering
\begin{tabular}{c|cc}
    \hline
     Apps Metrics & \textbf{Drebin Mixed} & \textbf{Genome Mixed} \\
    \hline
     \textbf{Detected C1}      & 188 & 61   \\
     \textbf{Pure Detected C2} & 179 & 61   \\
     \textbf{2-Mixed C3}       & 9   & 0    \\
     \textbf{Benign C4}        & 18  & 16   \\
    \hline
\end{tabular}
\caption{Evaluation Using Community Metrics}
\label{tab:mixed_communtiy_metrics}
\end{table}

\begin{table}
\centering
\begin{tabular}{c|cc}
    \hline
     Community Metrics & \textbf{Drebin} & \textbf{Genome} \\
    \hline
     \textbf{True Malware A1} & 2223 &   449 \\
    \hline
\end{tabular}
\caption{Evaluation Using Apps Metrics (Malware)}
\label{tab:only_malware_apps_metrics}
\end{table}

\begin{table}
\centering
\begin{tabular}{c|cc}
    \hline
     Apps Metrics & \textbf{Drebin} & \textbf{Genome} \\
    \hline
     \textbf{Detected C1}      & 170 & 45   \\
     \textbf{Pure Detected C2} & 161 & 45   \\
     \textbf{2-Mixed C3}       & 9   & 0    \\
    \hline
\end{tabular}
\caption{Evaluation Using Community Metrics}
\label{tab:only_malware_communtiy_metrics}
\end{table}

\subsection{Results of Malware-only Datasets}

Tables \ref{tab:only_malware_apps_metrics} and \ref{tab:only_malware_communtiy_metrics} show the results of \textsf{Cypider} in terms of the \textit{app metrics} and \textit{community metrics}, and under malware only datasets. As the \textit{mixed dataset} is used, i.e., by only filtering out the benign apps, we achieved almost similar results. \textsf{Cypider} succeeds in detecting about $50\%$ of all malware in one iteration. Besides, nearly all the obtained communities are pure. This result shows that \textsf{Cypider} can provide the security analyst with the necessary tool that can analyze the maliciousness of a given suspicious community by only matching one or two samples. Moreover, the analysis process complexity is significantly reduced from $2413$ detected malware to only $188$ found communities. Based on this result, we believe that the analysis window could be reduced, and it is possible to overcome the overwhelming number of detected Android malware daily. We can also observe that nine \textit{2-mixed} communities are extracted from the \textit{Drebin dataset}, which contain malware with different names but refer to the same family, as obtained previously. The \textit{similarity network} of the Drebin malware dataset is shown in Figure \ref{fig:net_drebin_only_raw}. After executing the community detection algorithm, a set of \textit{malicious communities} are identified, as depicted in Figure \ref{fig:net_drebin_only_clustering}.

\begin{scriptsize}
\begin{figure}
     \begin{center}        
        \subfigure[Simialrity Network]{%
            \includegraphics[width=0.70\textwidth, trim=0.0cm 0.0cm 0.0cm 0.0cm, clip]
            {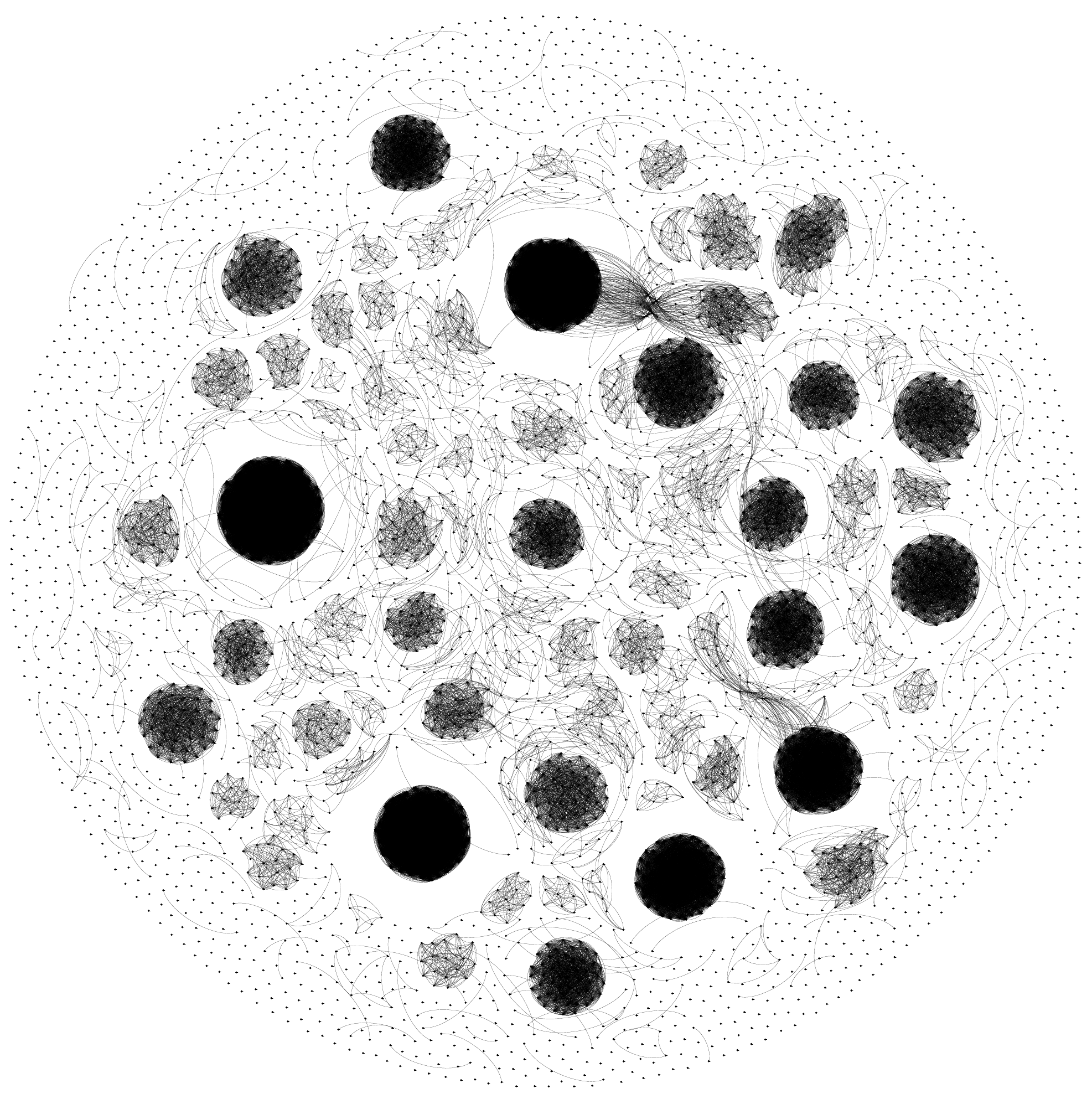}
            \label{fig:net_drebin_only_raw}
        }\\
        \subfigure[Detected Communities] {%
           \includegraphics[width=0.70\textwidth, trim=0.0cm 0.0cm 0.0cm 0.0cm, clip]
           {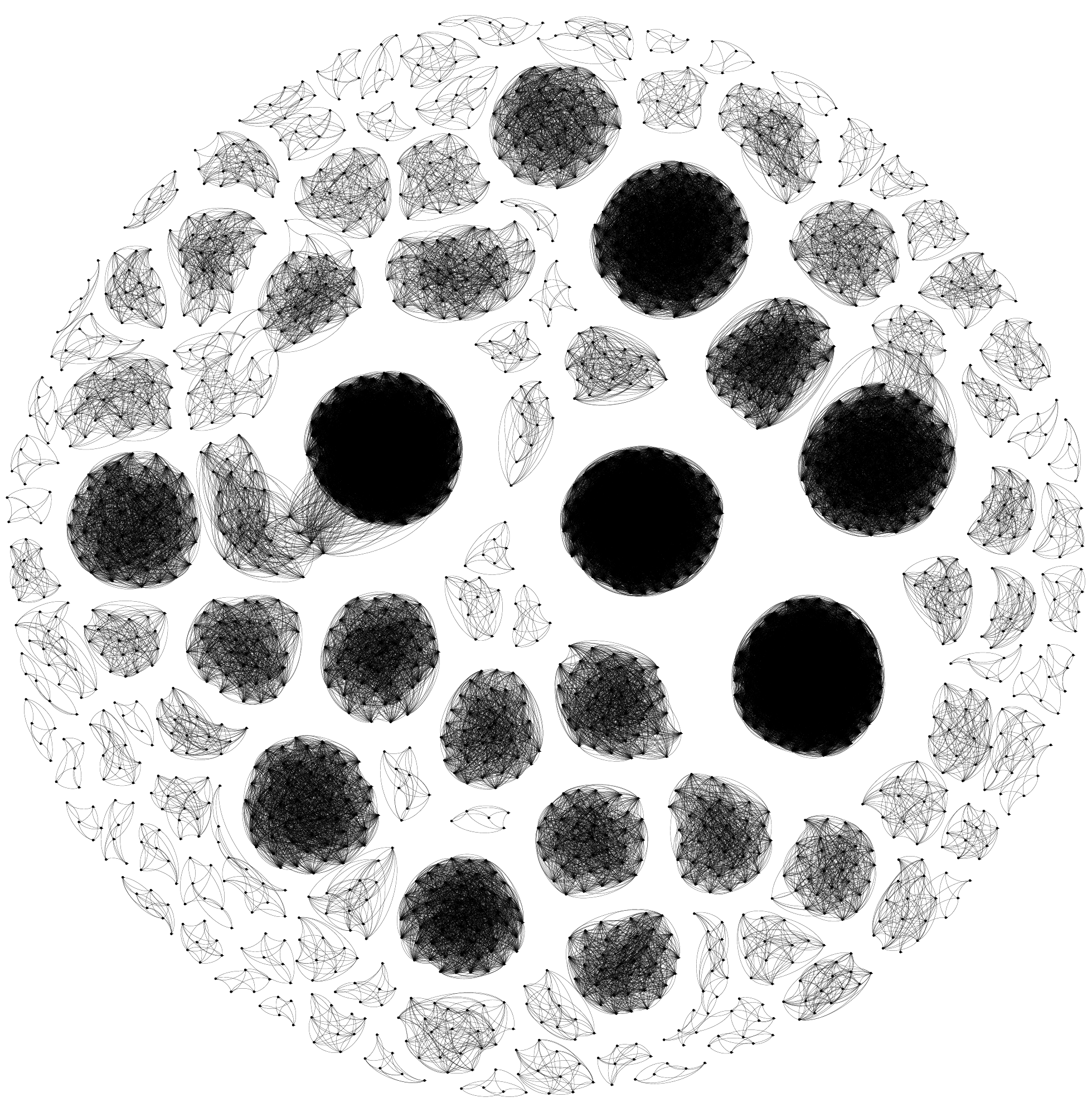}
           \label{fig:net_drebin_only_clustering}
        }
    \end{center}
    \caption{\textsf{Cypider} Network of \textit{Drebin} Malware Dataset}
    \label{fig_SimNetMalwareDrebin}
\end{figure}
\end{scriptsize}

\begin{scriptsize}
\begin{figure}
     \begin{center}        
        \subfigure[Simialrity Network]{%
            \includegraphics[width=0.70\textwidth, trim=0.0cm 0.0cm 0.0cm 0.0cm, clip]
            {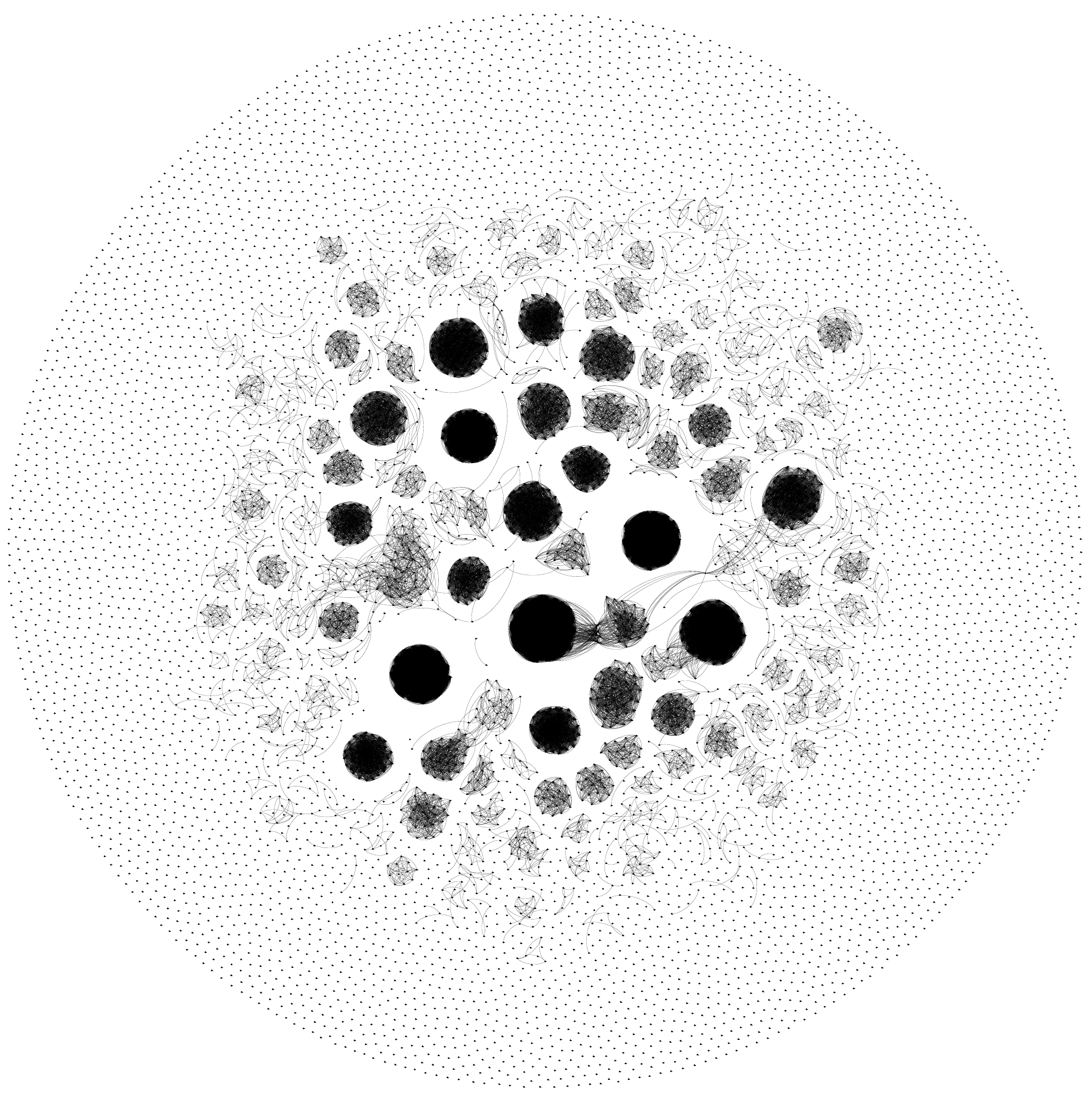}
            \label{fig:net_drebin_mixed_raw}
        }\\
        \subfigure[Detected Communities] {%
           \includegraphics[width=0.70\textwidth, trim=0.0cm 0.0cm 0.0cm 0.0cm, clip]
           {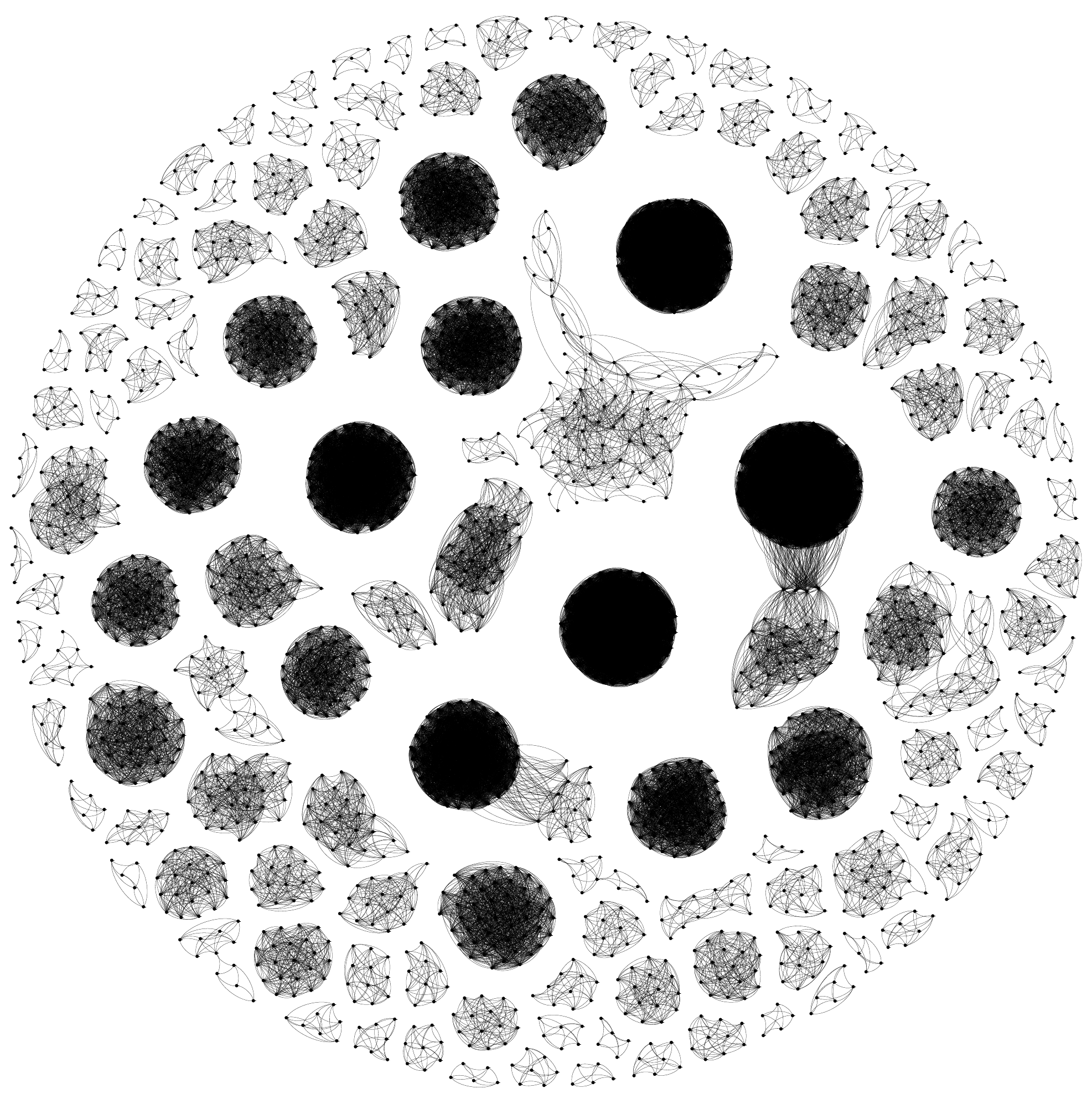}
           \label{fig:net_drebin_mixed_clustering}
        }
    \end{center}
    \caption{\textsf{Cypider} Network of \textit{Drebin} Malware Dataset}
    \label{fig_SimNetMixedDrebin}
\end{figure}
\end{scriptsize}

\subsection{Community Fingerprint Results} 
\label{sec:community_fingerprint_result}

Table \ref{tab_CommunityFingerprintAccuracy} shows the evaluation results for the \textit{community fingerprint}, which is applied to different detected communities with various Android malware families. The community fingerprint model (One-Class SVM) achieves $87\%$ f1-score in detecting malware from the same malware family that is used in the training phase. In the \textit{signature database}, these new malware samples share the same family with a given \textit{community fingerprint}. Furthermore, the compressed format of this fingerprint, i.e., learning model in a binary format, could fingerprint an entire Android family, which results in a far more compacted \textit{signature database}. 

\begin{table}
\centering
\begin{scriptsize}
\resizebox{0.999\linewidth}{!}{
\begin{tabular}{l|cccccc}
\hline
  Family      & Community Size & Family Detection (Acc) & Malware (Acc) & Benign (Acc) & General (F1) \\
\hline
 HiddenAds    & 1774           & 77.43                  & 93.40         & 94.93        & 87.83        \\
 BridgeMaster & 1041           & 76.99                  & 93.08         & 93.22        & 87.05        \\
 InfoStealer  & 2973           & 86.60                  & 88.37         & 91.26        & 75.50        \\
 Plankton     & 495            & 77.76                  & 100.0         & 100.0        & 75.10        \\
 BaseBrigge   & 1499           & 76.76                  & 88.51         & 92.47        & 73.28        \\
 Utchi        & 973            & 76.19                  & 99.98         & 100.0        & 72.02        \\
\hline
\end{tabular}
}
\caption{Community Fingerprint Accuracy on Different Families} 
\label{tab_CommunityFingerprintAccuracy}
\end{scriptsize}
\end{table}

The performance of the community fingerprint mainly depends on the number of malware in the detected community. A higher detection performance is achieved when more malware instances exist in the community.  To achieve this aim, we plan to determine a threshold for the community cardinal, which is required to compute the fingerprint and store it in the \textit{signature database}. 

As shown in Table \ref{tab_CommunityFingerprintAccuracy}, one of the main characteristics that are provided by the community fingerprint, is its ability to differentiate between general malware apps and benign apps with high accuracy. The reason behind this high accuracy is the high similarity between general malware and the trained family. Notice that the one-class SVM model is trained on samples from only one malware family. In other words, malicious apps tend to have similar features, although they do not belong to the same malware family.  Thus, benign samples and general malware are highly dissimilar; hence, benign samples cannot match the community fingerprint.

\section{Hyper-parameters Analyses}
\label{sec_hyperparametersAnalyses}

In this section, we analyze the effect of \textsf{Cypider} hyper-parameters on the overall performance measured using \textbf{Purity}, \textbf{Coverage}, and \textbf{Community Numbers} metrics. Specifically, we investigate the \textbf{similarity threshold}, the \textbf{content threshold}, and the \textbf{community size}, as presented in Section \ref{sec:lsh_similartiy} and Section \ref{sec:community_detection}.

\subsection{Purity Analysis}

In the purity analysis, we compute the overall percentage of clustered malware samples of the groups belonging to the same Android malware family. A perfect purity metric means that each detected community (cluster) contains samples from the same Android malware family. Figure \ref{fig_detectionPurityDrebin} and \ref{fig_detectionPurityZoo} show the effect of \textsf{Cypider} hyper-parameters on the purity of the detected malware communities in the similarity network of \textit{Drebin} and {AndroZoo} datasets respectively. It is worth noting that the \textbf{content threshold} is the most affecting hyper-parameter on the overall purity. A small content threshold results in a lower purity percentage, as shown in the evaluation of both \textit{Drebin} and \textit{AndroZoo} datasets. This finding is intuitive because the outcome of \textsf{Cypider} grouping is more accurate when using more content types threshold in the \textit{majority voting} similarity computation.

  On the other hand, the \textbf{similarity threshold} has a secondary effect compared to the \textbf{content threshold}. This means that a tight distant threshold outputs less false (malware family) samples in the detected communities.  Finally, we notice a very minor effect of the \textbf{community size} on the overall purity metric for both \textit{Drebin} and \textit{AndroZoo} evaluations, as shown in Figure \ref{fig_detectionPurityDrebin} and Figure \ref{fig_detectionPurityZoo} respectively.

\begin{scriptsize}
\begin{figure}[ht!]
     \begin{center}        
        \subfigure[Similairty \& Content]{%
            \label{fig:effectiveness_drebin_malware_purep_both}
            \includegraphics[width=0.50\textwidth, trim=1.4cm .5cm 2.1cm 2.0cm, clip]
            {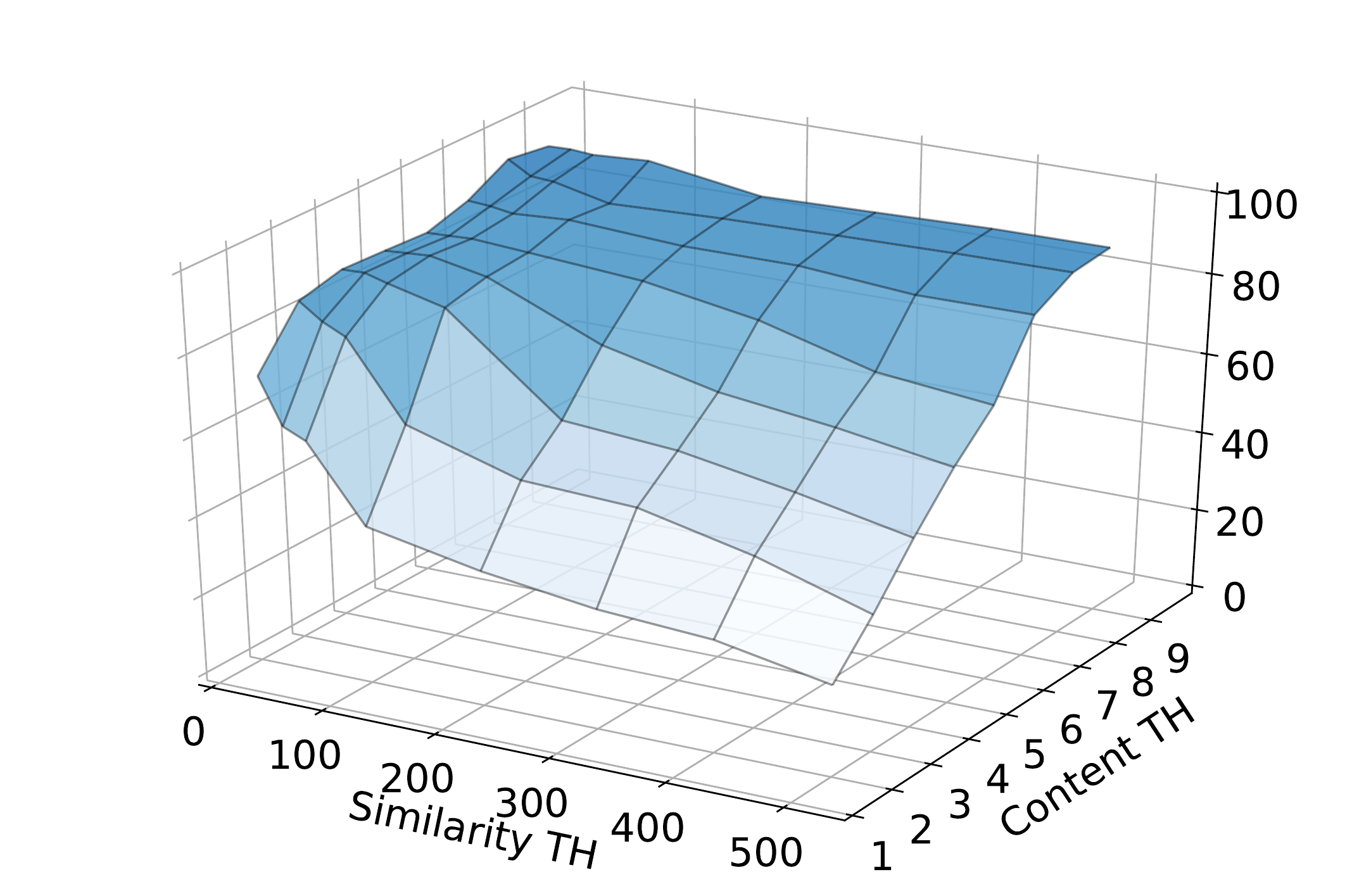}
        }
        \subfigure[Similarity Threshold] {%
           \label{fig:effectiveness_drebin_malware_purep_simth}
           \includegraphics[width=0.40\textwidth, trim=.1cm 1.0cm 0.1cm 0.0cm, clip]
           {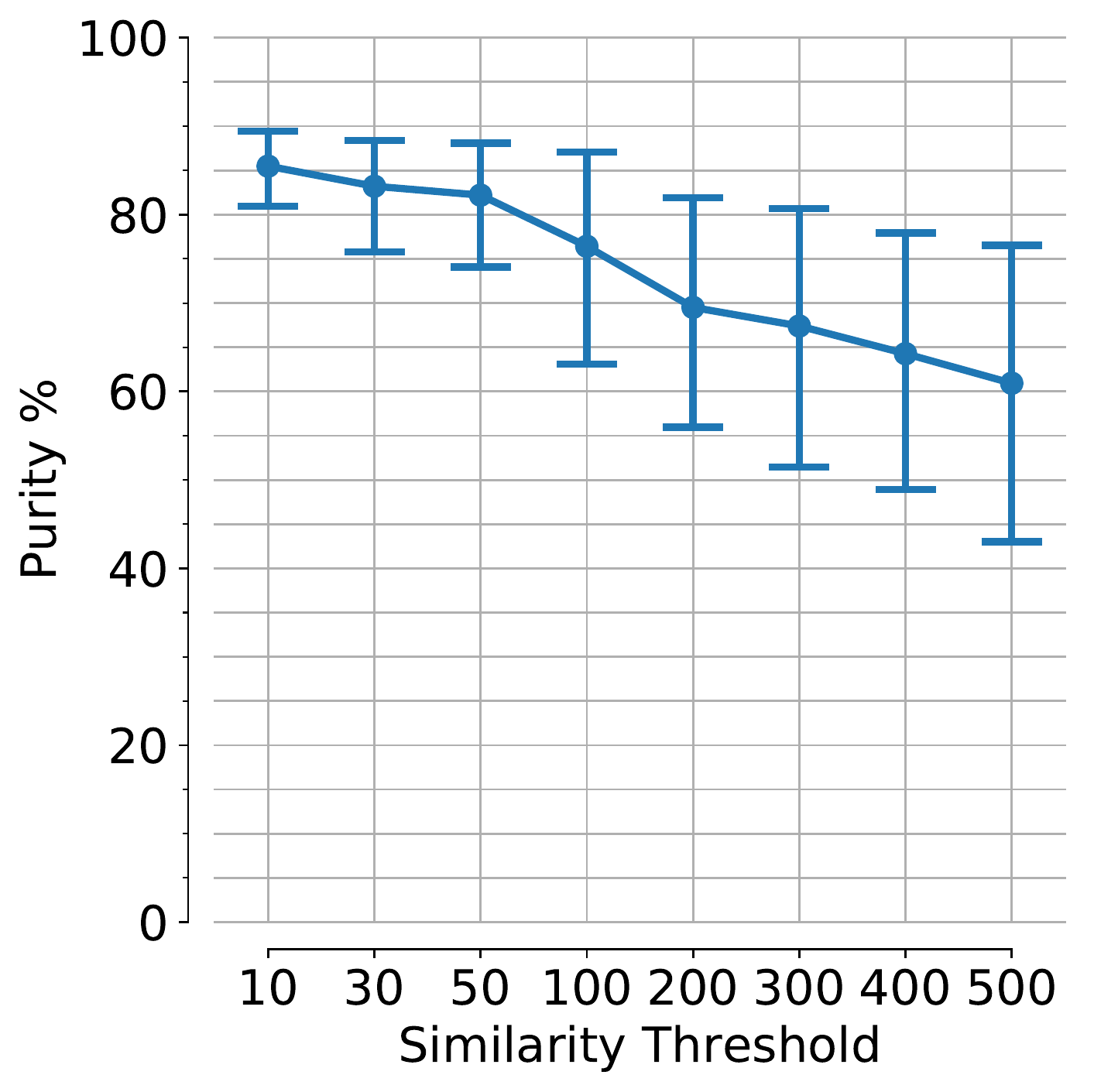}
        }\\
        \subfigure[Content Threshold]{%
            \label{fig:effectiveness_drebin_malware_purep_cntth}
           \includegraphics[width=0.42\textwidth, trim=.1cm 1.0cm 0.1cm 0.0cm, clip]
            {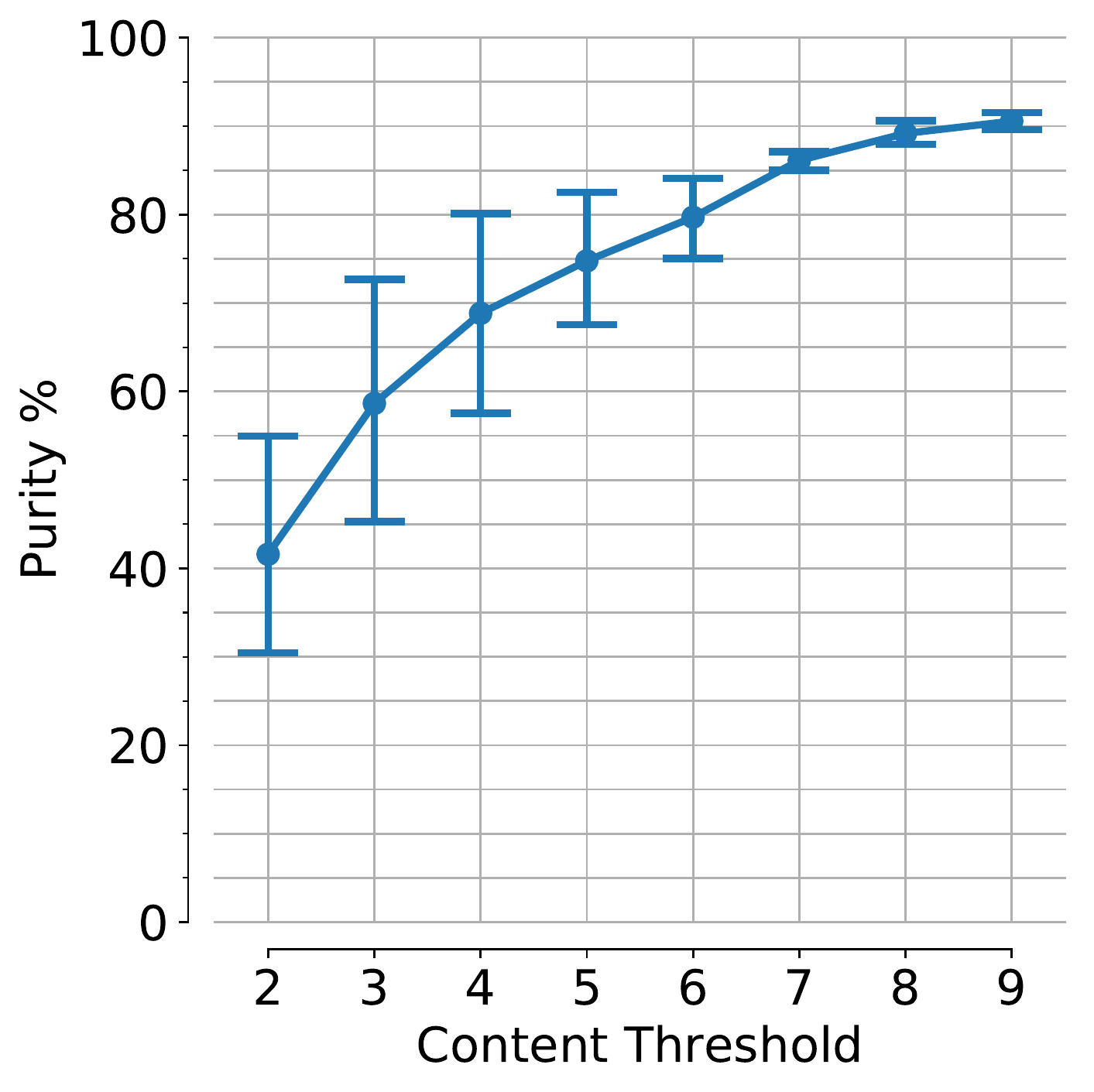}
        }
        \subfigure[Community Size] {%
           \label{fig:effectiveness_drebin_malware_purep_sizeth}
           \includegraphics[width=0.42\textwidth, trim=.1cm 1.0cm 0.1cm 0.0cm, clip]
           {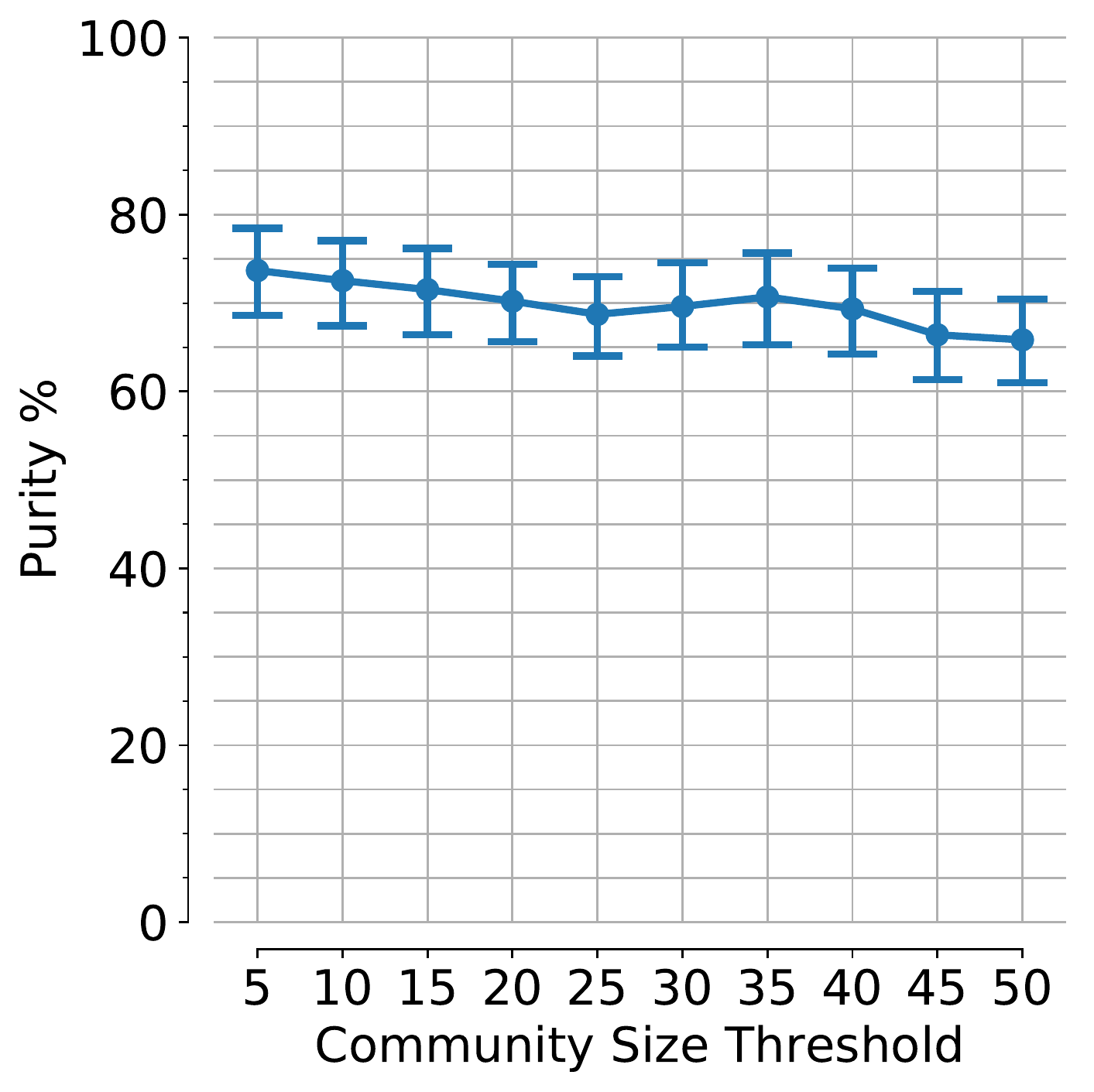}
        }                                                    
    \end{center}
    \caption{Detection Purity, Hyper-parameters Analysis Results of Drebin
    Dataset}
   \label{fig_detectionPurityDrebin}
\end{figure}
\end{scriptsize}

\begin{scriptsize}
\begin{figure}[ht!]
     \begin{center}        
        \subfigure[Similairty \& Content]{%
            \label{fig:effectiveness_zoo_malware_purep_both}
            \includegraphics[width=0.5\textwidth, trim=1.4cm .5cm 2.1cm 2.0cm, clip]
           {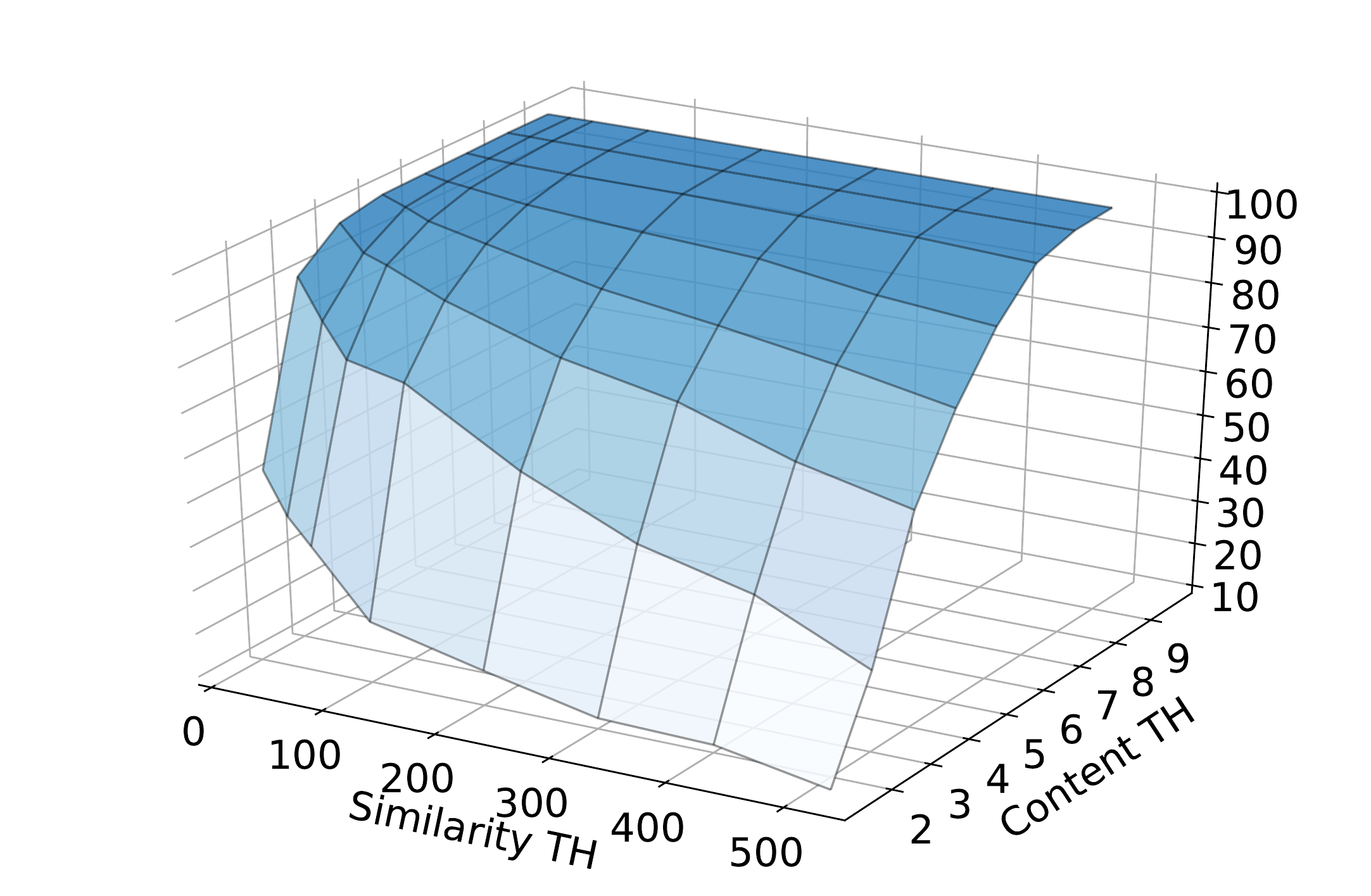}
        }
        \subfigure[Similarity Threshold] {%
           \label{fig:effectiveness_zoo_malware_purep_simth}
           \includegraphics[width=0.40\textwidth, trim=.1cm 1.0cm 0.1cm 0.0cm, clip]
           {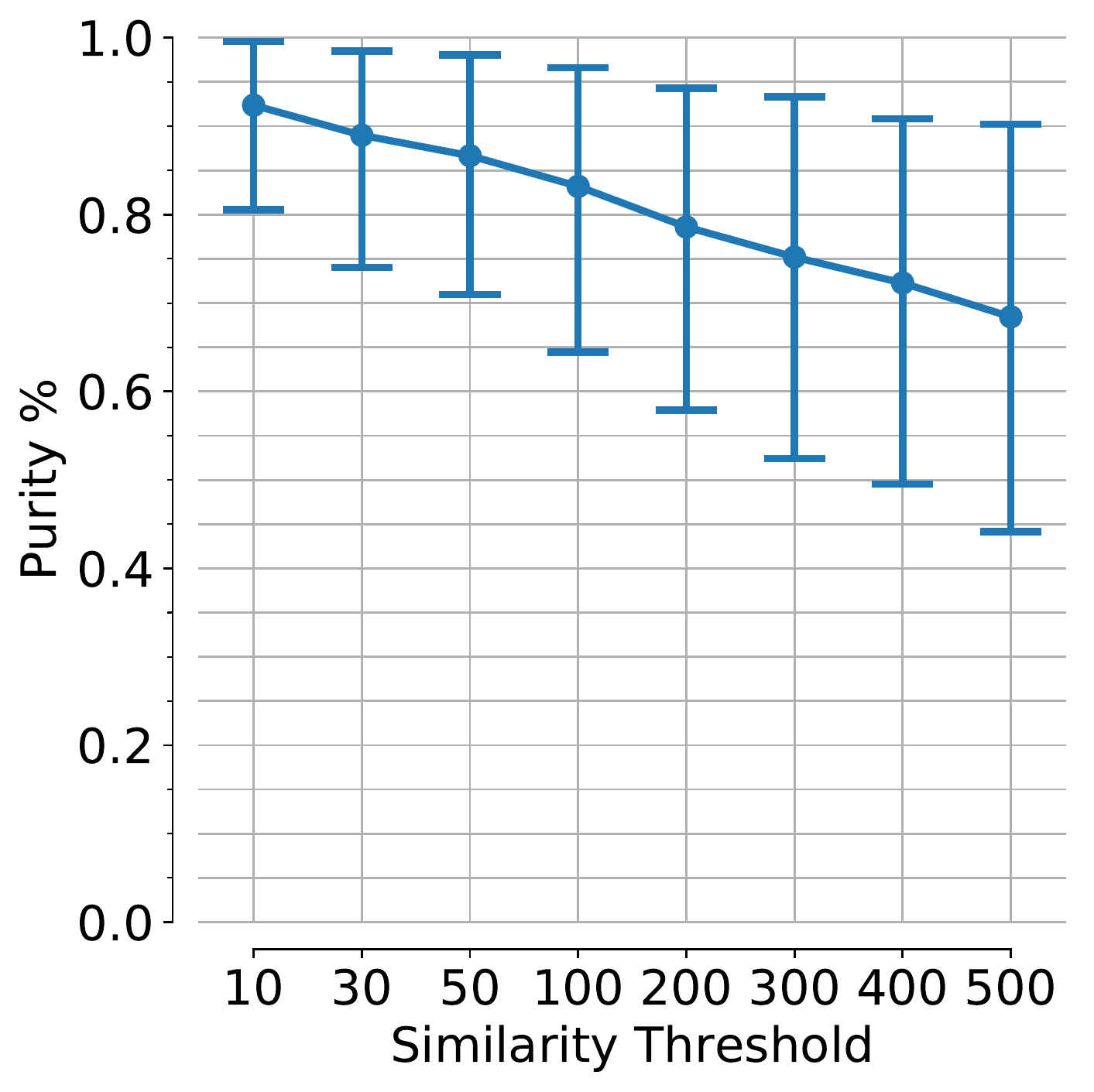}
        }\\
        \subfigure[Content Threshold]{%
            \label{fig:effectiveness_zoo_malware_purep_cntth}
           \includegraphics[width=0.42\textwidth, trim=.1cm 1.0cm 0.1cm 0.0cm, clip]
           {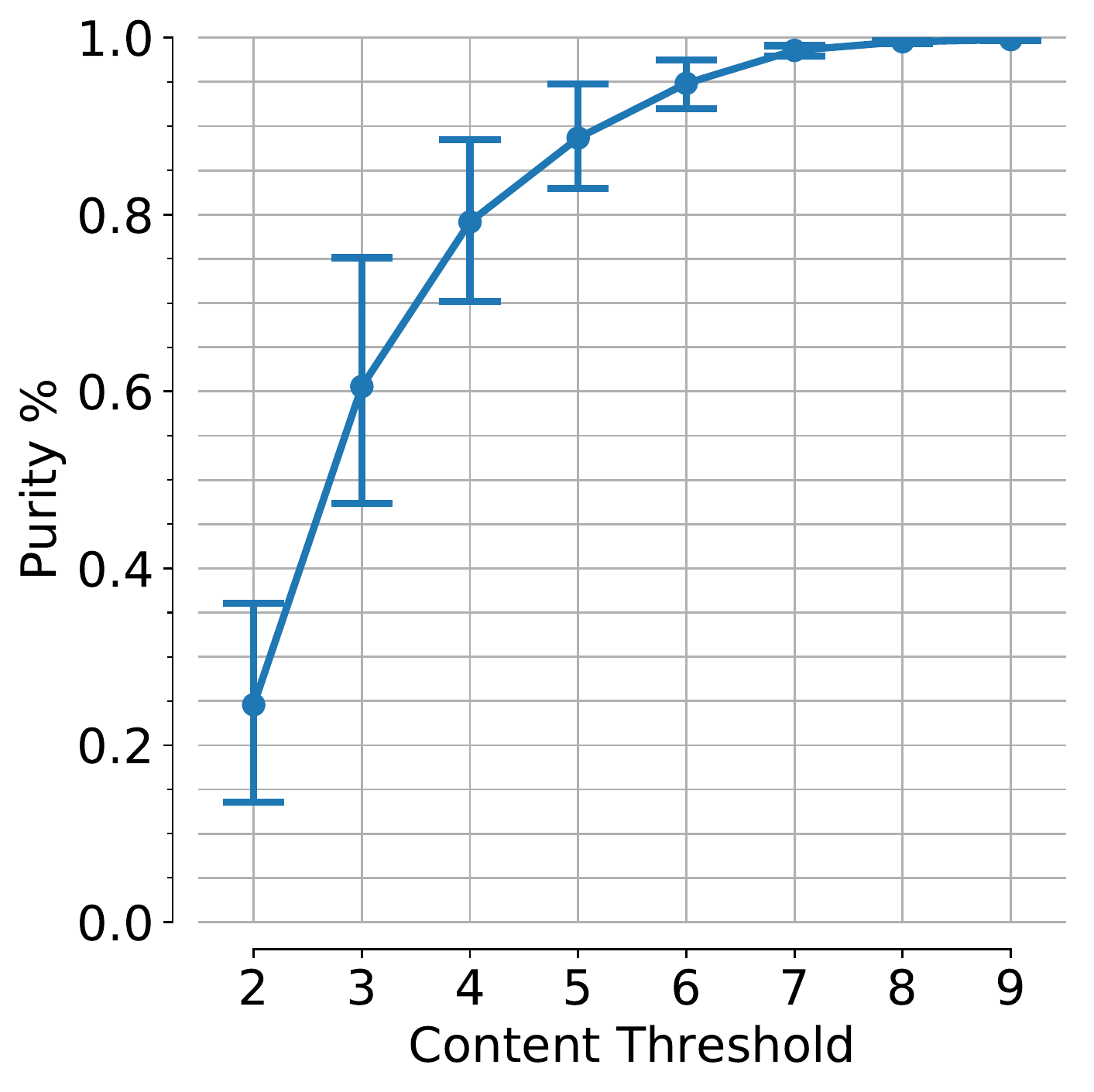}
        }
        \subfigure[Community Size] {%
           \label{fig:effectiveness_zoo_malware_purep_sizeth}
           \includegraphics[width=0.42\textwidth, trim=.1cm 1.0cm 0.1cm 0.0cm, clip]
           {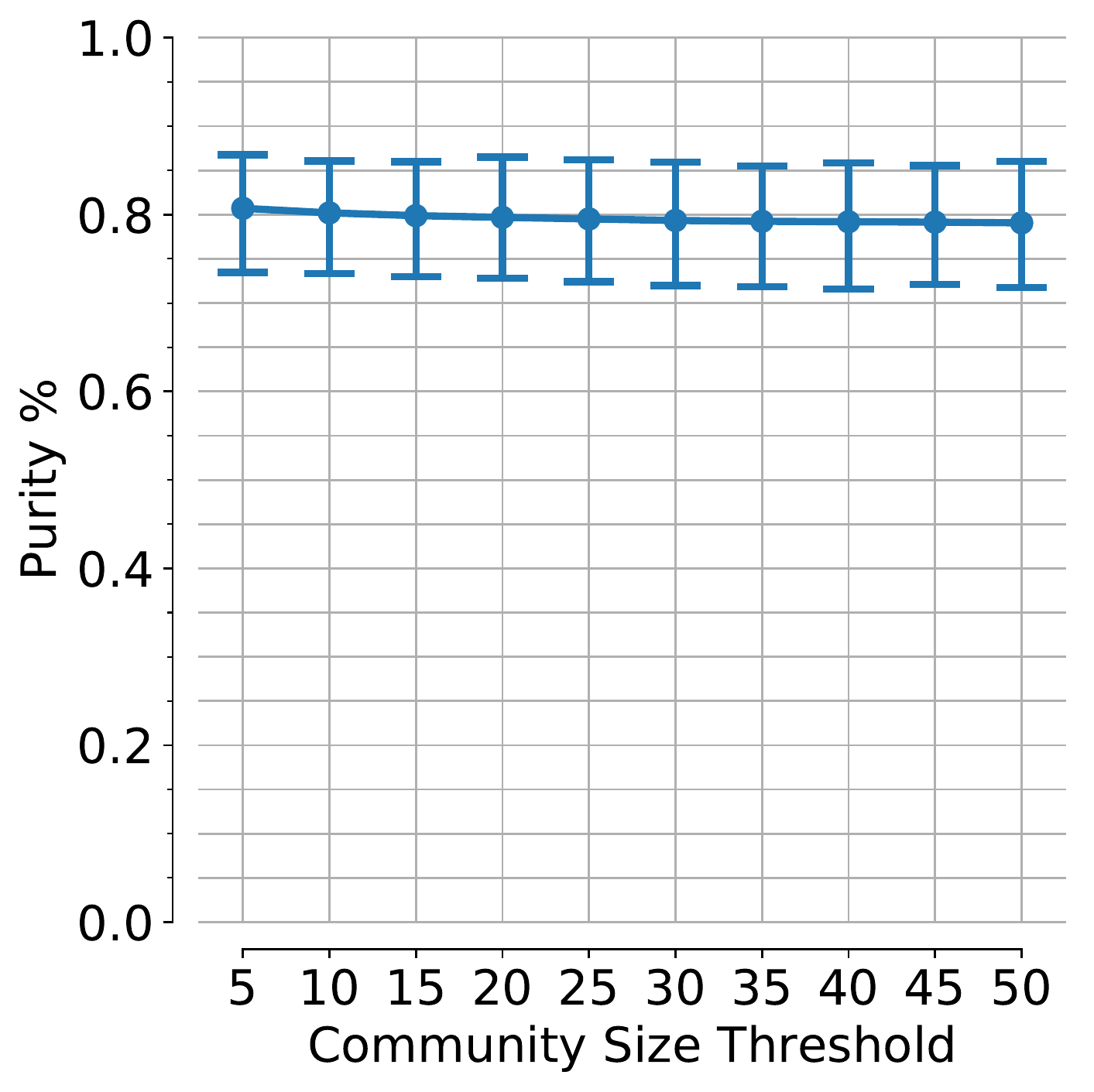}
        }                                                    
    \end{center}
    \caption{Detection Purity, Hyper-parameters Analysis Results of AndroZoo
    Dataset}
   \label{fig_detectionPurityZoo}
\end{figure}
\end{scriptsize}

\subsection{Coverage Analysis}

In the coverage analysis, we assess the percentage of the detected malware from the overall input dataset. A perfect coverage means that \textsf{Cypider} detects the malware samples in the produced malware communities. Figure \ref{fig_detectionCoverageDrebin} and Figure \ref{fig_detectionCoverageZoo} depict the change in the coverage percentage with  \textsf{Cypider} hyper-parameters for \textit{Drebin} and \textit{AndroZoo} datasets respectively. We notice that the \textbf{content threshold} is the most affecting hyper-parameter on the overall coverage metric. This means that a high content threshold in the majority voting (Section \ref{sec:lsh_similartiy}) detects fewer malware samples in the produced malware communities. Therefore, the coverage metric decreases drastically with a high content threshold, as shown in Figure \ref{fig_detectionCoverageDrebin} and Figure \ref{fig_detectionCoverageZoo}. 

The \textbf{similarity threshold} and the \textbf{community size} have a secondary effect on the coverage metric. For the similarity threshold, a wide distance threshold yields a higher detection rate, and therefore a high coverage metric. For the community size threshold, a large community size threshold ignores many small malware communities, which affects the detection coverage metric negatively.

\begin{scriptsize}
\begin{figure}[ht!]
     \begin{center}        
        \subfigure[Similairty \& Content]{%
            \label{fig:effectiveness_drebin_malware_dtcp_both}
            \includegraphics[width=0.50\textwidth, trim=1.4cm .5cm 2.1cm 2.0cm, clip]
            {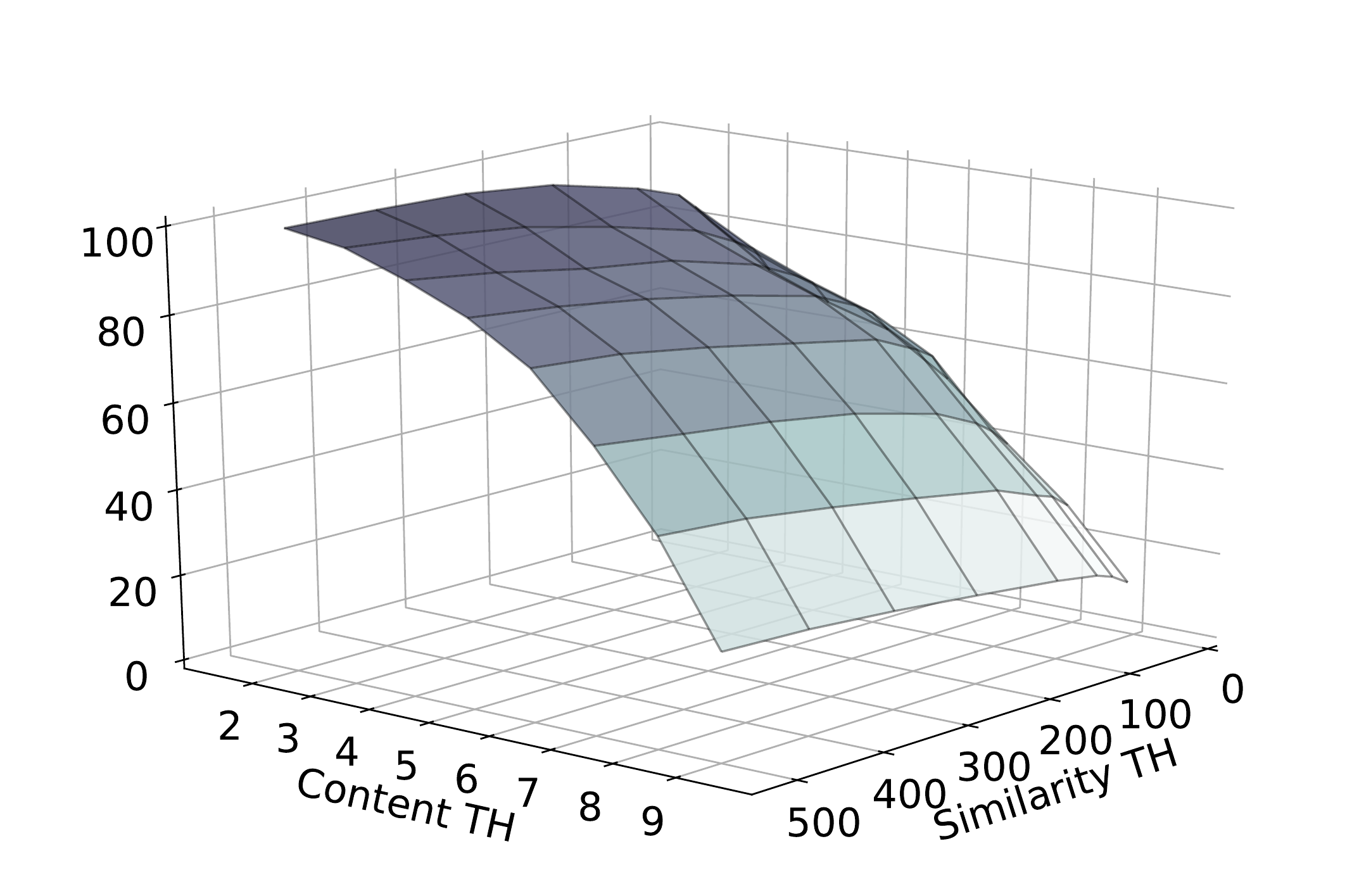}
        }
        \subfigure[Similarity Threshold] {%
           \label{fig:effectiveness_drebin_malware_dtcp_simth}
           \includegraphics[width=0.40\textwidth, trim=.1cm 1.0cm 0.1cm 0.0cm, clip]
           {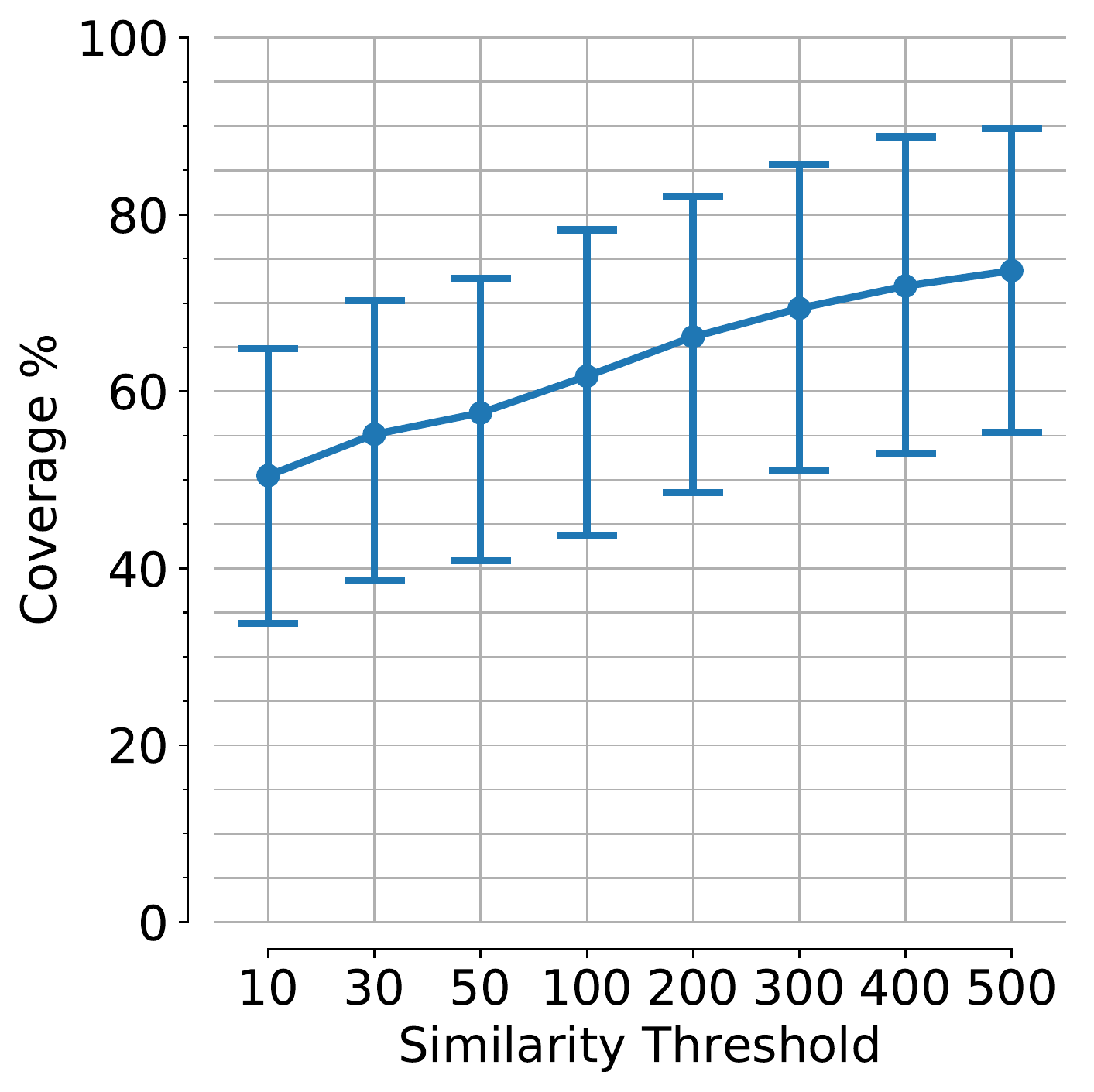}
        }\\
        \subfigure[Content Threshold]{%
            \label{fig:effectiveness_drebin_malware_dtcp_cntth}
            \includegraphics[width=0.42\textwidth, trim=.1cm 1.0cm 0.1cm 0.0cm, clip]
            {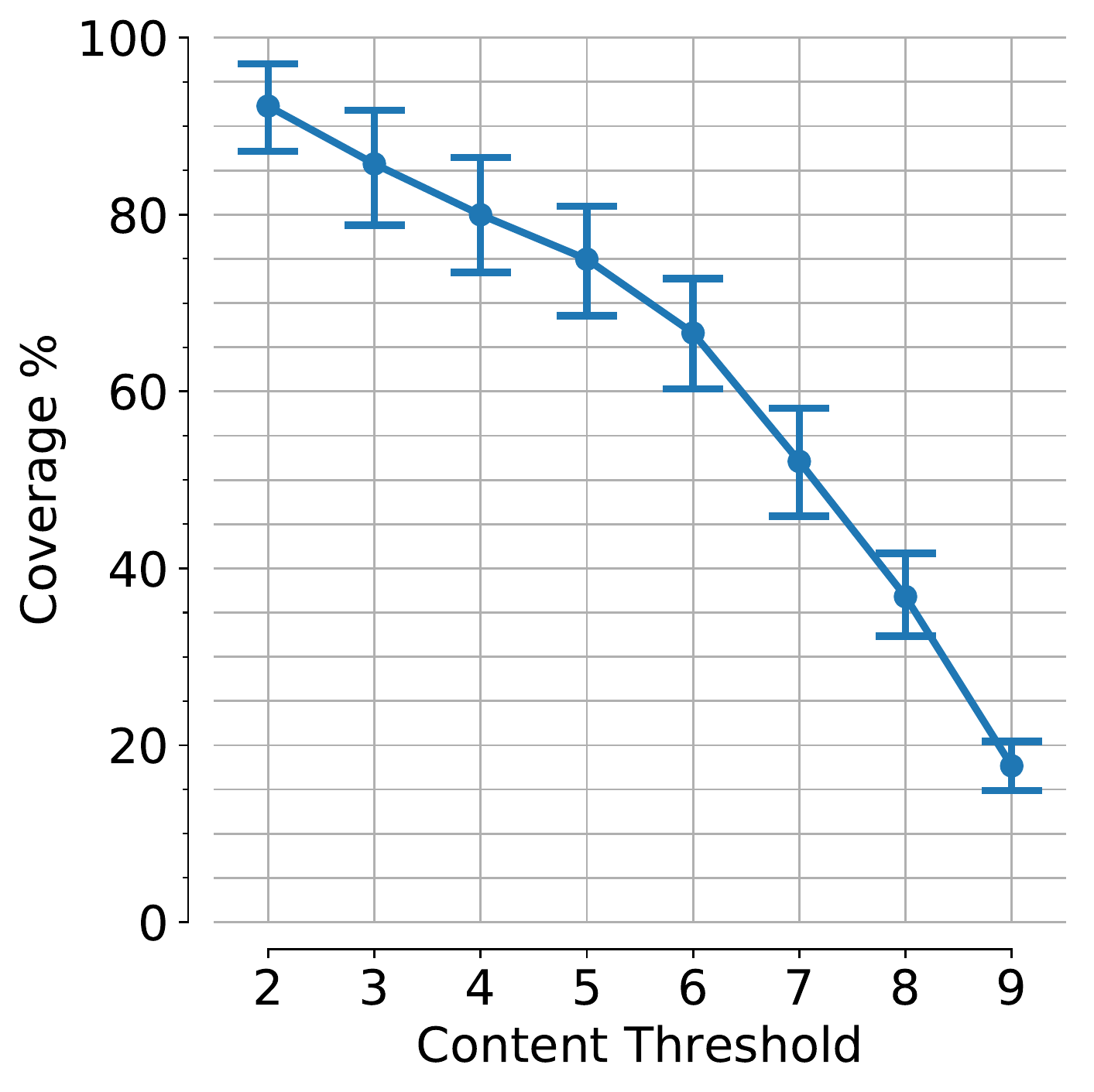}
        }
        \subfigure[Community Size] {%
           \label{fig:effectiveness_drebin_malware_dtcp_sizeth}
           \includegraphics[width=0.42\textwidth, trim=.1cm 1.0cm 0.1cm 0.0cm, clip]
           {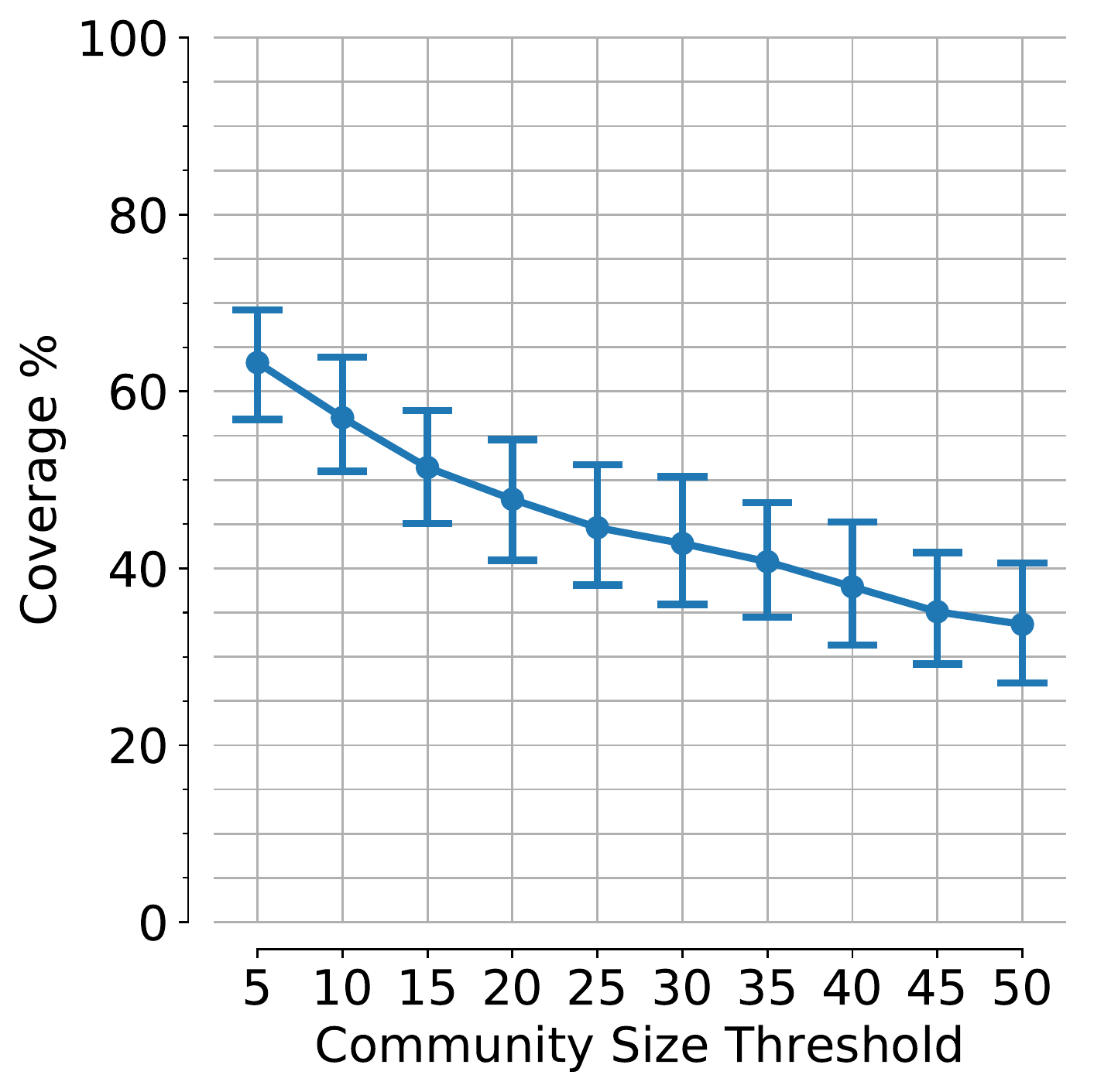}
        }                                                    
    \end{center}
    \caption{Detection Coverage, Hyper-parameters Analysis Results of Drebin
    Dataset}
   \label{fig_detectionCoverageDrebin}
\end{figure}
\end{scriptsize}

\begin{scriptsize}
\begin{figure}[ht!]
     \begin{center}        
        \subfigure[Similairty \& Content]{%
            \label{fig:effectiveness_zoo_malware_dtcp_both}
            \includegraphics[width=0.50\textwidth, trim=1.4cm .5cm 2.1cm 2.0cm, clip]
           {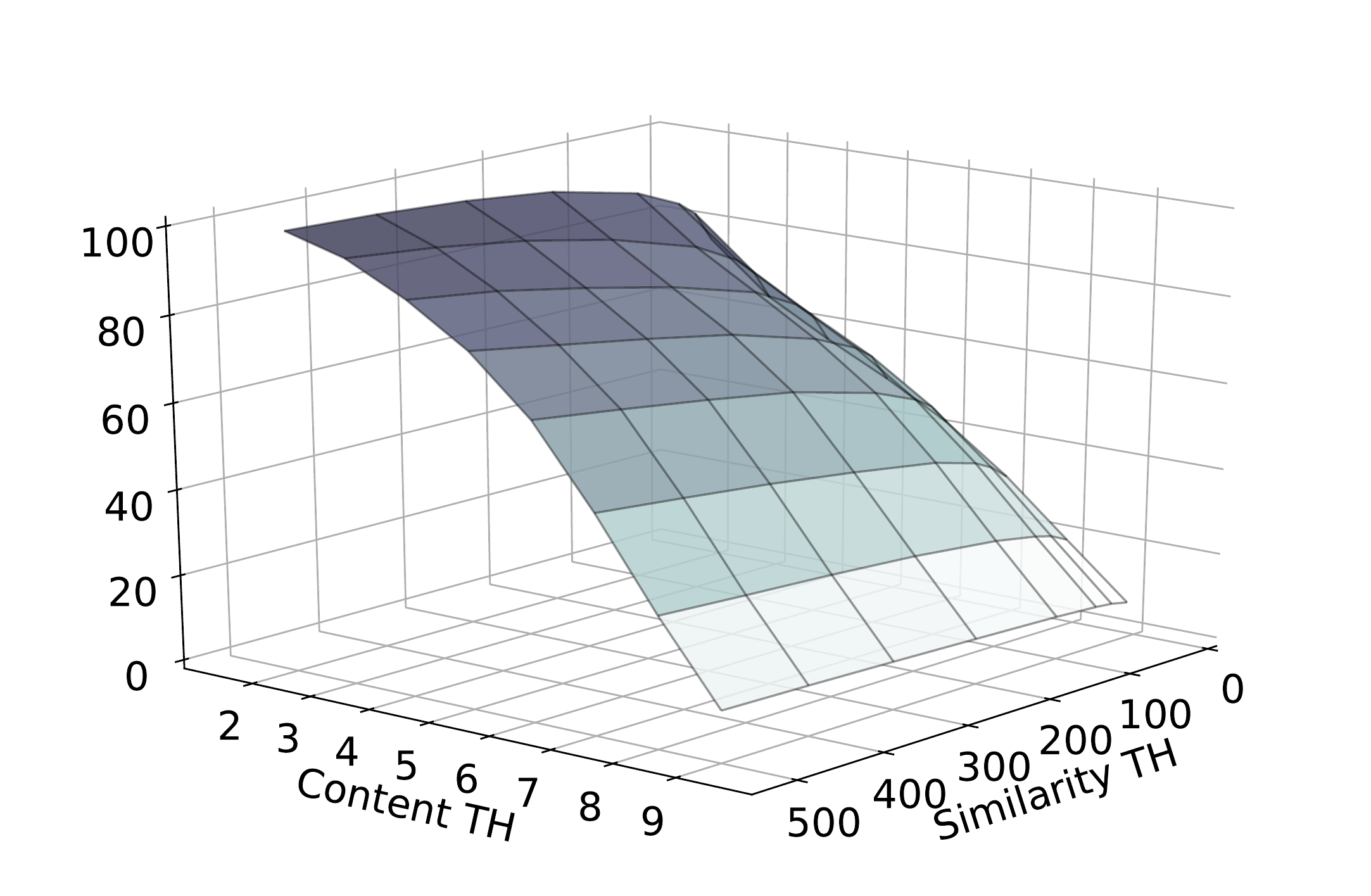}
        }
        \subfigure[Similarity Threshold] {%
           \label{fig:effectiveness_zoo_malware_dtcp_simth}
           \includegraphics[width=0.40\textwidth, trim=.1cm 1.0cm 0.1cm 0.0cm, clip]
           {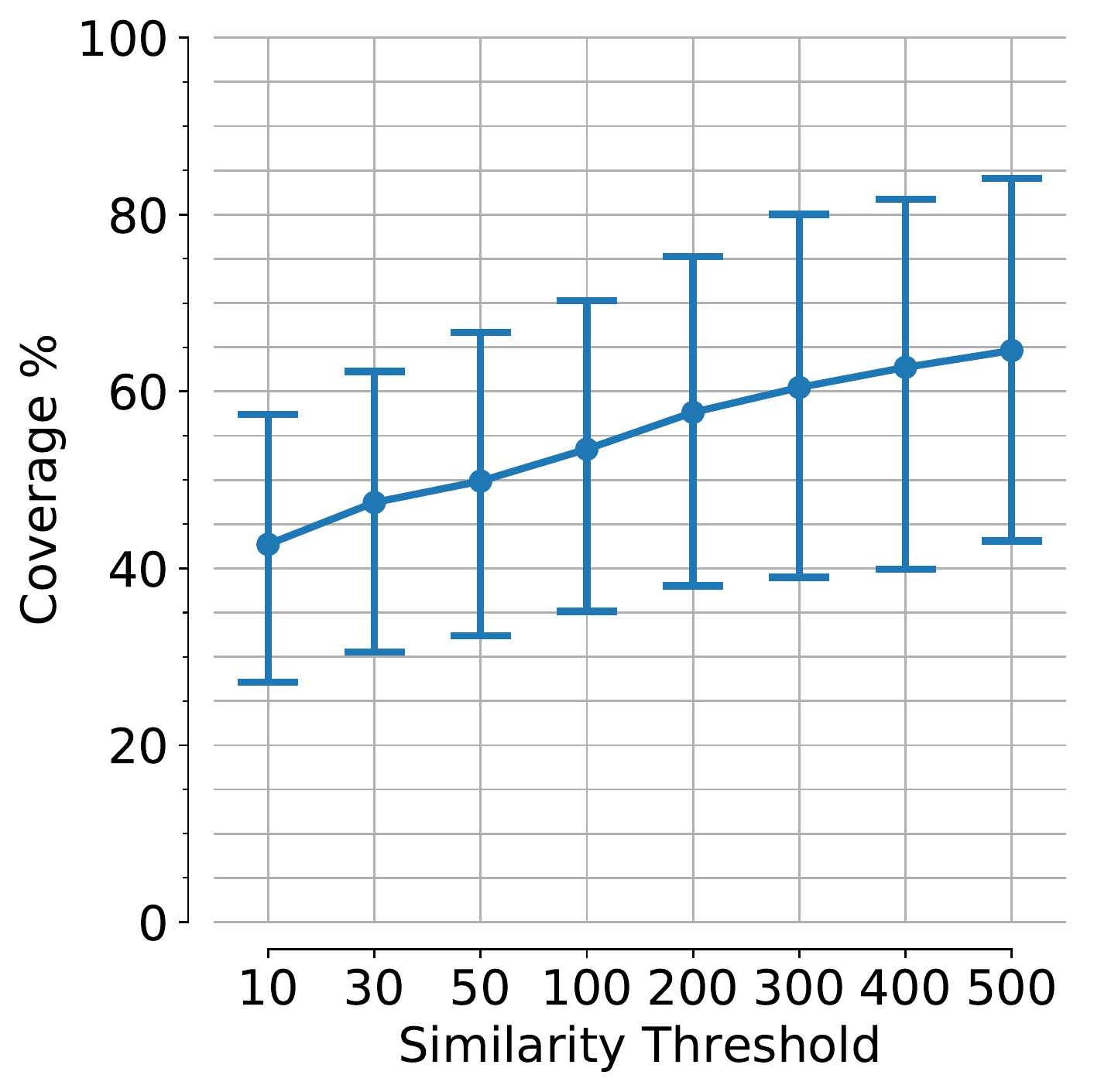}
        }\\
        \subfigure[Content Threshold]{%
            \label{fig:effectiveness_zoo_malware_dtcp_cntth}
           \includegraphics[width=0.42\textwidth, trim=.1cm 1.0cm 0.1cm 0.0cm, clip]
            {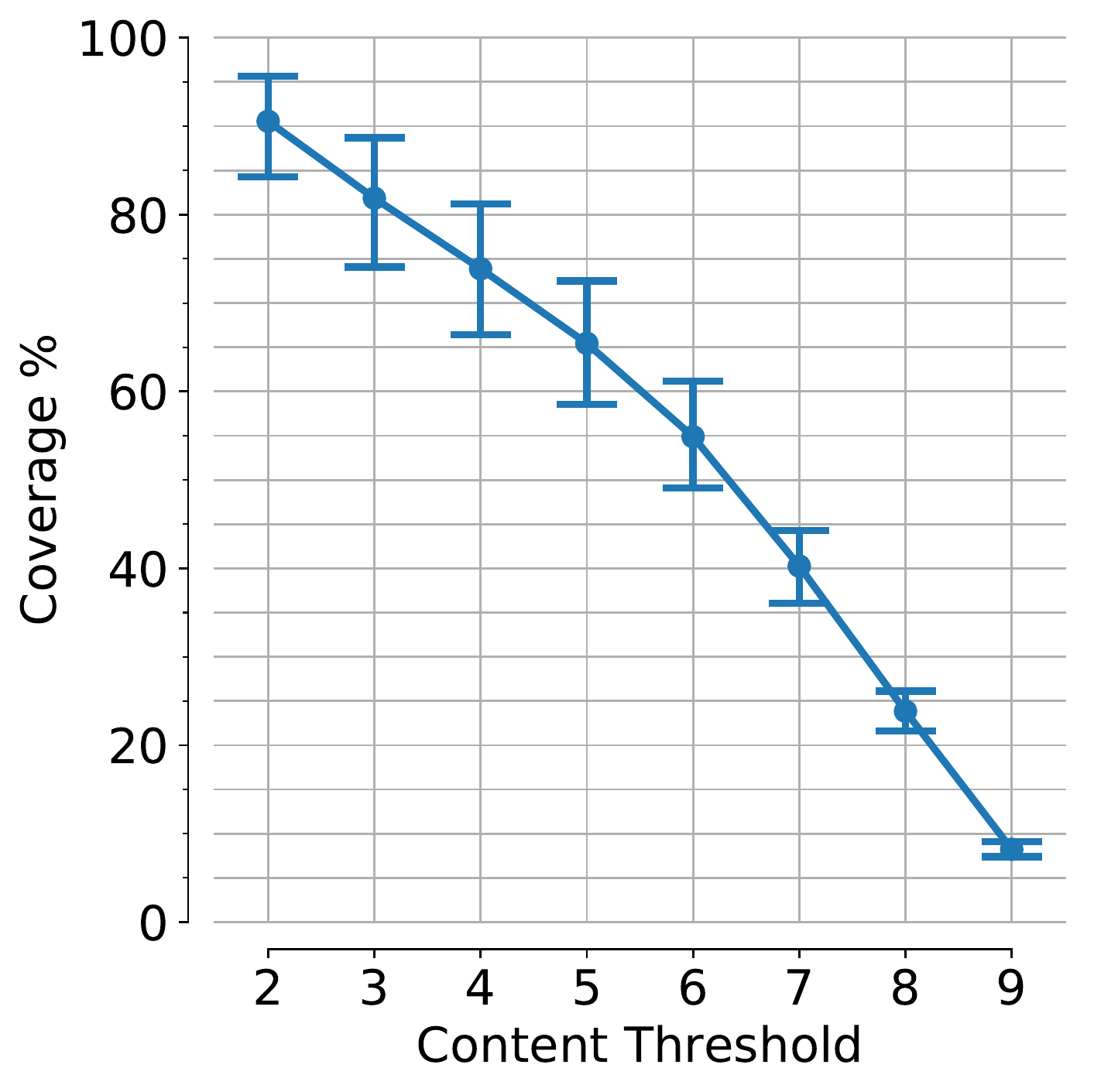}
        }
        \subfigure[Community Size] {%
           \label{fig:effectiveness_zoo_malware_dtcp_sizeth}
           \includegraphics[width=0.42\textwidth, trim=.1cm 1.0cm 0.1cm 0.0cm, clip]
           {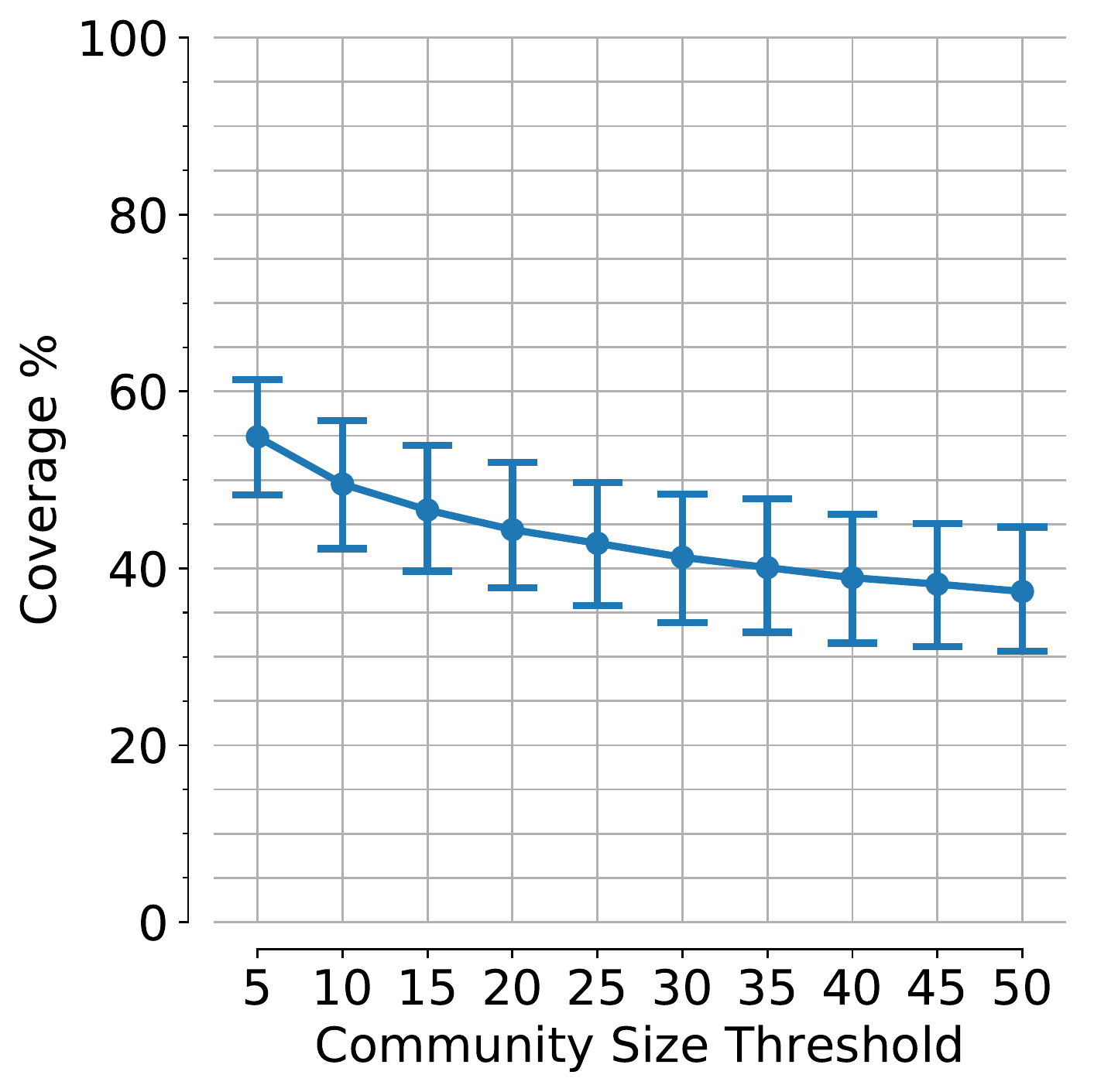}
        }                                                    
    \end{center}
    \caption{Detection Coverage, Hyper-parameters Analysis Results of AndroZoo
    Dataset}
   \label{fig_detectionCoverageZoo}
\end{figure}
\end{scriptsize}

\subsection{Number of Communities Analysis}

In this section, we analyze the total number of the detected communities produced by \textsf{Cypider}. A perfect \textsf{Cypider} clustering yields a result in which the number of communities is equal to the actual number of malware families in the input dataset. Figure \ref{fig_detectionCommunityDrebin} and Figure \ref{fig_detectionCommunityZoo} depict the effect of \textsf{Cypider} hyper-parameters on the number of the detected communities on \textit{Drebin} and \textit{AndroZoo} datasets respectively. It is important to mention that the community size has a strong influence on the number of communities. A significant community size threshold will filter many small malware communities, which influences the number of detected communities. 

For the content threshold, the majority voting with a small content threshold causes many communities to merge. This is because the samples have to be similar in only two content thresholds to maintain a similarity link in the similarity network.  On the other hand, the majority voting with a high content threshold will detect fewer malware samples and, therefore, less overall malicious communities.  Finally, we notice a minor effect of the similarity threshold on the overall number of the detected communities, as depicted in Figure \ref{fig_detectionCommunityDrebin} and Figure \ref{fig_detectionCommunityZoo} for \textit{Drebin} and \textit{AndroZoo} datasets respectively.

\begin{scriptsize}
\begin{figure}[ht!]
     \begin{center}        
        \subfigure[Similairty \& Content]{%
            \label{fig:effectiveness_drebin_malware_dtccomn_both}
            \includegraphics[width=0.50\textwidth, trim=1.4cm .5cm 2.1cm 2.0cm, clip]
           {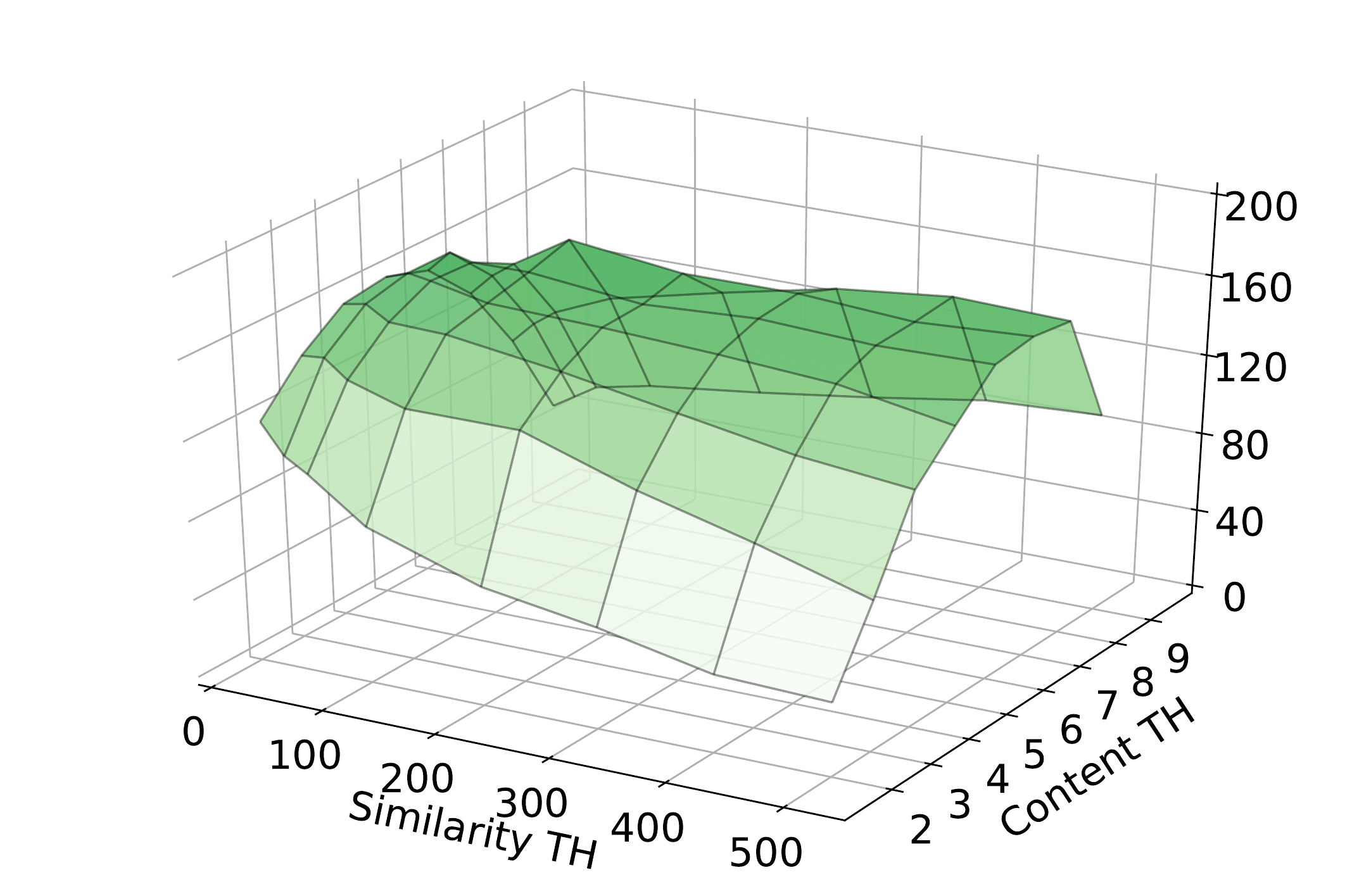}
        }
        \subfigure[Similarity Threshold] {%
           \label{fig:effectiveness_drebin_malware_dtccomn_simth}
           \includegraphics[width=0.40\textwidth, trim=.1cm 1.0cm 0.1cm 0.0cm, clip]
           {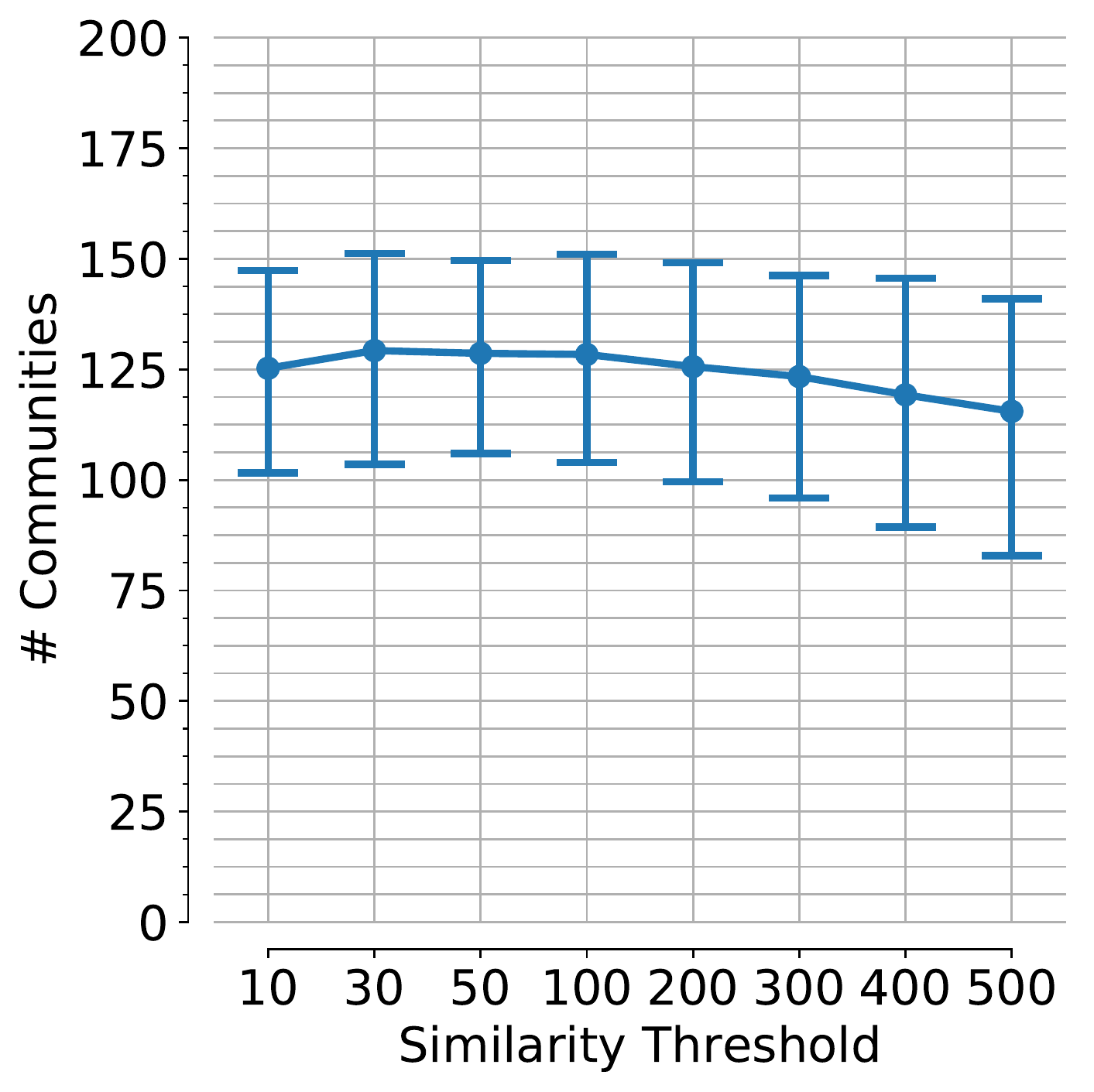}
        }\\
        \subfigure[Content Threshold]{%
            \label{fig:effectiveness_drebin_malware_dtccomn_cntth}
           \includegraphics[width=0.42\textwidth, trim=.1cm 1.0cm 0.1cm 0.0cm, clip]
           {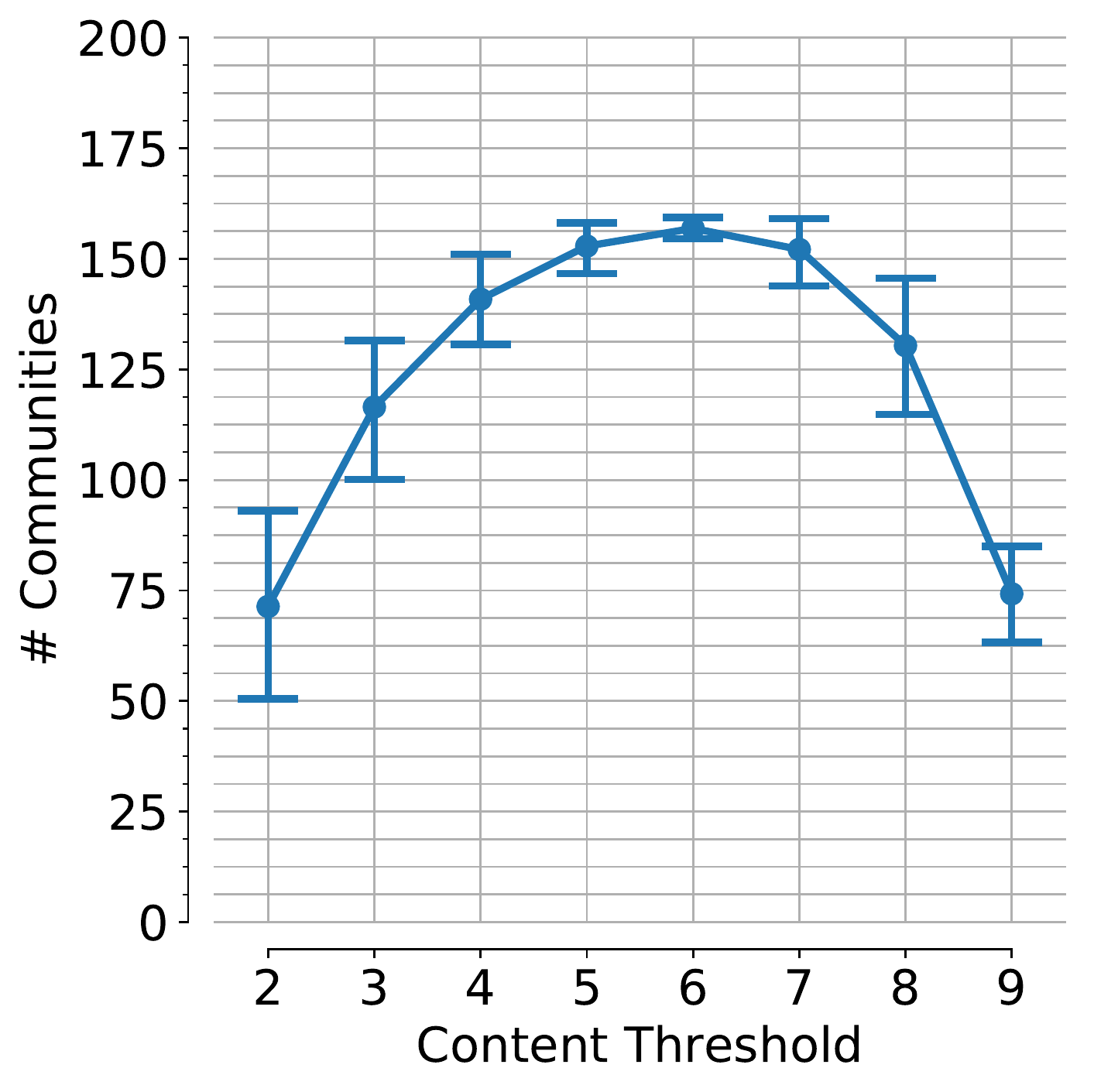}
        }
        \subfigure[Community Size] {%
           \label{fig:effectiveness_drebin_malware_dtccomn_sizeth}
           \includegraphics[width=0.42\textwidth, trim=.1cm 1.0cm 0.1cm 0.0cm, clip]
           {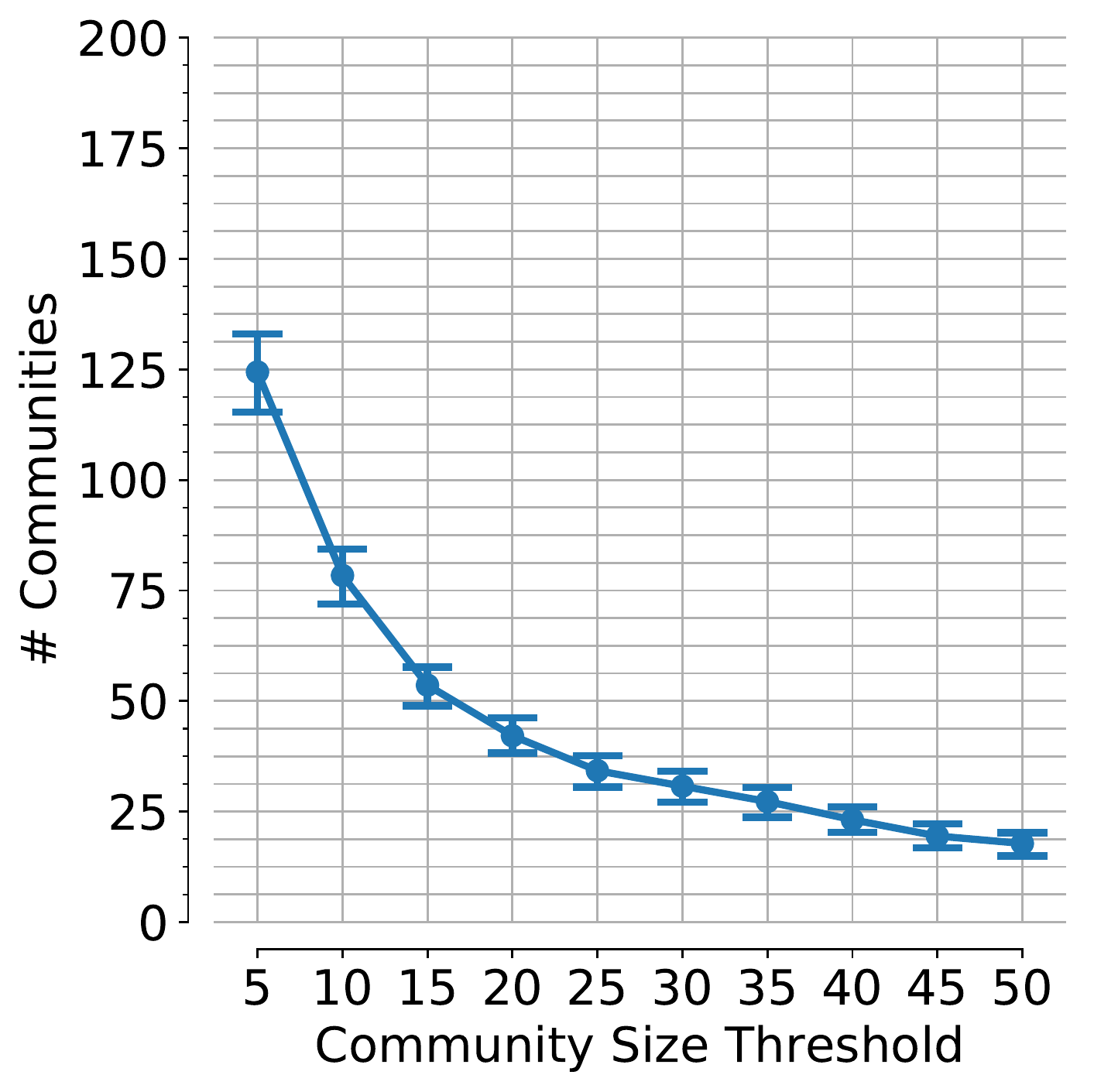}
        }                                                    
    \end{center}
    \caption{Detected Communities, Hyper-parameters Analysis Results of Drebin
    Dataset}
   \label{fig_detectionCommunityDrebin}
\end{figure}
\end{scriptsize}

\begin{scriptsize}
\begin{figure}[ht!]
     \begin{center}        
        \subfigure[\scriptsize Similairty \& Content]{%
            \label{fig:effectiveness_zoo_malware_dtccomn_both}
            \includegraphics[width=0.50\textwidth, trim=1.4cm .5cm 2.1cm 2.0cm, clip]
            {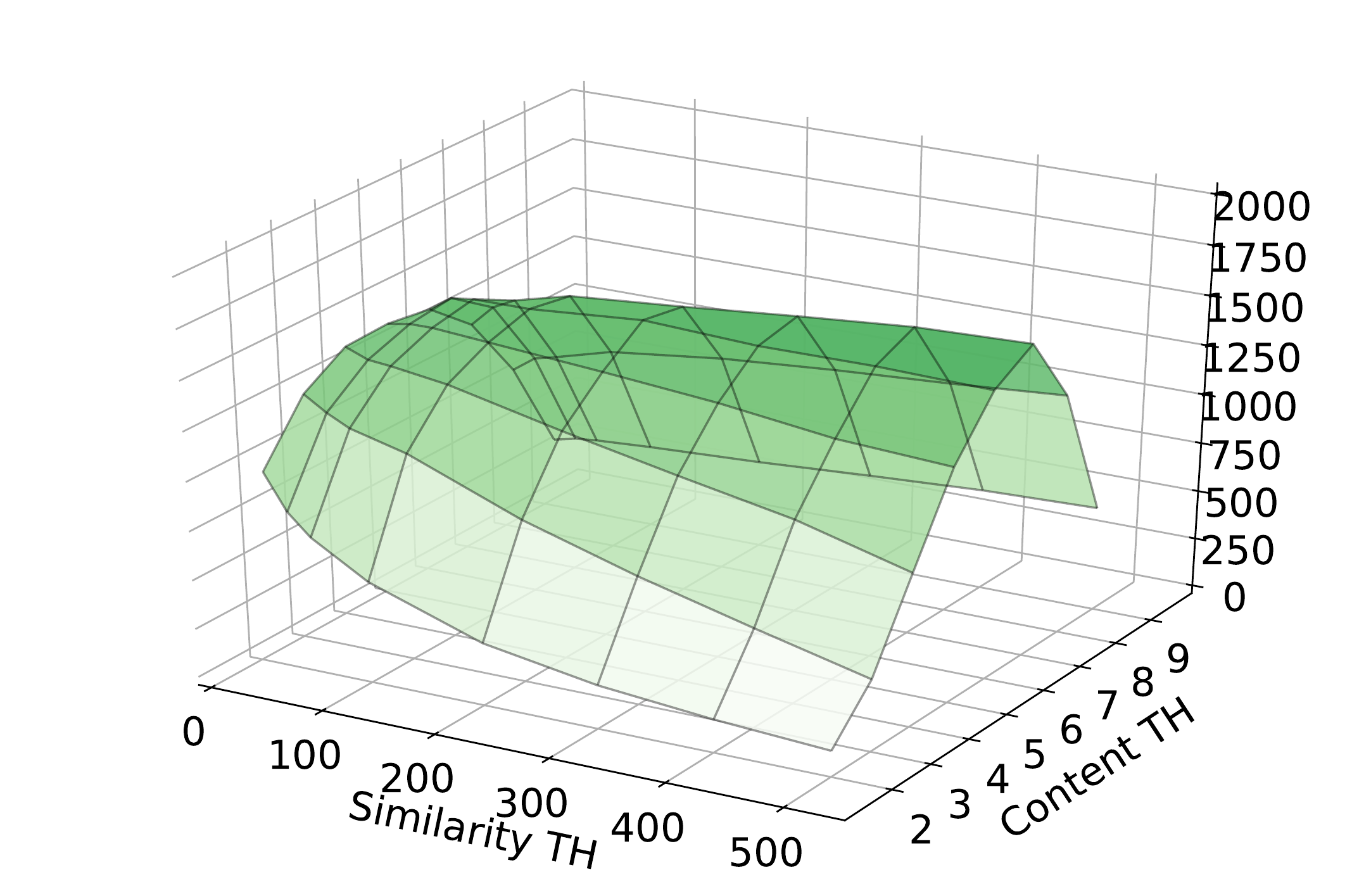}
        }
        \subfigure[\scriptsize Similarity Threshold] {%
           \label{fig:effectiveness_zoo_malware_dtccomn_simth}
           \includegraphics[width=0.40\textwidth, trim=.1cm 1.0cm 0.1cm 0.0cm, clip]
           {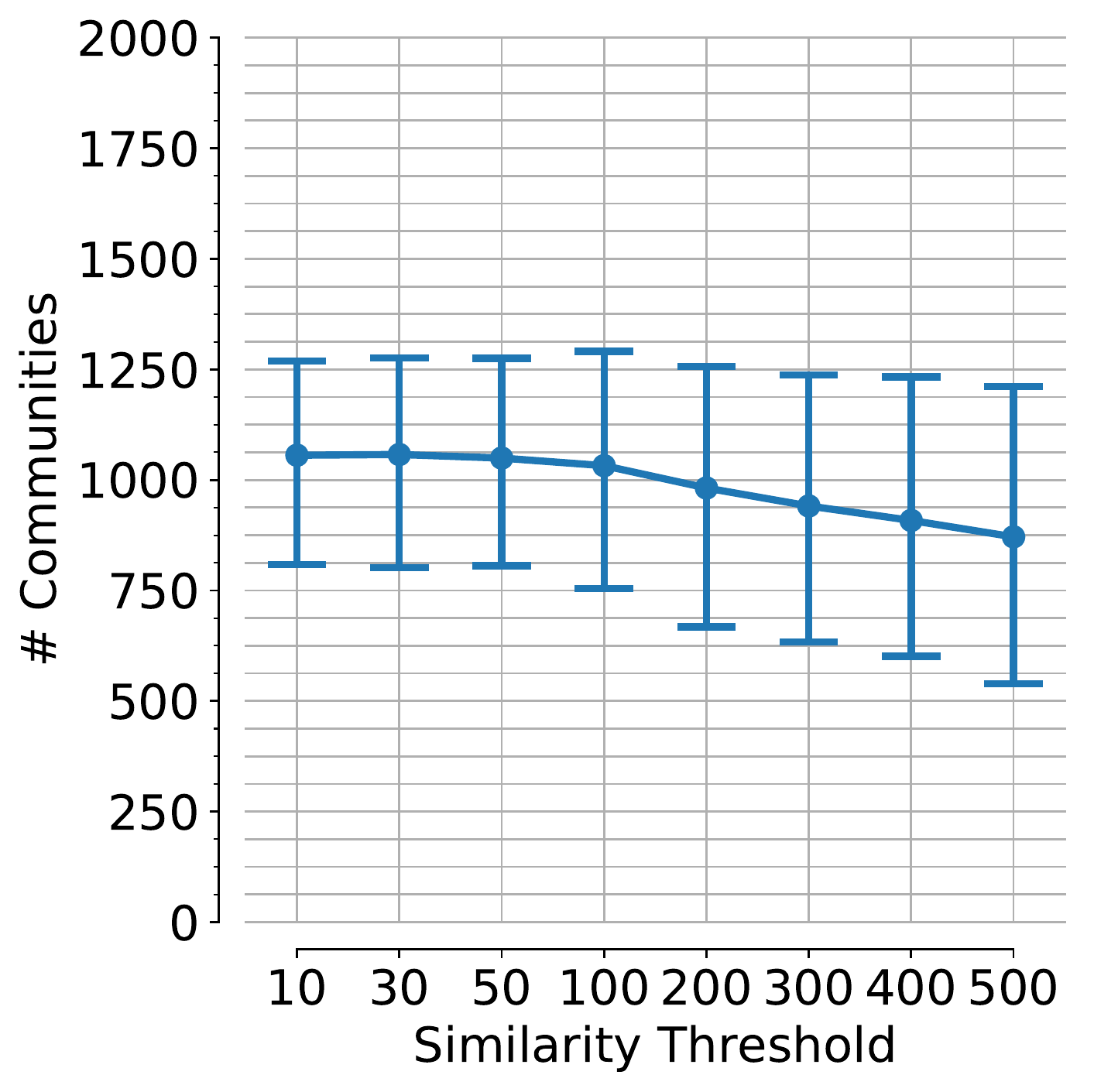}
        }\\
        \subfigure[\scriptsize Content Threshold]{%
            \label{fig:effectiveness_zoo_malware_dtccomn_cntth}
           \includegraphics[width=0.42\textwidth, trim=.1cm 1.0cm 0.1cm 0.0cm, clip]
            {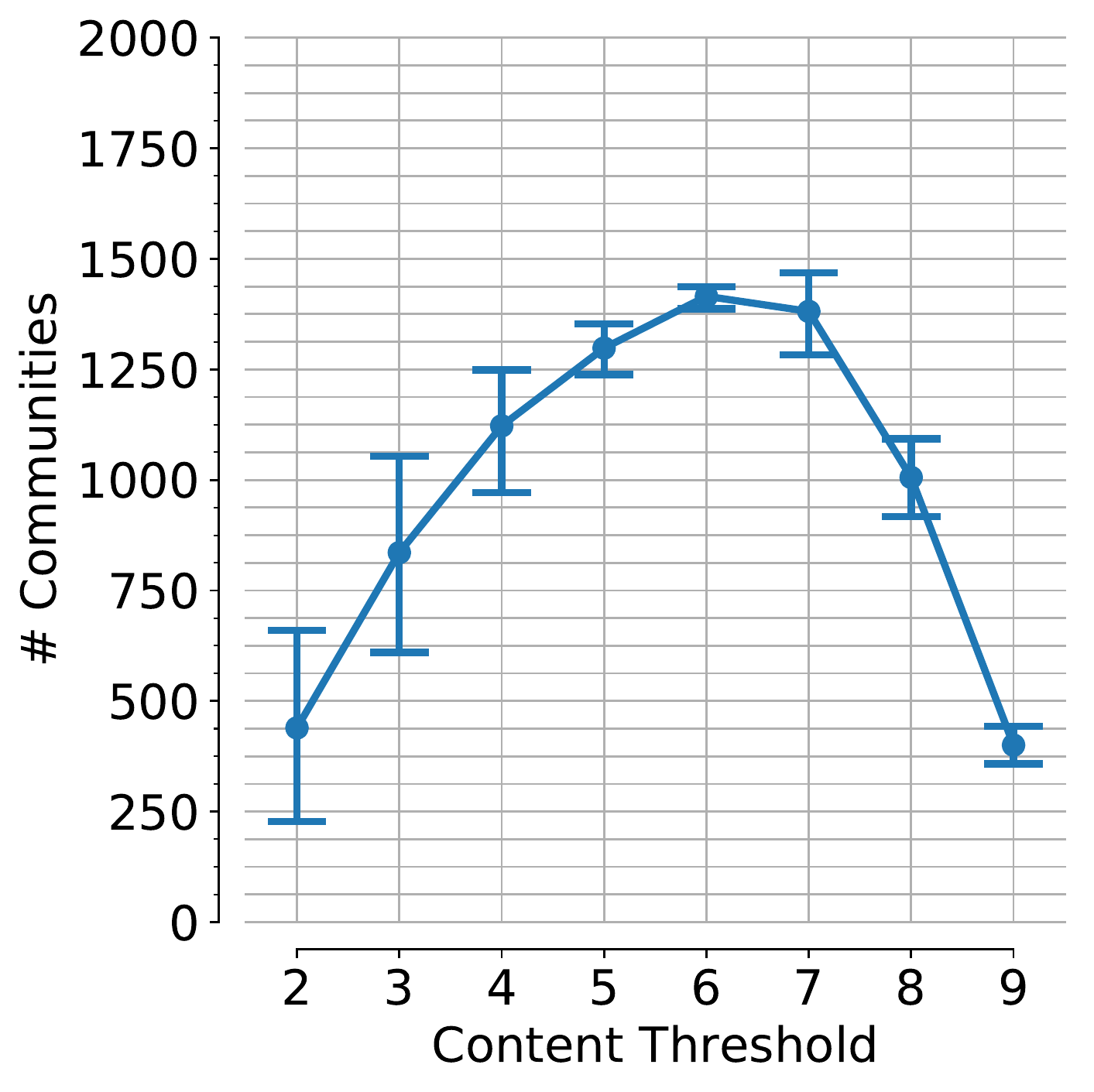}
        }
        \subfigure[\scriptsize Community Size] {%
           \label{fig:effectiveness_zoo_malware_dtccomn_sizeth}
           \includegraphics[width=0.42\textwidth, trim=.1cm 1.0cm 0.1cm 0.0cm, clip]
           {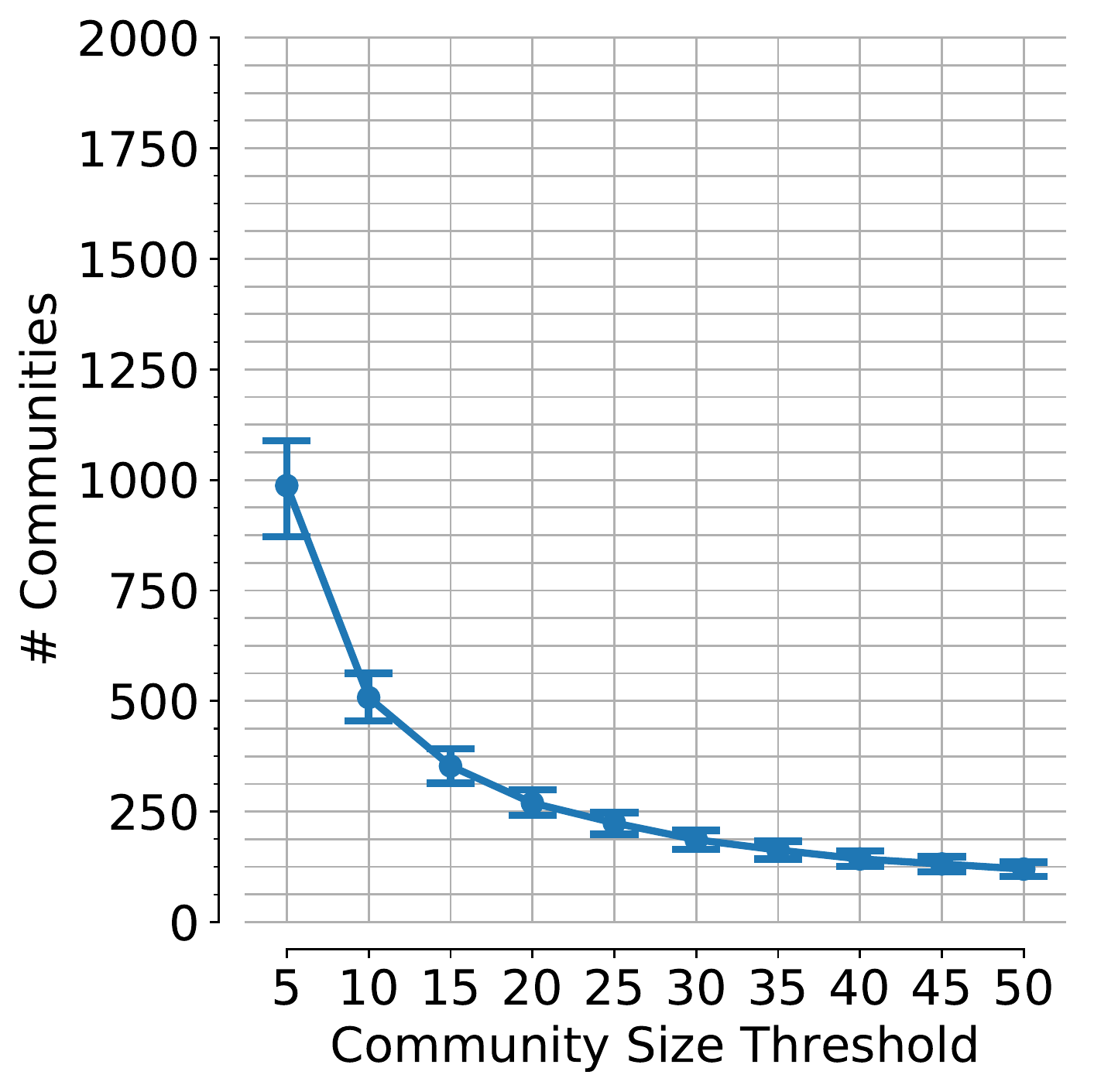}
        }                                                    
    \end{center}
    \caption{Detected Communities, Hyper-parameters Analysis Results of
    AndroZoo Dataset}
   \label{fig_detectionCommunityZoo}
\end{figure}
\end{scriptsize}


\subsection{Efficiency}
\label{sec_efficiency}

In this section, we investigate the overall efficiency of \textsf{Cypider} framework. Specifically, we present the runtime in seconds on the core computation of our framework: (i) the \textbf{similarity computation} to build the similarity network, (ii) the \textbf{community detection} to partition the similarity network into a set of malicious communities.

\paragraph{Efficiency Metrics} \label{sec_efficiency_metrics}

The following metrics aim to answer the following questions: How much time is needed to build the similarity network? How much time is required to extract malicious communities?

\begin{itemize}

\item \textbf{Similarity Network Building Time}: It is the amount of time needed to build the malware similarity network using an approximate similarity computation on the malware FH digests.

\item \textbf{Community Detection Time}: It is the amount of time needed to detect and extract malicious communities from the malware similarity network.

\item \textbf{Total Time}: It is the sum of the times given by the previous metrics.

\end{itemize}

\paragraph{Similarity Computation}

\begin{figure}
  \centering
      \includegraphics[width=0.75\textwidth]{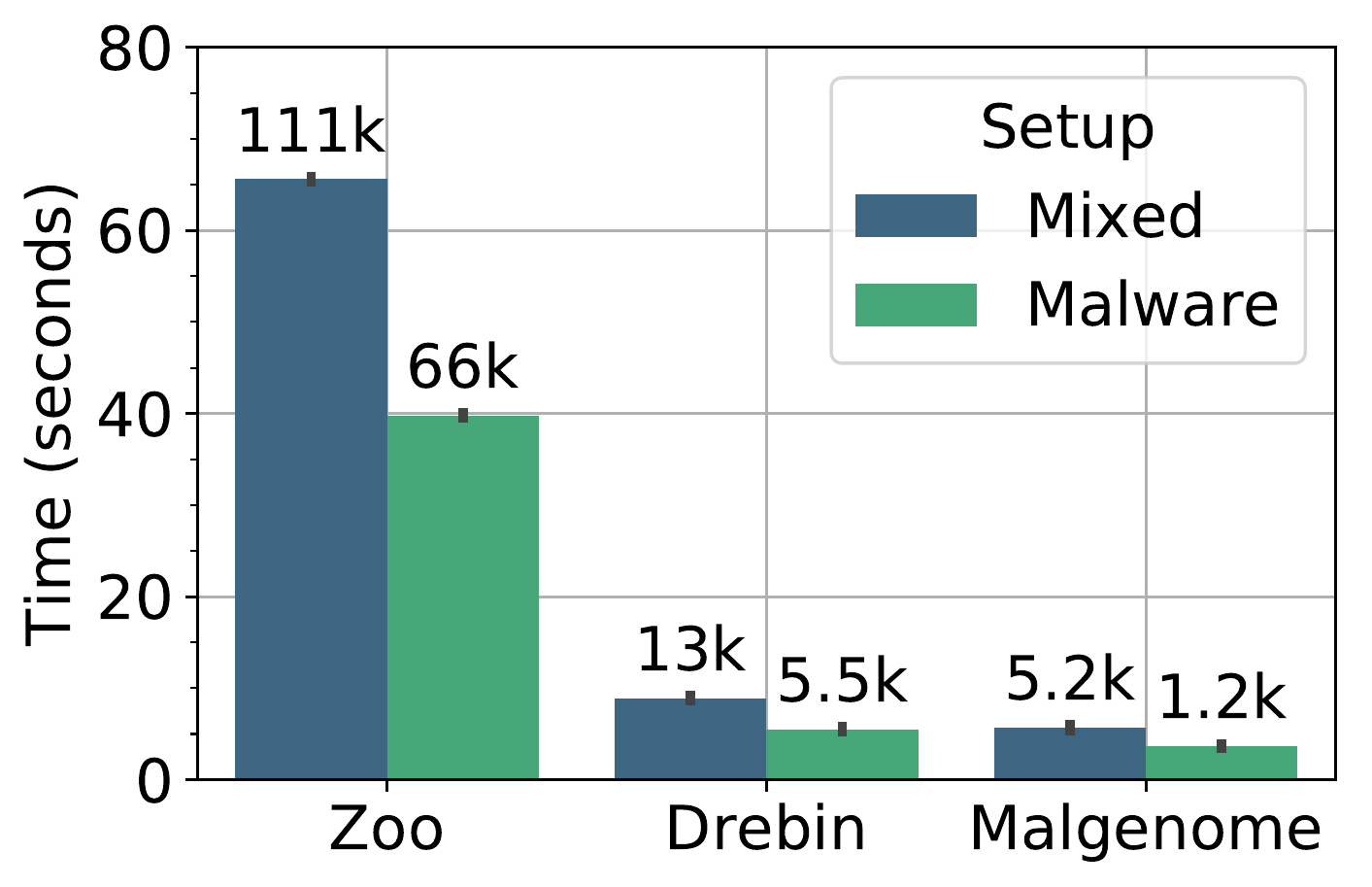}
  \caption{Overall Runtime Under Different Datasets}
  \label{fig_runtimeSimialrity}
\end{figure}

Figure \ref{fig_runtimeSimialrity} depicts the similarity computation time in seconds to build \textsf{Cypider} similarity network. We notice that \textsf{Cypider} framework is very efficient in producing the similarity network because we employ locality sensitive hashing techniques to speed up the pairwise similarity computation between the feature hashing vectors. For example, \textsf{Cypider} took only few seconds to compute the similarity between $111k$ samples feature hashing vectors.

\paragraph{Community Detection}

Figures \ref{fig_runtimeDrebin} and \ref{fig_runtimeZoo} present the community detection time in seconds on \textit{Drebin} and \textit{AndroZoo} datasets respectively. We analyze the effect of \textit{similarity} and \textit{content} thresholds on the overall community detection time. In Figures \ref{fig_runtimeDrebin} and \ref{fig_runtimeZoo}, we notice that \textsf{Cypider} spends more time by decreasing the content threshold and decreasing the similarity threshold in \textit{Drebin} and \textit{AndroZoo} experiments. The previous thresholds setup increases the density of \textsf{Cypider} similarity network, and therefore the community detection processing takes more time in the partition process of the network.  On the other hand, increasing the content threshold while decreasing the similarity threshold produces a very sparse similarity network. The latter takes a negligible time in the partition process.

\begin{scriptsize}
\begin{figure}[ht!]
     \begin{center}        
        \subfigure[Similairty \& Content]{%
            \label{fig:performance_partition_drebin_malware_both}
            \includegraphics[width=0.70\textwidth, trim=.1cm 0.5cm 2.cm .0cm, clip]
            {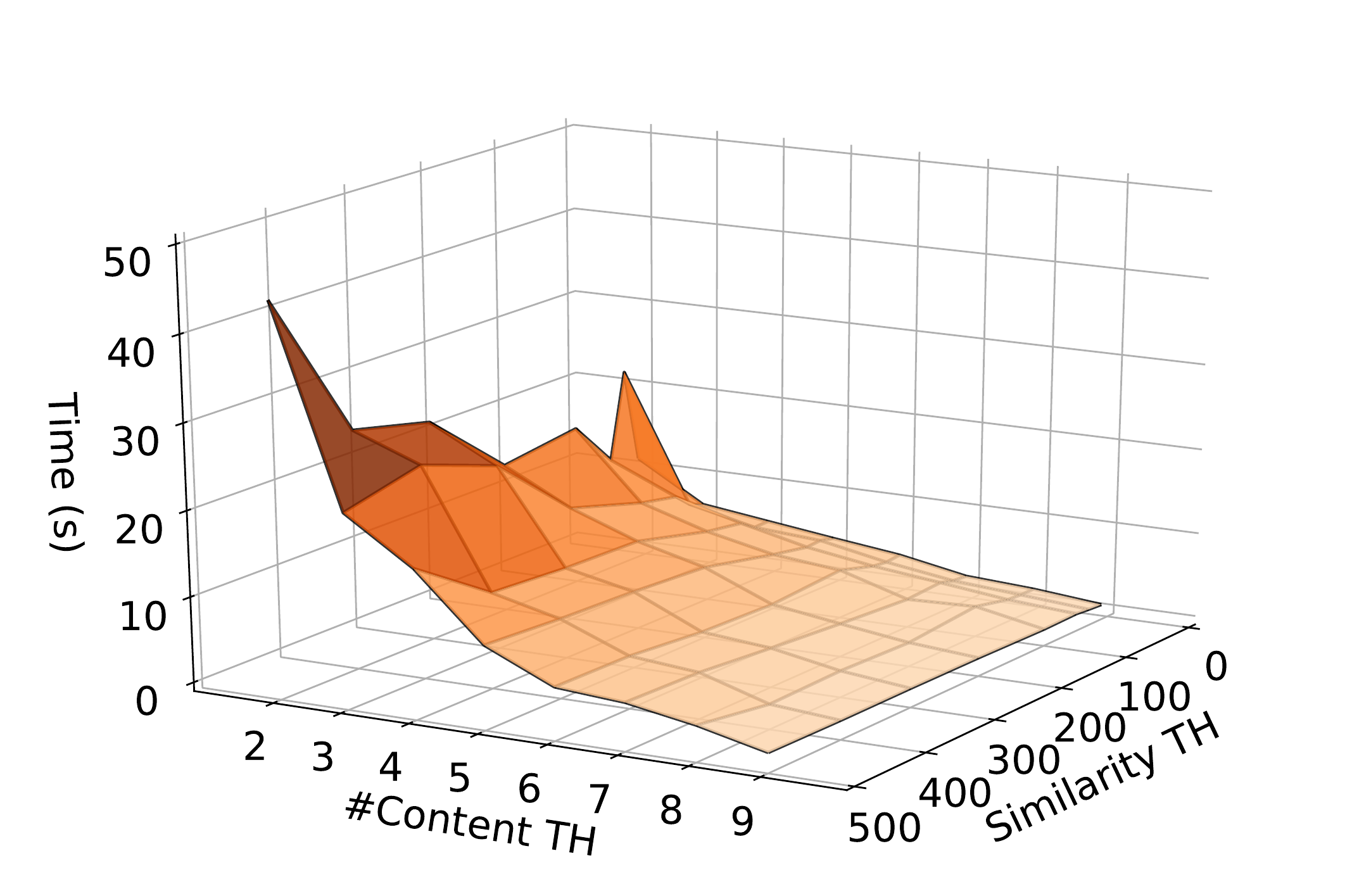}
        }\\
        \subfigure[Similarity Threshold] {%
           \label{fig:performance_partition_drebin_malware_both}
           \includegraphics[width=0.45\textwidth, trim=.1cm 1.0cm .0cm 0.0cm, clip]
           {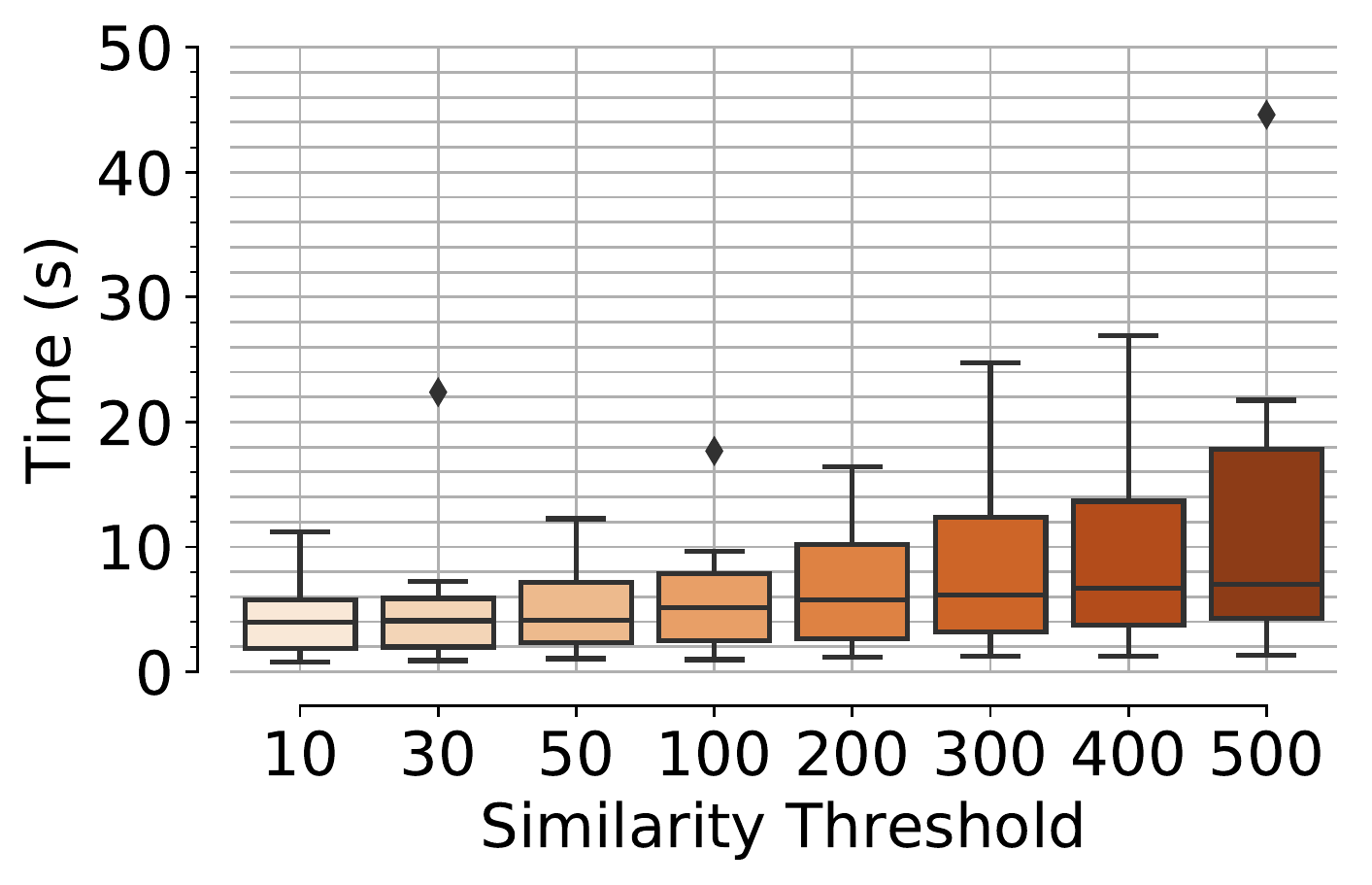}
        }
        \subfigure[Content Threshold]{%
            \label{fig:performance_partition_drebin_malware_both}
            \includegraphics[width=0.45\textwidth, trim=.1cm 1.0cm .0cm 0.0cm, clip]
            {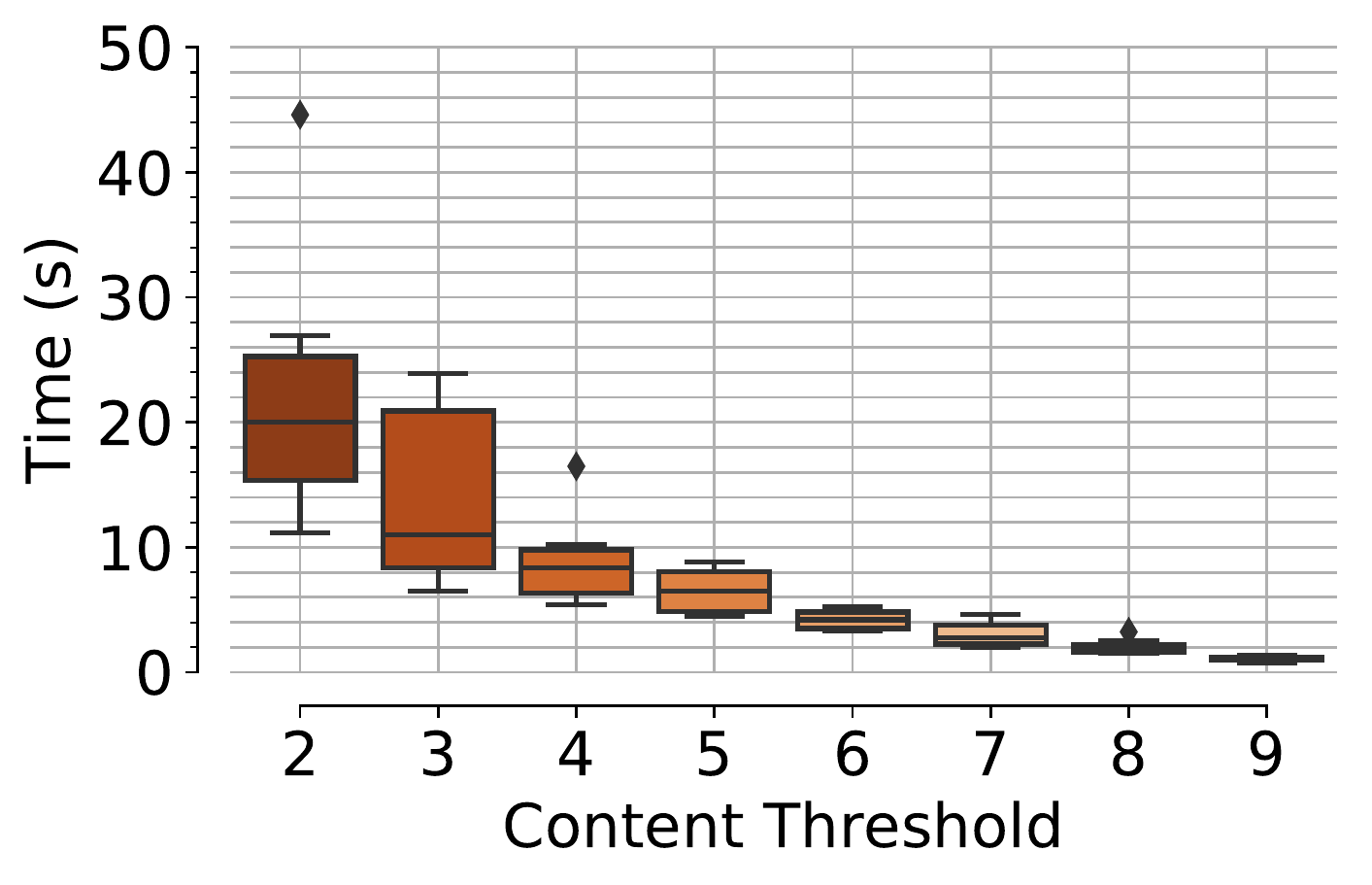}
        }
    \end{center}
    \caption{\textsf{Cypider} Community Detection Time, Hyper-parameters
    Analysis (Drebin Malware Scenario)}
   \label{fig_runtimeDrebin}
\end{figure}
\end{scriptsize}

\begin{scriptsize}
\begin{figure}[ht!]
     \begin{center}        
        \subfigure[Similairty \& Content]{%
            \label{fig:performance_partition_zoo_malware_both}
            \includegraphics[width=0.70\textwidth, trim=.1cm 0.5cm 2.cm .0cm, clip]
            {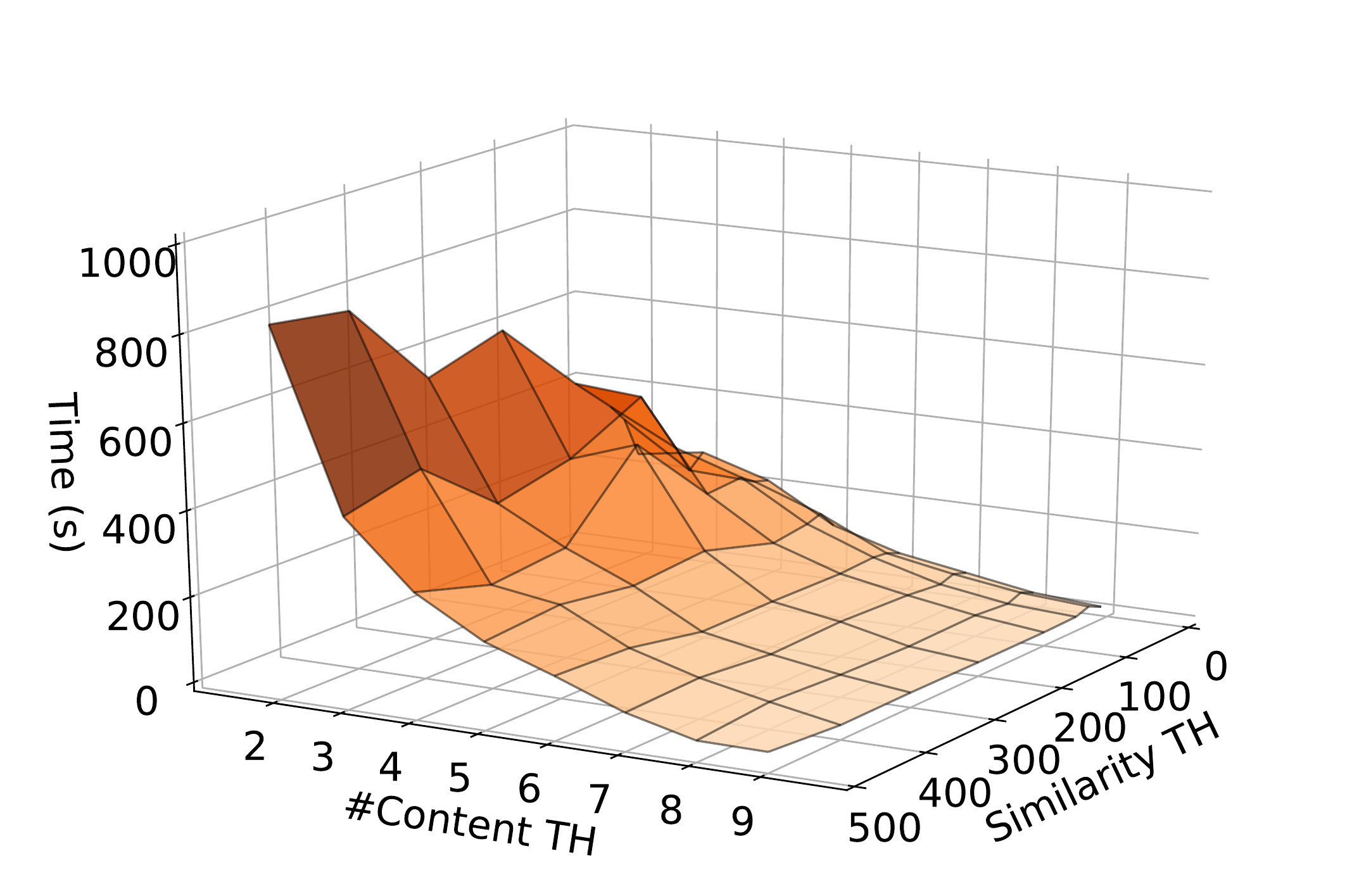}
        }\\
        \subfigure[Similarity Threshold] {%
           \label{fig:performance_partition_zoo_malware_both}
           \includegraphics[width=0.45\textwidth, trim=.1cm 1.0cm .0cm 0.0cm, clip]
           {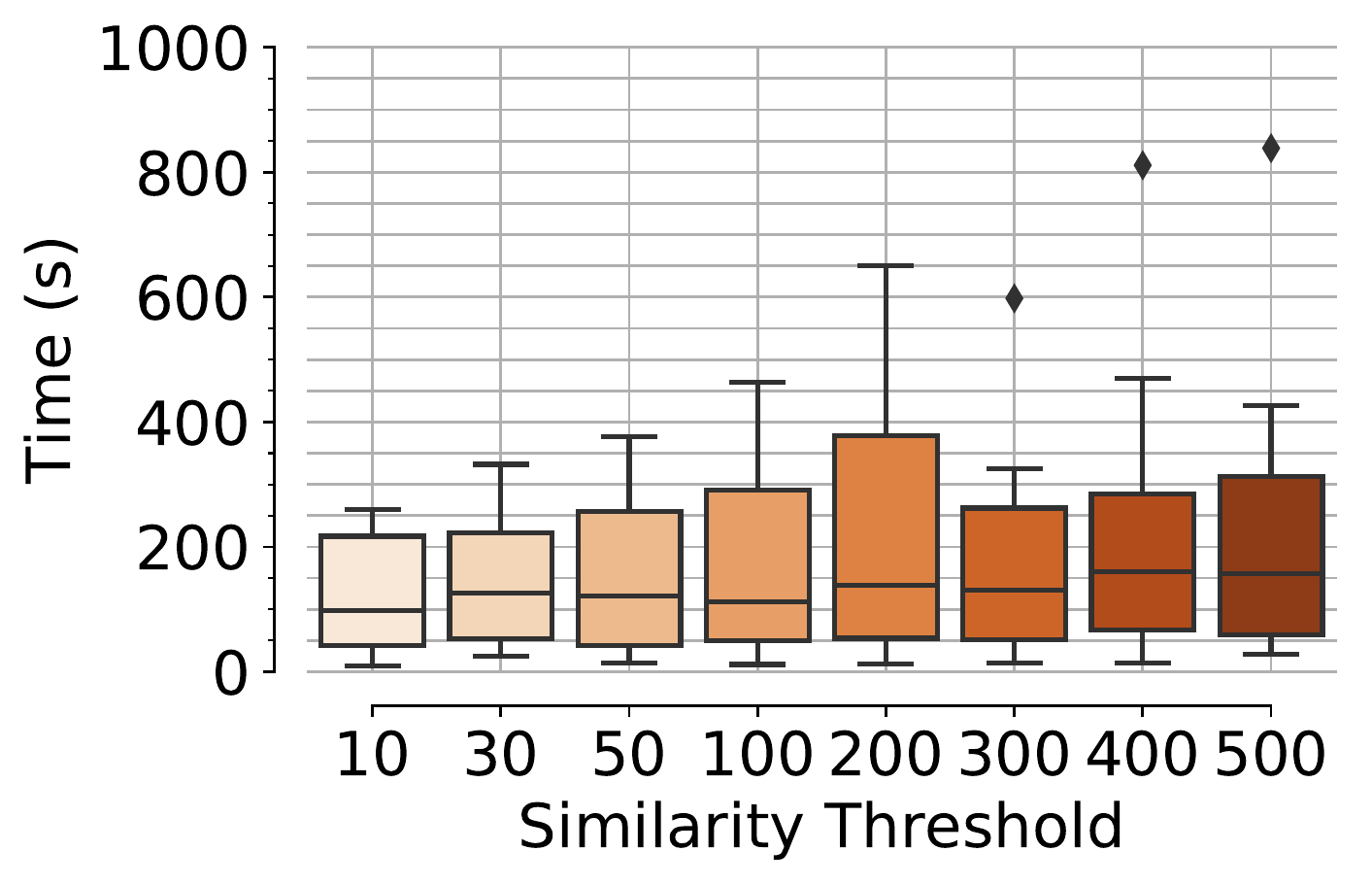}
        }
        \subfigure[Content Threshold]{%
            \label{fig:performance_partition_zoo_malware_both}
            \includegraphics[width=0.45\textwidth, trim=.1cm 1.0cm .0cm 0.0cm, clip]
            {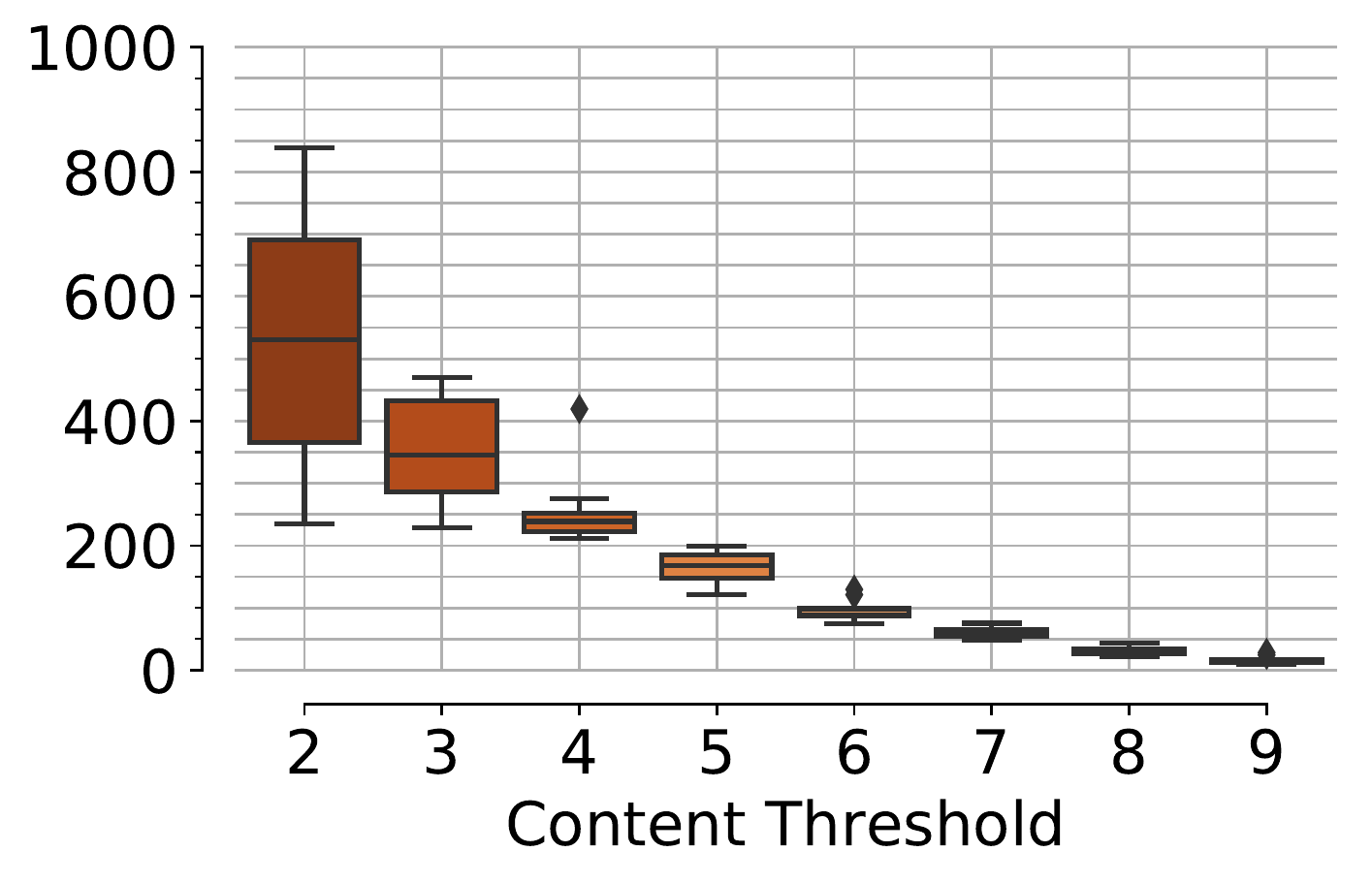}
        }
    \end{center}
    \caption{\textsf{Cypider} Community Detection Time, Hyper-parameters
    Analysis (Zoo Malware Scenario)}
   \label{fig_runtimeZoo}
\end{figure}
\end{scriptsize}

\section{Case Study: Recall and Precision Settings}
\label{sec_recallPrecisionSettings}

In this section, we evaluate \textsf{Cypider} framework under recall and precision settings. We aim to assess \textsf{Cypider} performance in terms of \textit{purity} and \textit{coverage} in case the security practitioner focuses on having: (i) maximum recall (a minimum false detection), or (ii) maximum precision (maximum coverage). Both recall and precision settings are common in real deployments. We achieved the recall and precision settings by adjusting \textsf{Cypider} hyperparameters to reach the set goal.
Figure \ref{fig_malwareResultRPS} presents \textsf{Cypider} malware only performance on different datasets (\textit{MalGenome, Drebin, AndroZoo}) under recall and precision settings. In the recall setting, \textsf{Cypider} achieved $95\%-100\%$ purity while maintaining $15\%-26\%$ malware coverage. Therefore, we detect about $20\%$ (on average) of the input malware in the form of communities with $98\%$ purity (Figure \ref{fig_malwareResultRPS}, recall charts). On the other hand, \textsf{Cypider} achieved $63\%-73\%$ malware coverage while maintaining $84\%-93\%$ purity, as shown in Figure \ref{fig_malwareResultRPS} (precision charts). 

\begin{scriptsize}
\begin{figure}[ht!]
     \begin{center}        
        \subfigure[Recall]{%
            \label{fig:malgenome_mixed_effective_malware_recall_pie}
            \includegraphics[width=0.35\textwidth, trim=2.39cm 1.2cm 2.39cm 1.2cm, clip]
            {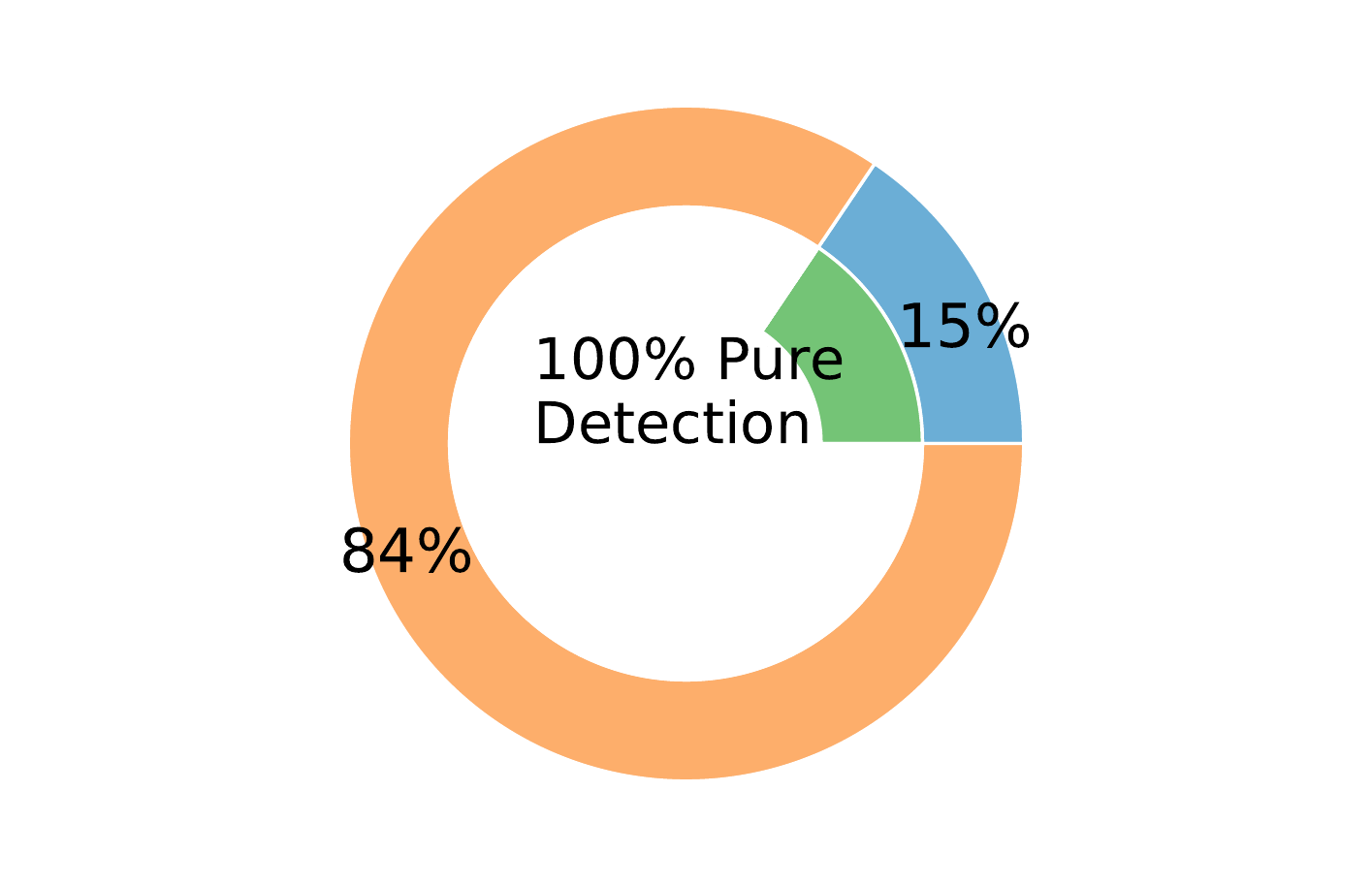}
        }
        \subfigure[Precision] {%
           \label{fig:malgenome_mixed_effective_malware_precision_pie}
           \includegraphics[width=0.35\textwidth, trim=2.39cm 1.2cm 2.39cm 1.2cm, clip]
           {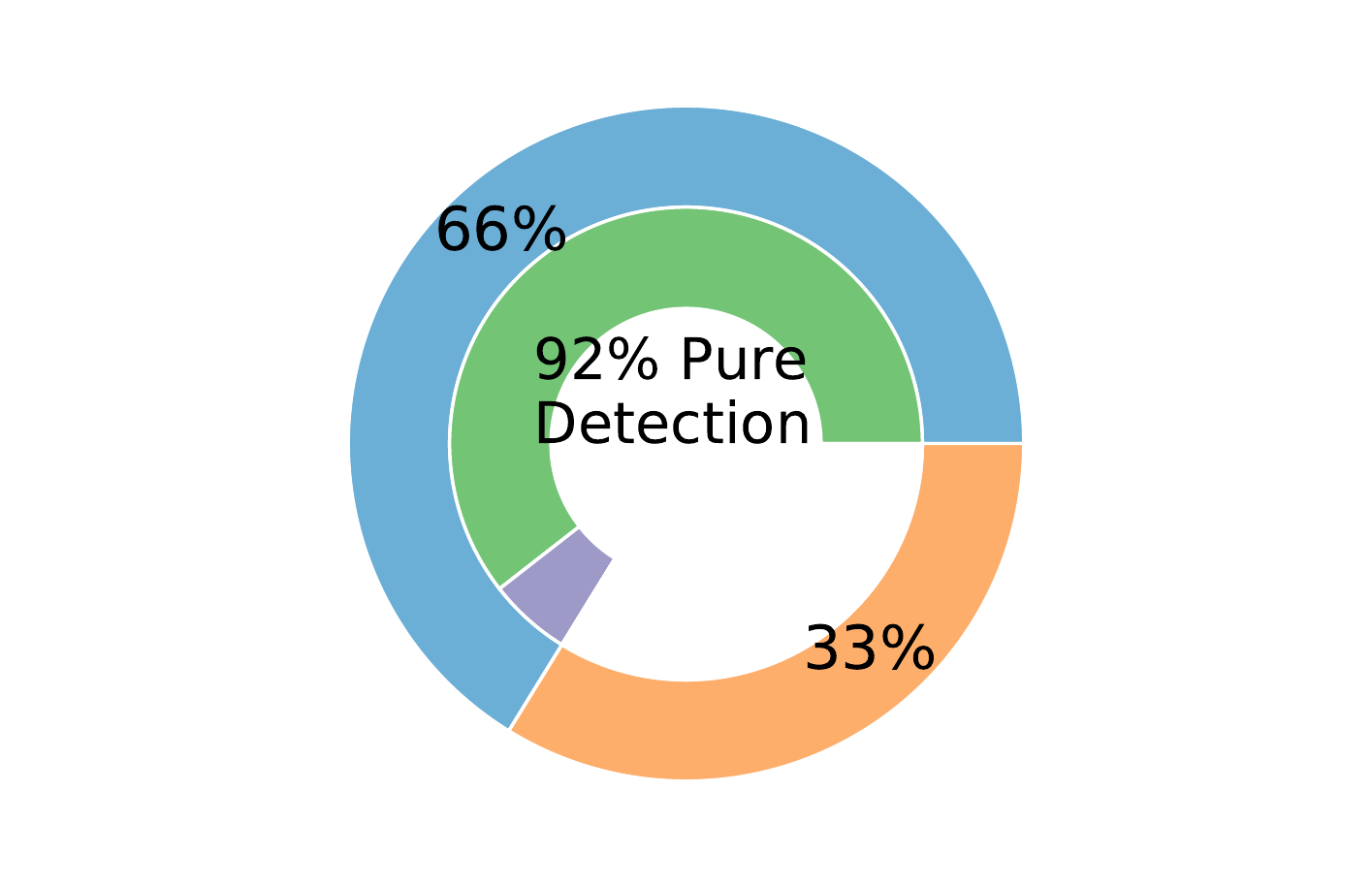}
        }
        \\
        {MalGenome Dataset}
        \\
        \subfigure[Recall]{%
            \label{fig:drebin_mixed_effective_malware_recall_pie}
            \includegraphics[width=0.35\textwidth, trim=2.39cm 1.2cm 2.39cm 1.2cm, clip]
            {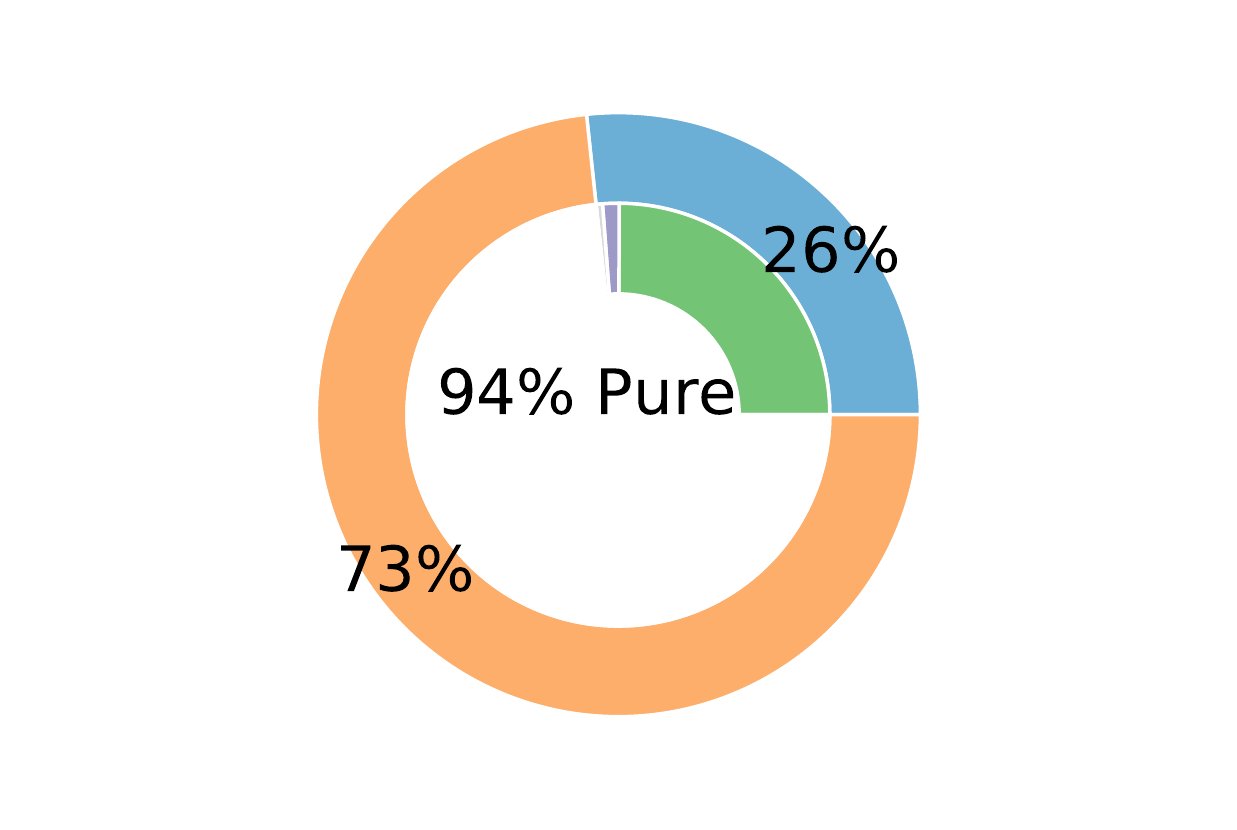}
        }
        \subfigure[Precision]{%
           \label{fig:drebin_mixed_effective_malware_precision_pie}
           \includegraphics[width=0.35\textwidth, trim=2.39cm 1.2cm 2.39cm 1.2cm, clip]
           {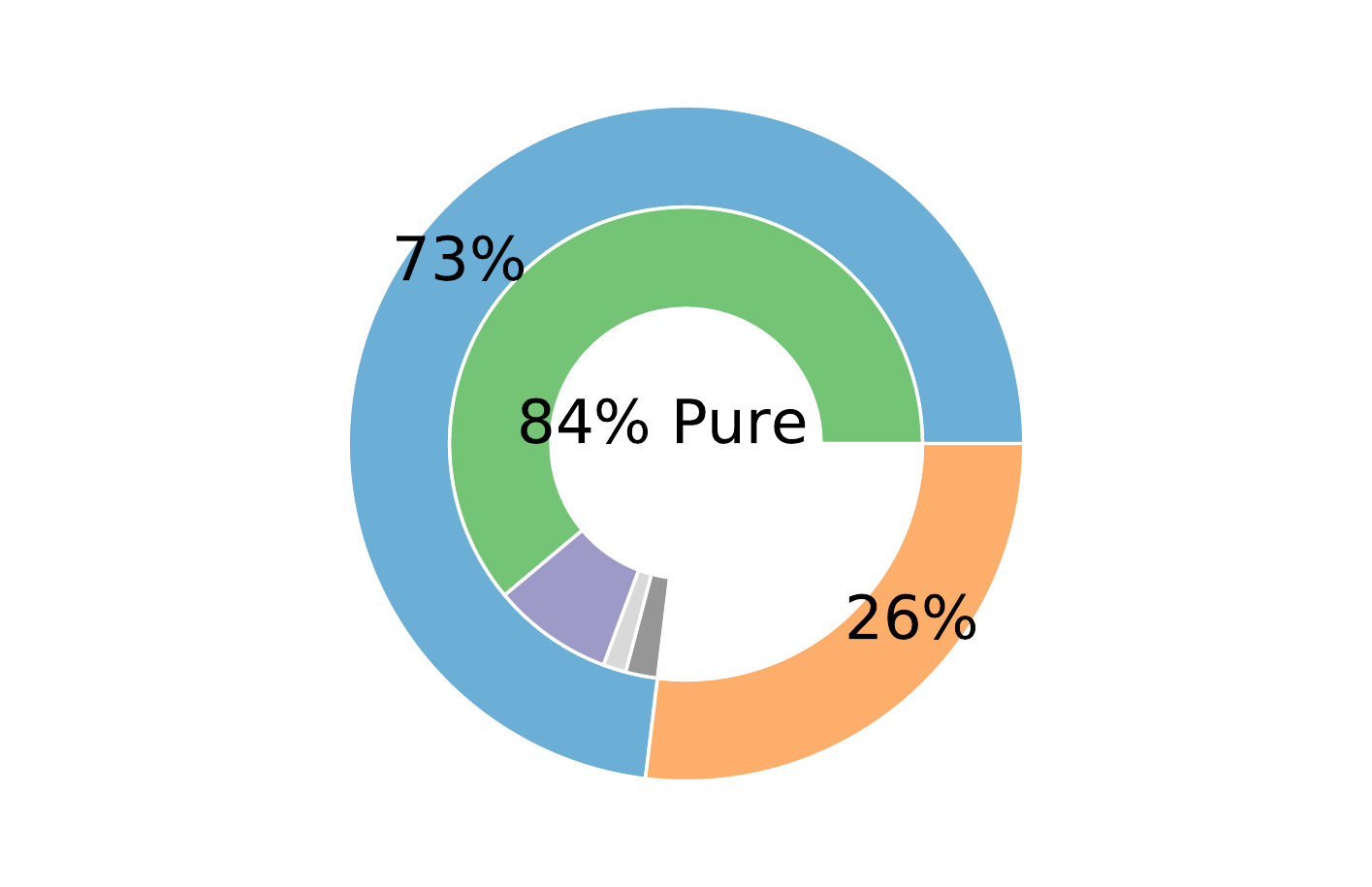}
        }
        \\
        {Drebin Dataset}
        \\
        \subfigure[Recall]{%
            \label{fig:zoo_mixed_effective_malware_recall_pie}
           \includegraphics[width=0.35\textwidth, trim=2.39cm 1.2cm 2.39cm 1.2cm, clip]
            {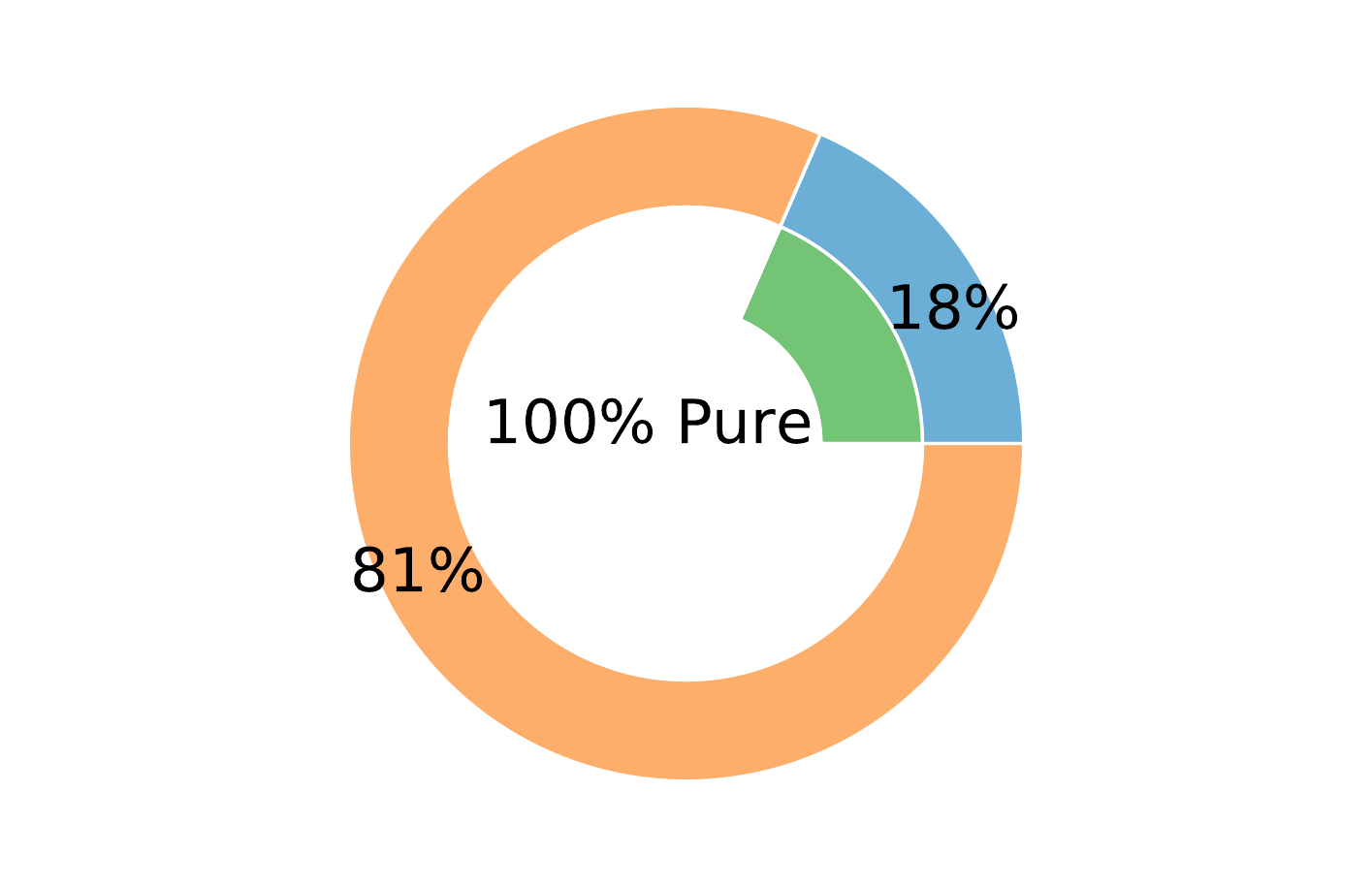}
        }
        \subfigure[Precision] {%
           \label{fig:zoo_mixed_effective_malware_precision_pie}
           \includegraphics[width=0.35\textwidth, trim=2.39cm 1.2cm 2.39cm 1.2cm, clip]
           {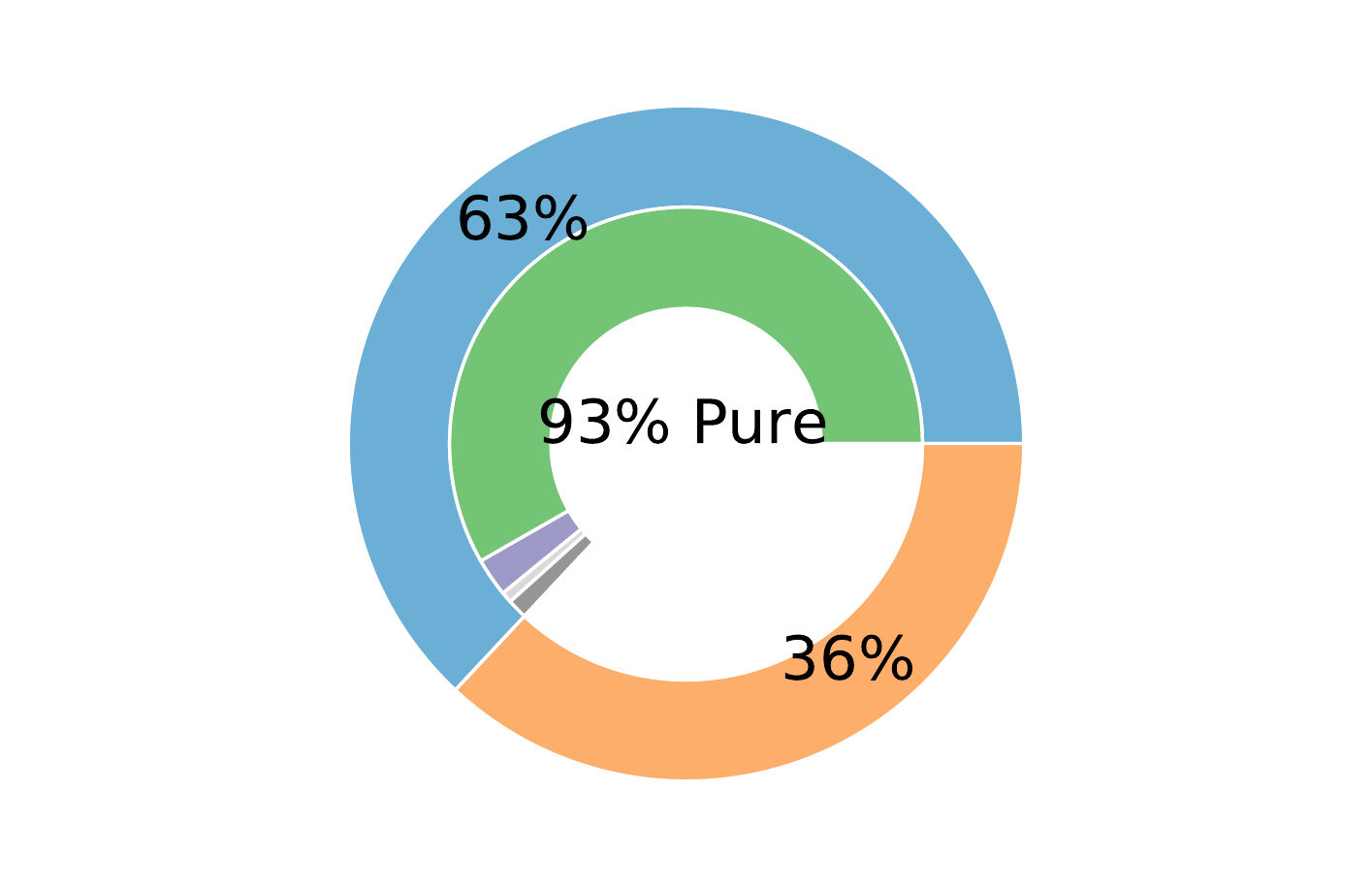}
        }
        \\
        {AndroZoo Dataset}
    \end{center}
    \caption{ \textsf{Cypider} Performance under Recall/Precision Settings
    On Malware Scenario}
    \label{fig_malwareResultRPS}
\end{figure}
\end{scriptsize}

\begin{scriptsize}
\begin{figure}[h!]
     \begin{center}
        \subfigure[Recall Settings]{%
            \label{fig:malgenome_malware_effective_recall_sim_network}
            \includegraphics[width=0.65\textwidth, trim=0.0cm 0.0cm 0.0cm 0.0cm, clip]
            {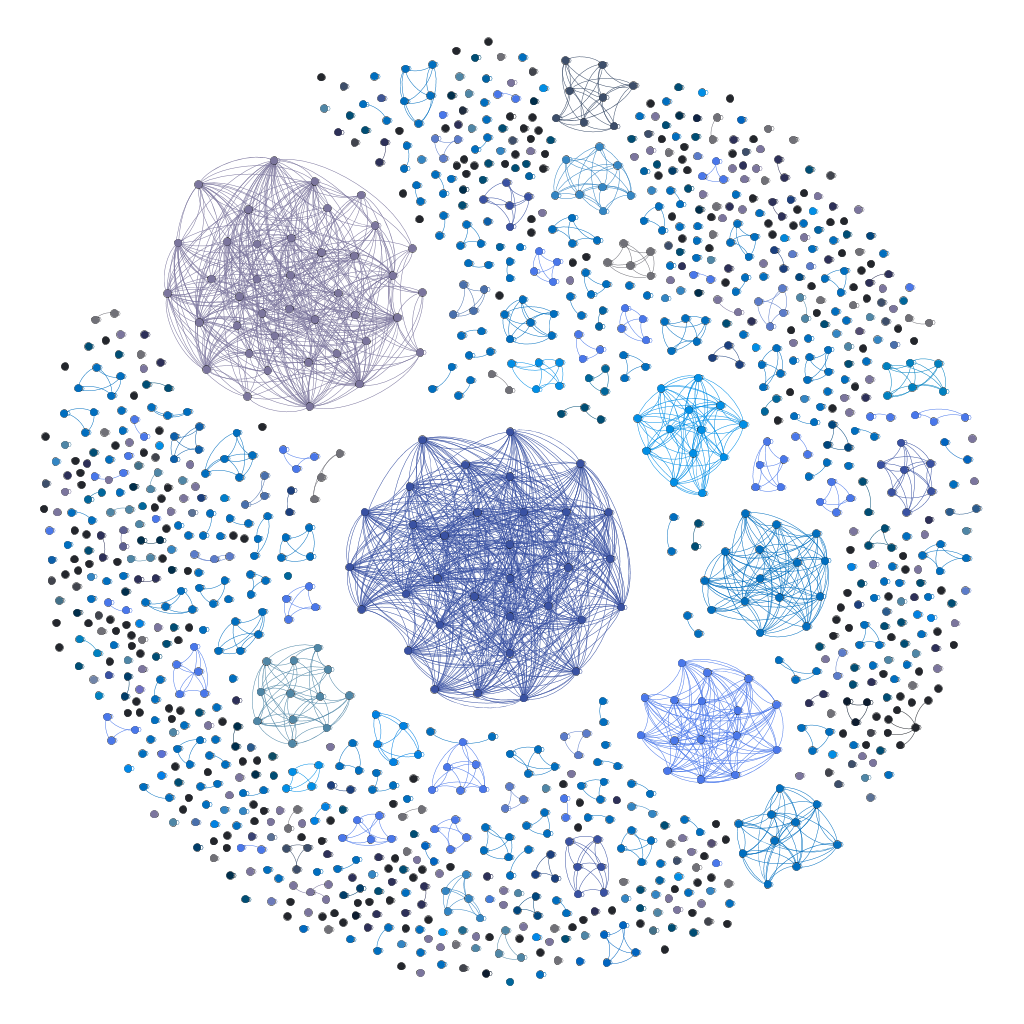}
        }\\
        \subfigure[Precision Settings] {%
           \label{fig:malgenome_malware_effective_precision_sim_network}
           \includegraphics[width=0.65\textwidth, trim=0.0cm 0.0cm 0.0cm 0.0cm, clip]
           {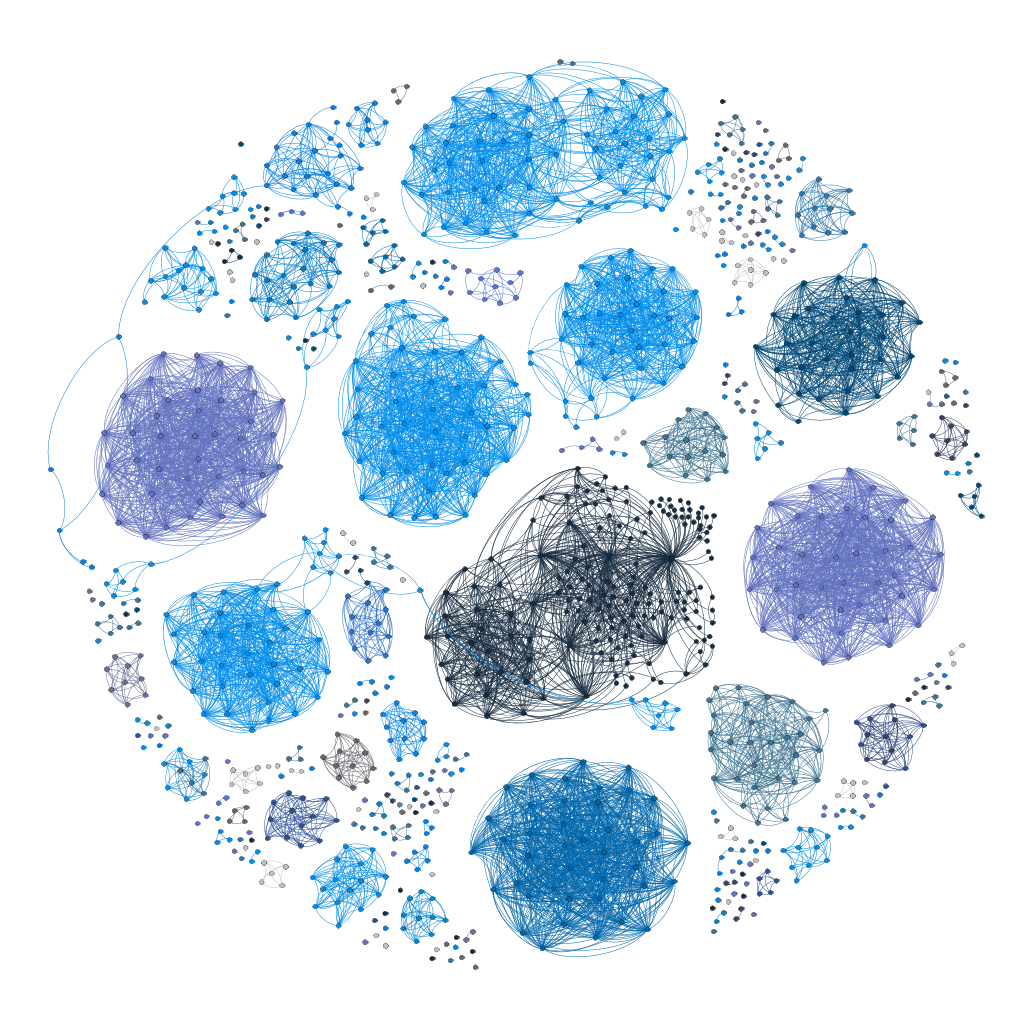}
        }
    \end{center}
  \caption{\textit{Malgenome} Malware Dataset, Similarity Network Under
  Recall/Precision Settings}
  \label{fig_malNetGenomeNetRPS}
\end{figure}
\end{scriptsize}

The contrast between the recall and precision settings is more clear visually in the similarity network, as shown in Figures \ref{fig_malNetGenomeNetRPS} and \ref{fig_malNetDrebinNetRPS} for \textit{MalGenome} and \textit{Drebin} datasets respectively.  Figure \ref{fig_malNetGenomeNetRPS} and Figure \ref{fig_malNetDrebinNetRPS} present each malware family in a different color.  Malware communities depicted with more than one color contain more than one malware family. Pure malware communities have only one color in the edges and nodes. We notice more detected malware communities in the similarity network in the precision settings. In contrast, in the recall similarity network, we notice fewer malware communities, and most of the nodes are part of any not-detected  community. 

Figure \ref{fig_mixedResultRPS} depicts \textsf{Cypider} mixed performance under recall and precision settings for \textit{MalGenome}, \textit{Drebin} and \textit{AndroZoo} datasets.  The most noticeable result is that all the detected benign communities have perfect purity metrics under both recall and precision settings. Moreover, benign coverage is less than the malware coverage under all settings. In other words, \textsf{Cypider} could result in benign samples in the clustering but gathered in pure communities, which is very helpful in case of manual investigations. 

\begin{scriptsize}
\begin{figure}[ht!]
     \begin{center}
        \subfigure[Recall Settings]{%
            \label{fig:drebin_malware_effective_recall_sim_network}
            \includegraphics[width=0.65\textwidth, trim=1.0cm 1.0cm 1.0cm 1.0cm, clip]
            {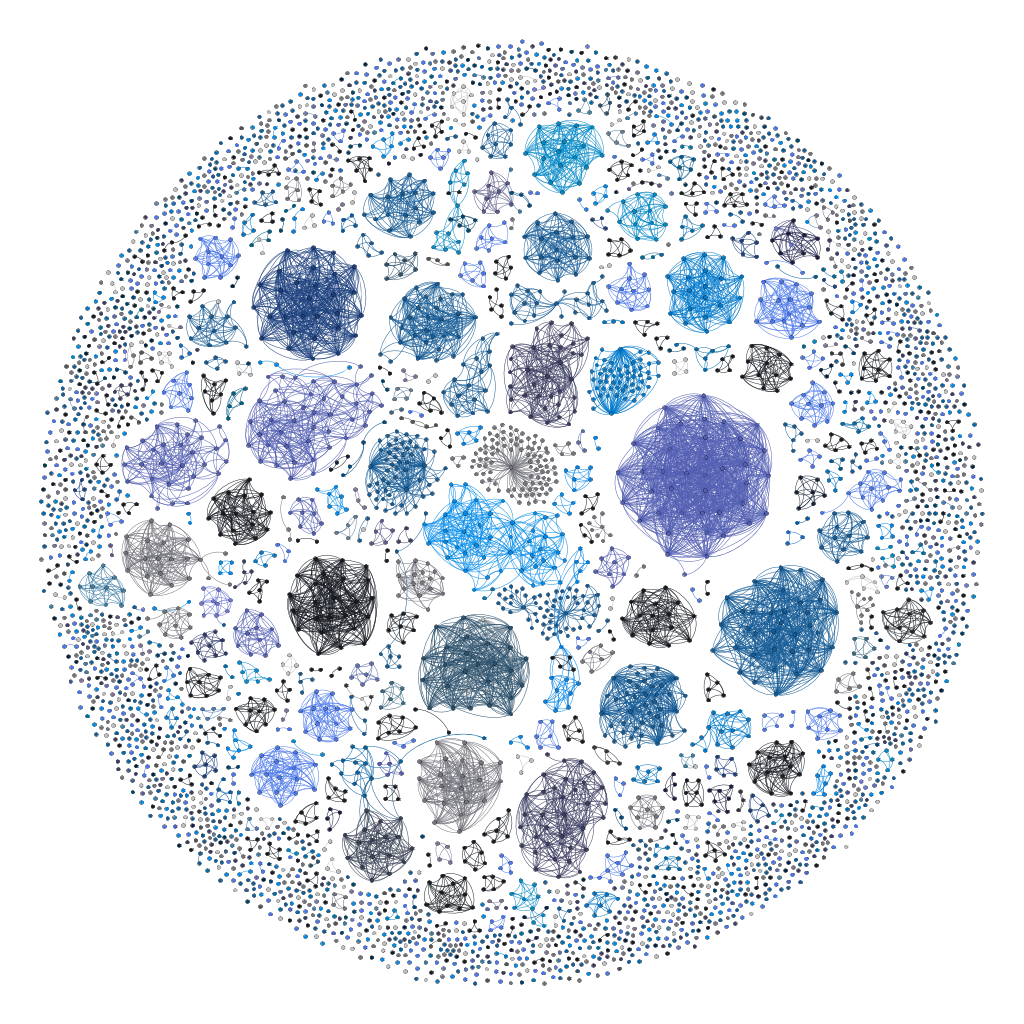}
        }\\
        \subfigure[Precision Settings] {%
           \label{fig:drebin_malware_effective_precision_sim_network}
           \includegraphics[width=0.65\textwidth, trim=0.0cm 0.0cm 0.0cm 0.0cm, clip]
           {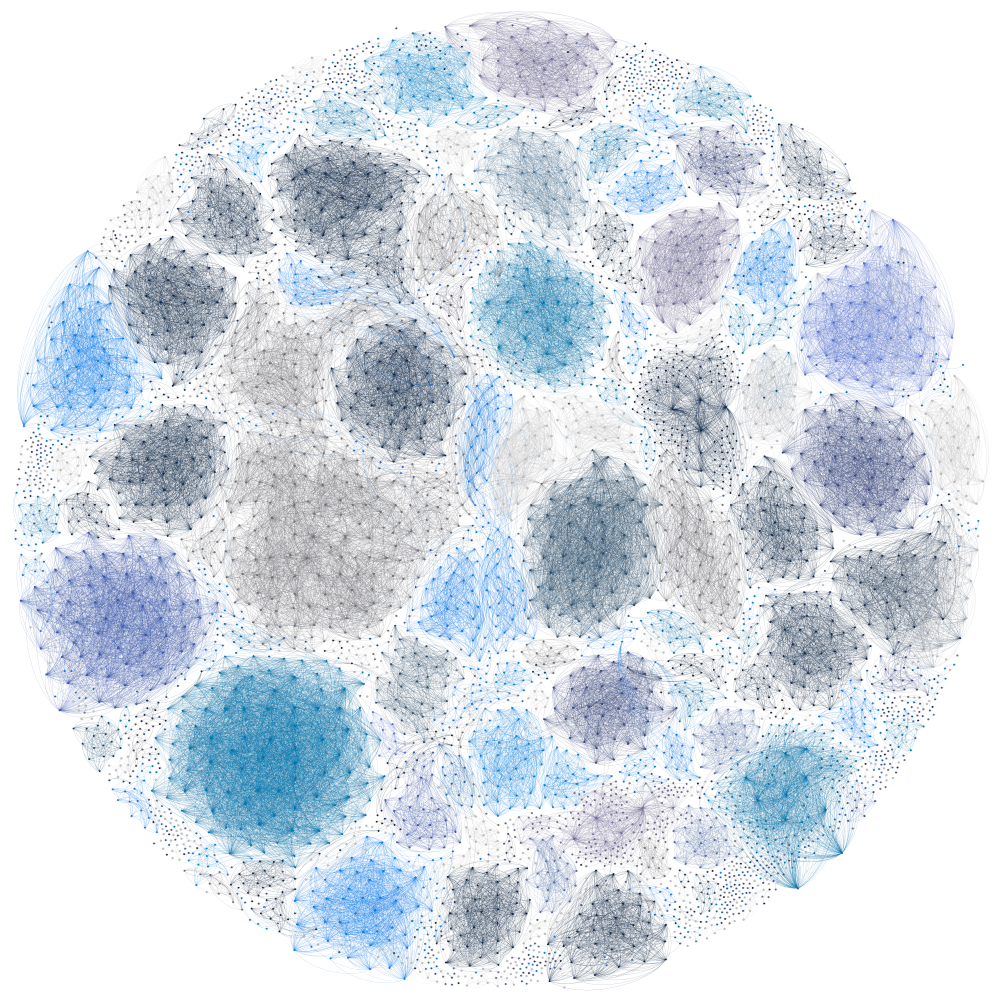}
        }
    \end{center}
    \caption{\textit{Drebin} Malware Dataset, Similarity Network Under Recall/Precision Settings}
  \label{fig_malNetDrebinNetRPS}
\end{figure}
\end{scriptsize}

\begin{scriptsize}
\begin{figure}[ht!]
     \begin{center}        
        \subfigure[\scriptsize Malware (Recall)]{%
            \label{fig:malgenome_mixed_effective_malware_recall_pie}
            \includegraphics[width=0.20\textwidth, trim=2.39cm 1.2cm 2.39cm 1.2cm, clip]
            {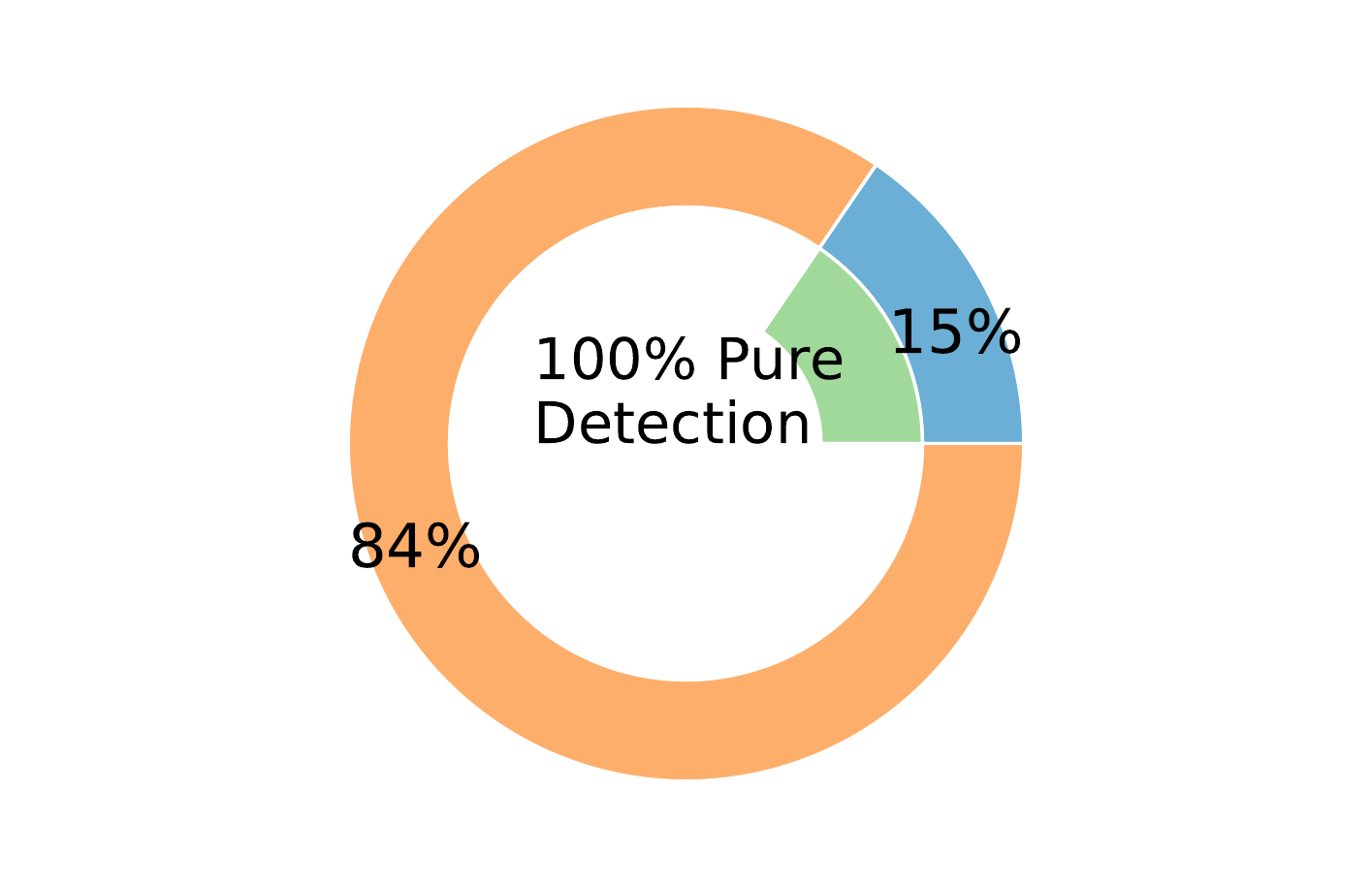}
        }
        \subfigure[\scriptsize Benign (Recall)] {%
           \label{fig:malgenome_mixed_effective_benign_recall_pie}
           \includegraphics[width=0.20\textwidth, trim=2.39cm 1.2cm 2.39cm 1.2cm, clip]
           {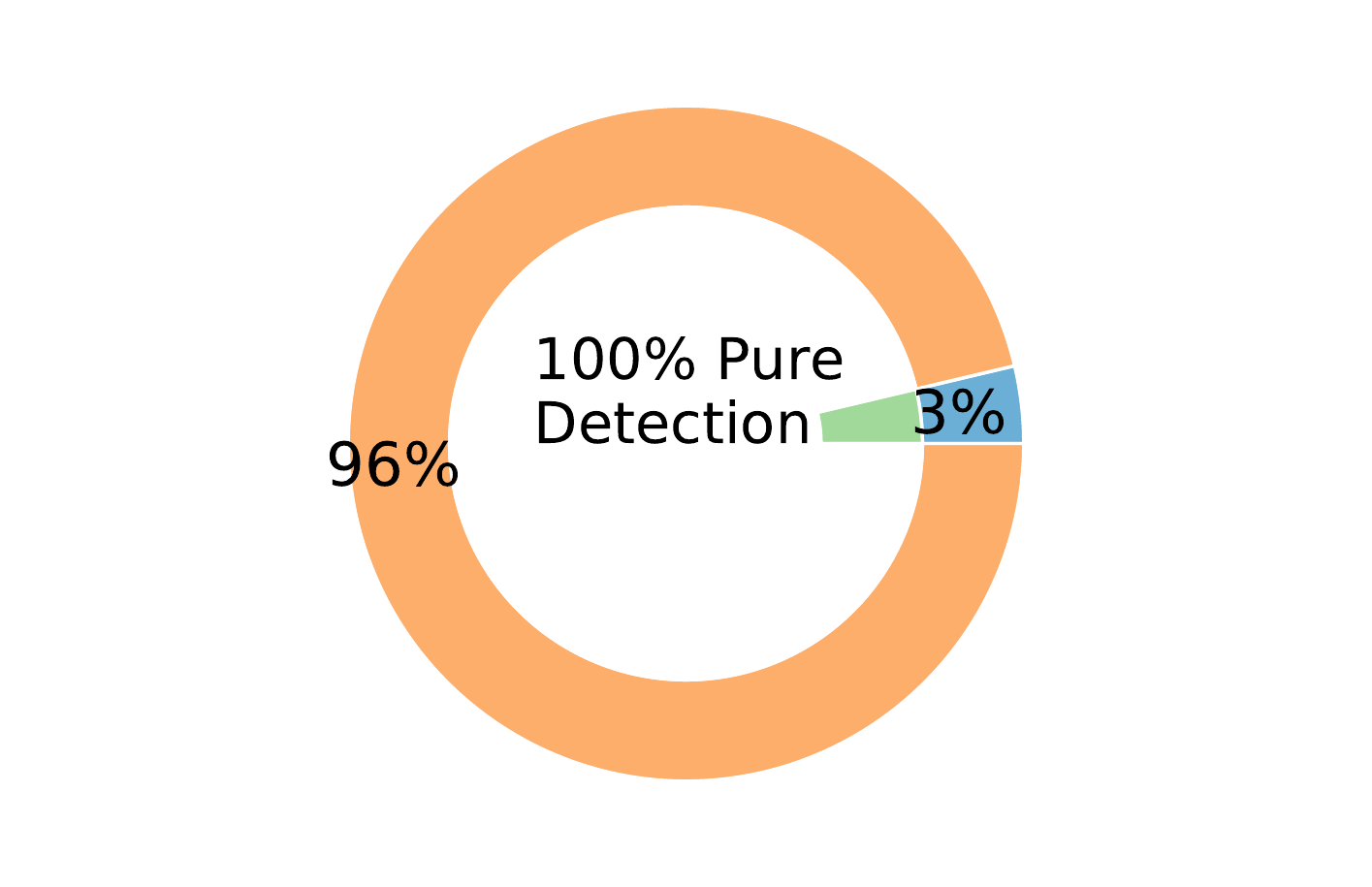}
        }
        \subfigure[\scriptsize Malware (Precision)]{%
            \label{fig:malgenome_mixed_effective_malware_precision_pie}
            \includegraphics[width=0.20\textwidth, trim=2.39cm 1.2cm 2.39cm 1.2cm, clip]
            {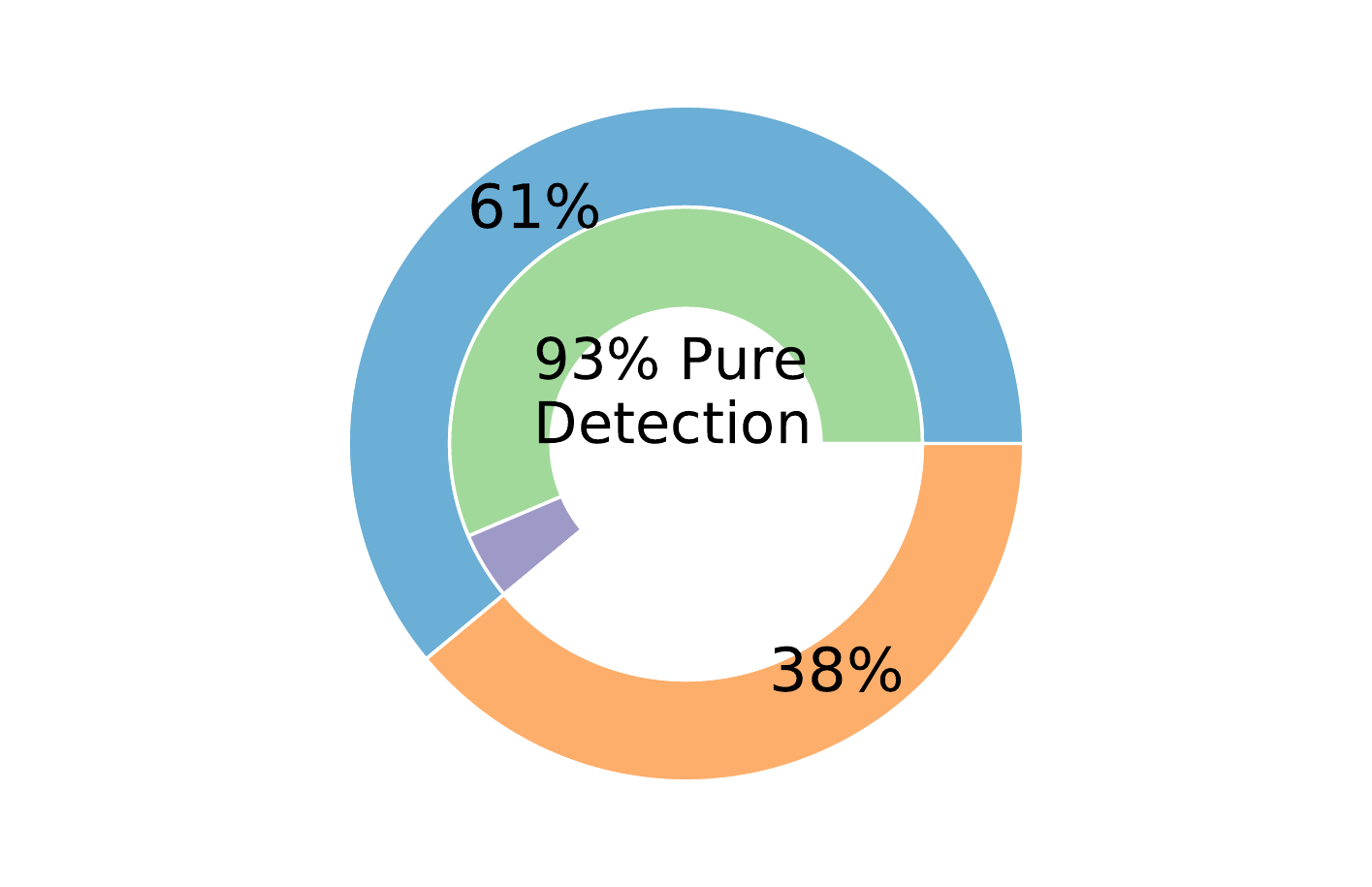}
        }
        \subfigure[\scriptsize Benign (Precision)] {%
           \label{fig:malgenome_mixed_effective_benign_precision_pie}
           \includegraphics[width=0.20\textwidth, trim=2.39cm 1.2cm 2.39cm 1.2cm, clip]
           {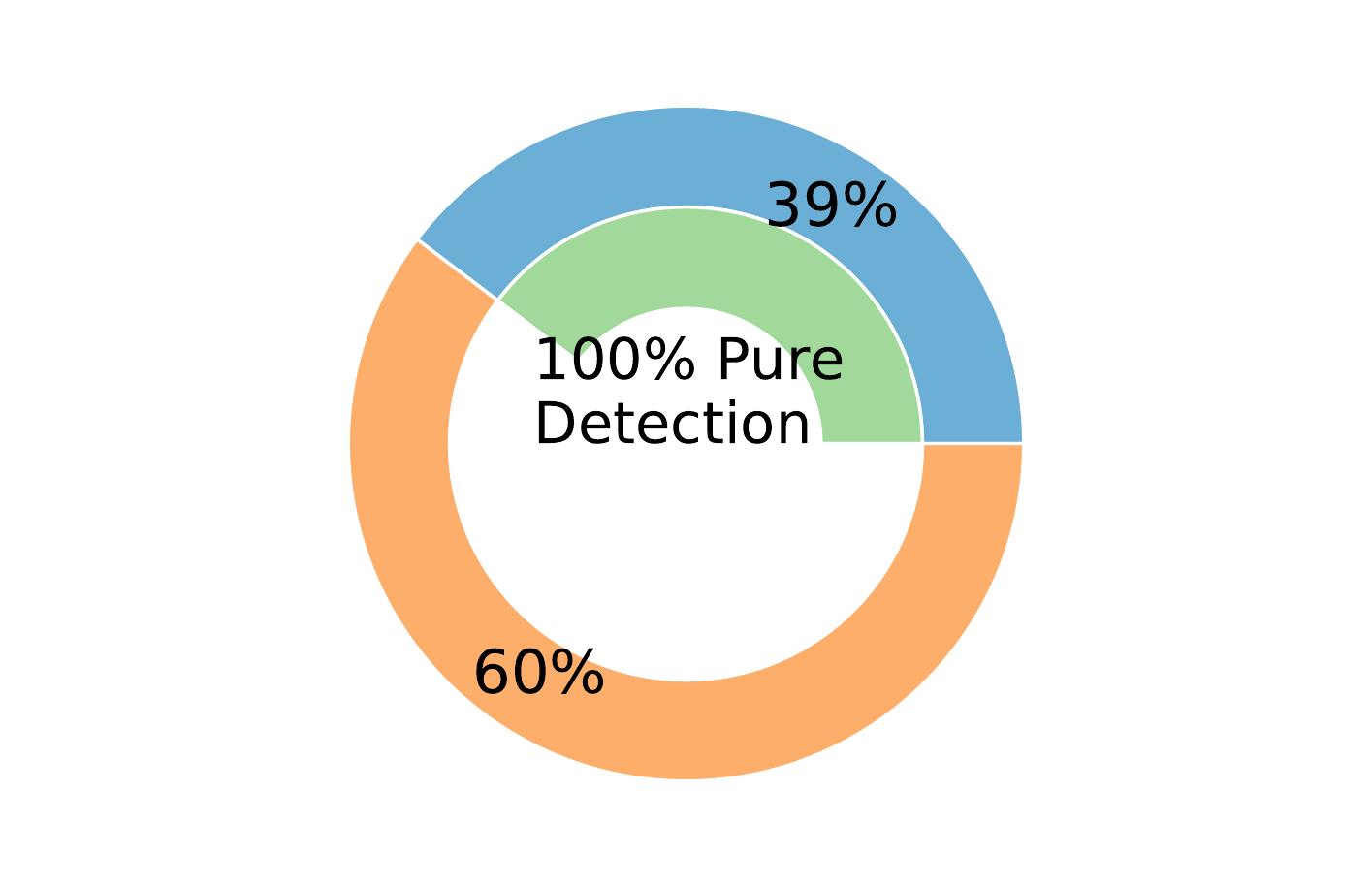}
        }
        \\
        {\small MalGenome Dataset}
        \\
        \subfigure[\scriptsize Malware (Recall)]{%
            \label{fig:drebin_mixed_effective_malware_recall_pie}
            \includegraphics[width=0.20\textwidth, trim=2.39cm 1.2cm 2.39cm 1.2cm, clip]
            {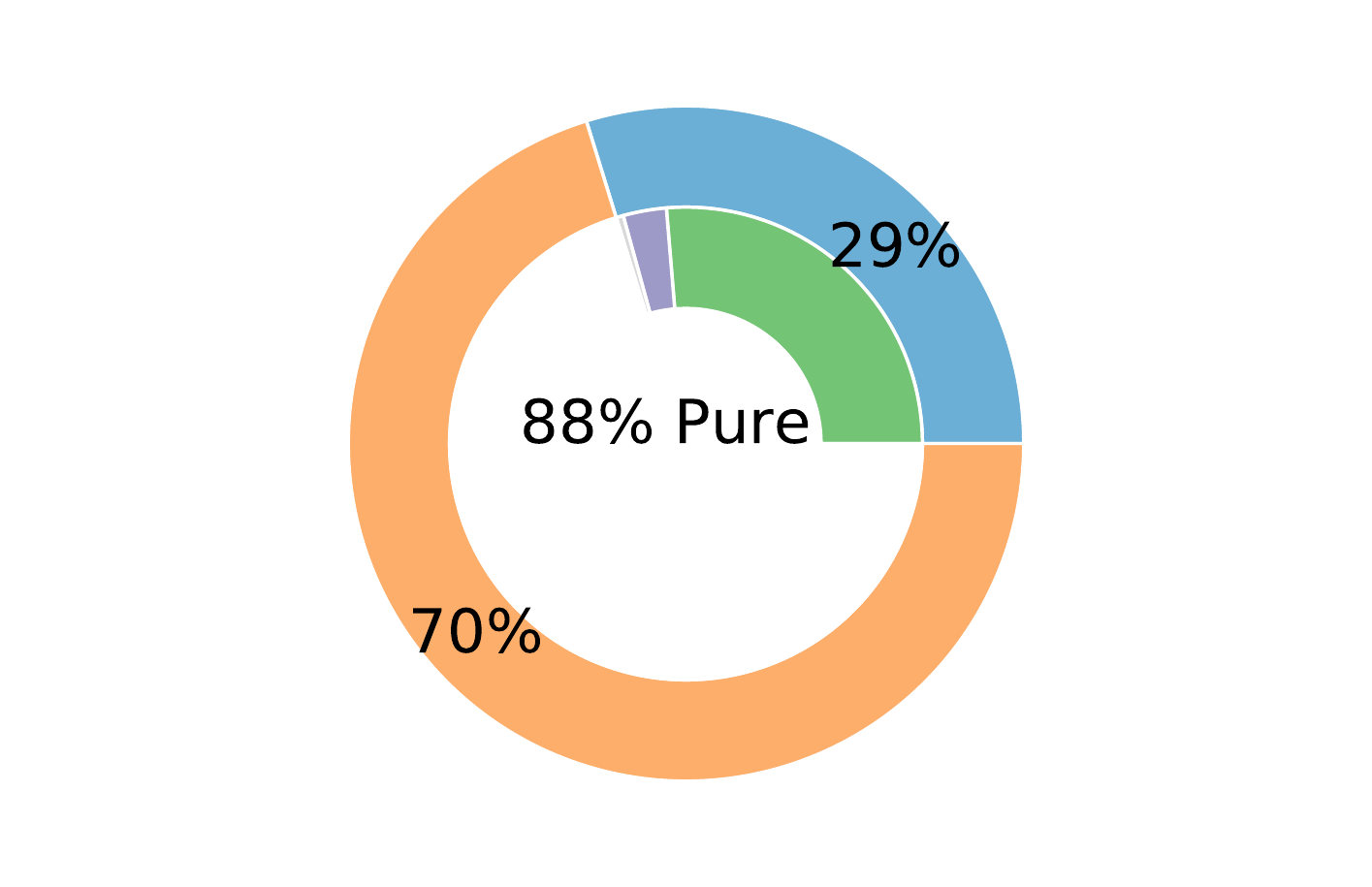}
        }
        \subfigure[\scriptsize Benign (Recall)] {%
           \label{fig:drebin_mixed_effective_benign_recall_pie}
           \includegraphics[width=0.20\textwidth, trim=2.39cm 1.2cm 2.39cm 1.2cm, clip]
           {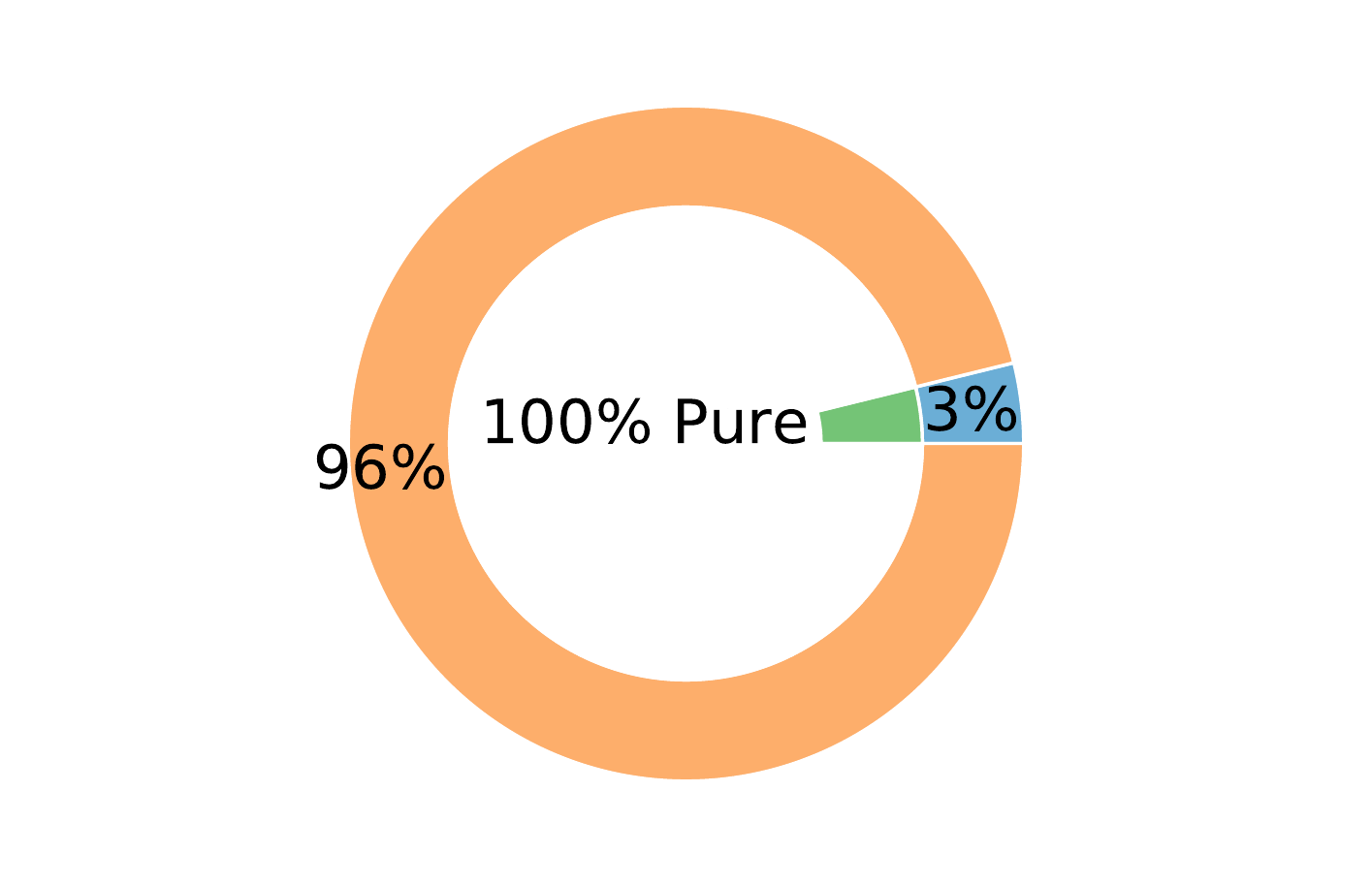}
        }
        \subfigure[\scriptsize Malware (Precision)]{%
            \label{fig:drebin_mixed_effective_malware_precision_pie}
            \includegraphics[width=0.20\textwidth, trim=2.39cm 1.2cm 2.39cm 1.2cm, clip]
            {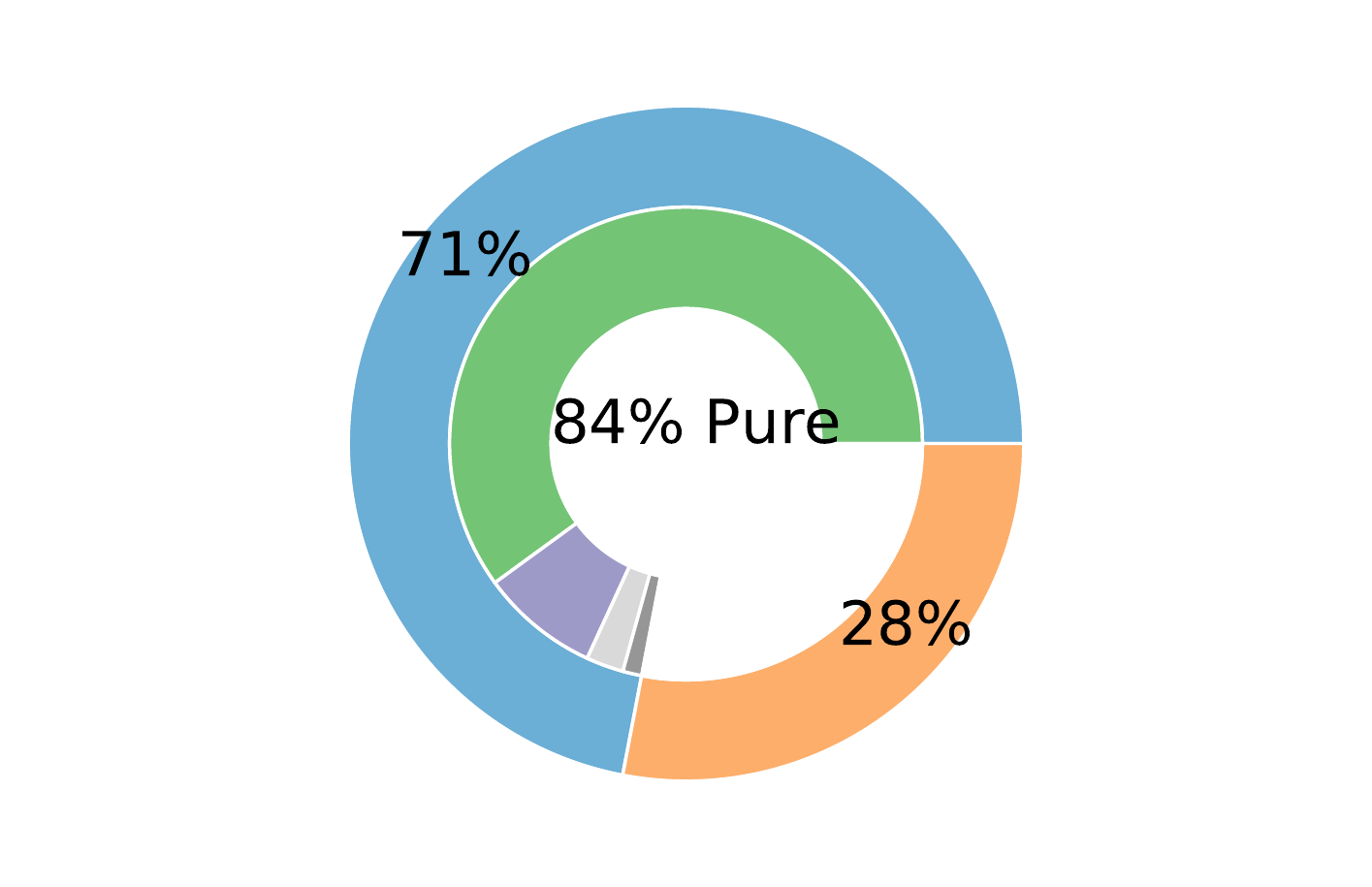}
        }
        \subfigure[\scriptsize Benign (Precision)] {%
           \label{fig:drebin_mixed_effective_benign_precision_pie}
           \includegraphics[width=0.20\textwidth, trim=2.39cm 1.2cm 2.39cm 1.2cm, clip]
           {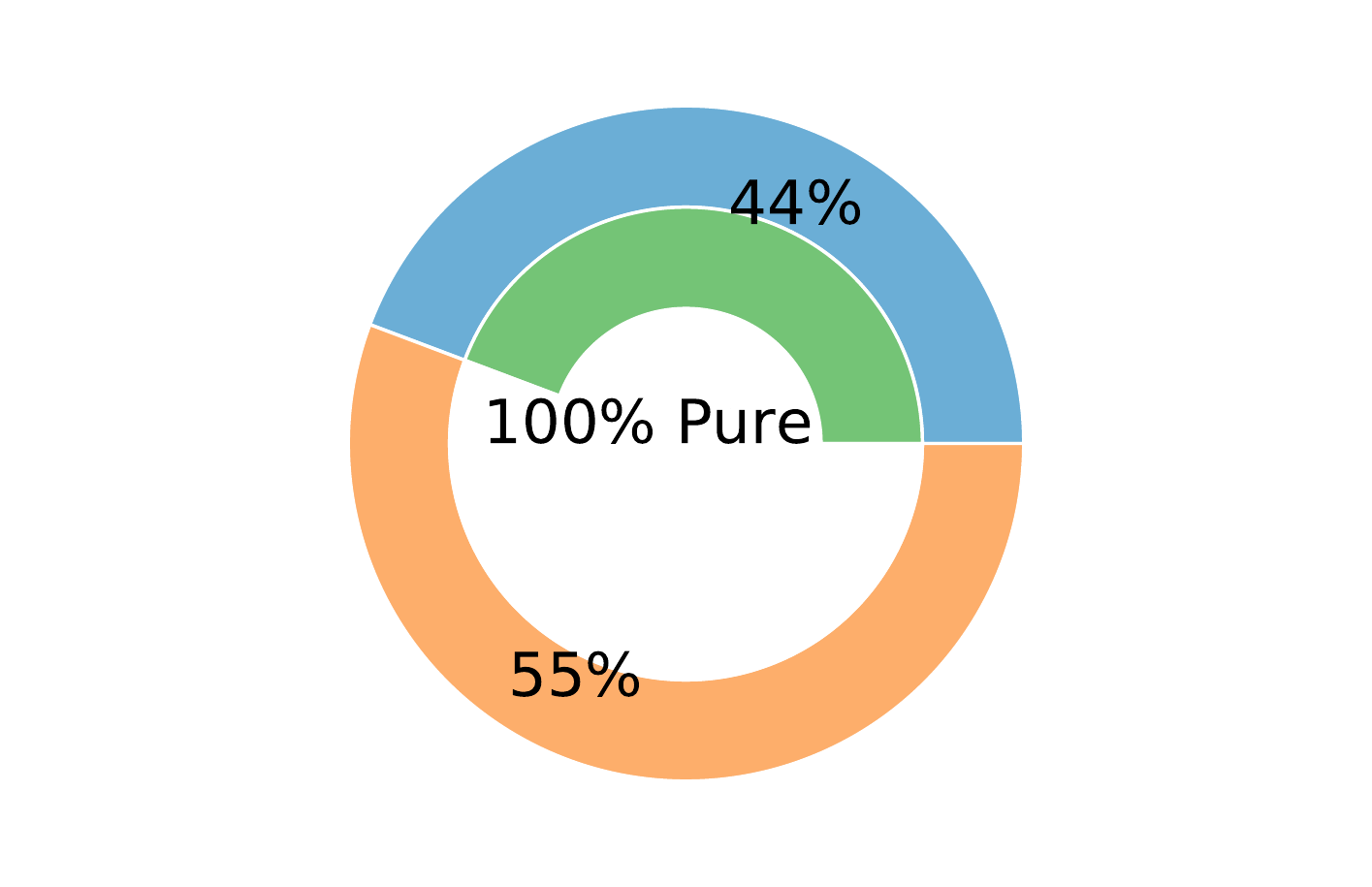}
        }
        \\
        {\small Drebin Dataset}
        \\
        \subfigure[\scriptsize Malware (Recall)]{%
            \label{fig:zoo_mixed_effective_malware_recall_pie}
           \includegraphics[width=0.20\textwidth, trim=2.39cm 1.2cm 2.39cm 1.2cm, clip]
            {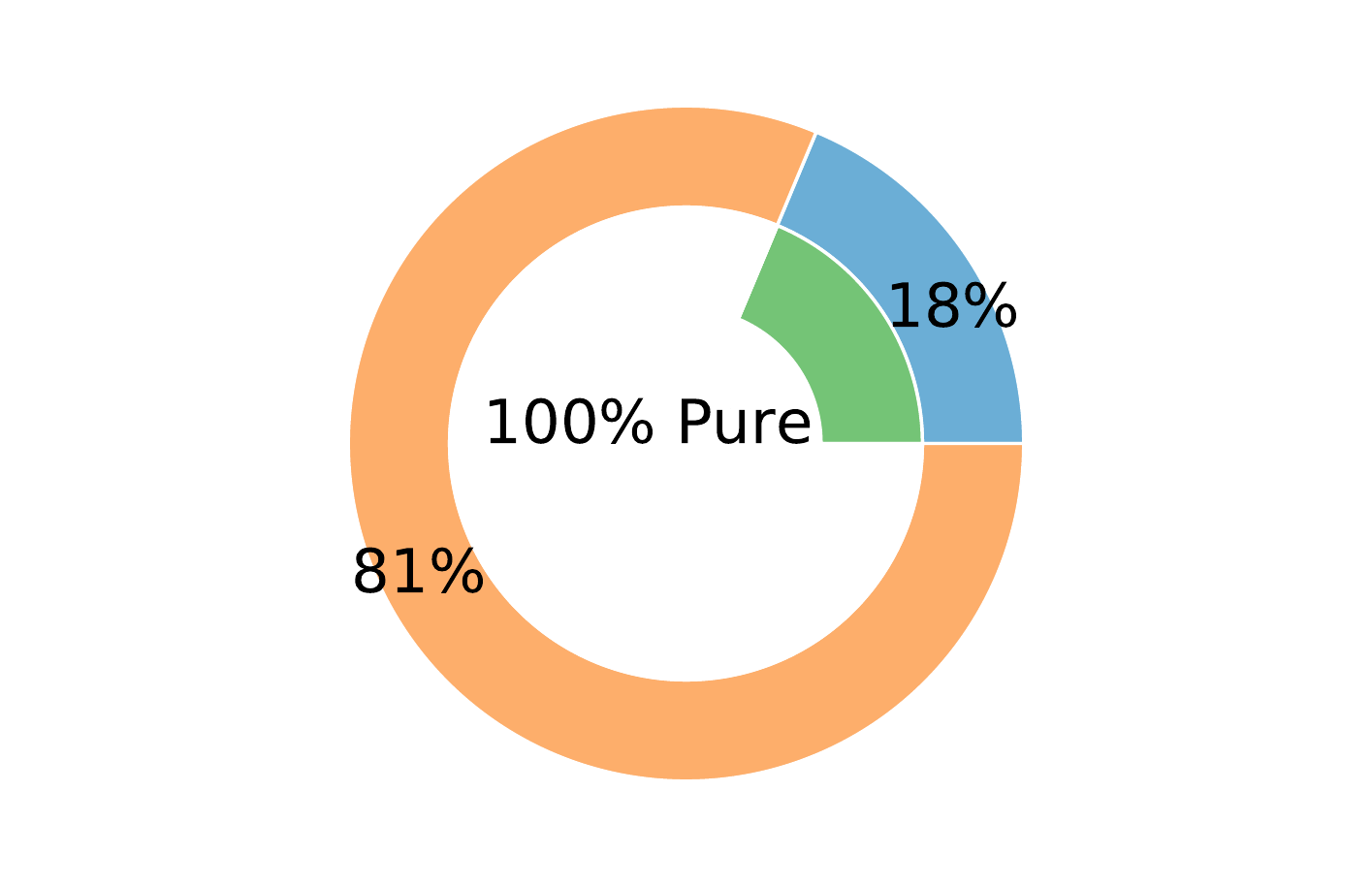}
        }
        \subfigure[\scriptsize Benign (Recall)] {%
           \label{fig:zoo_mixed_effective_benign_recall_pie}
           \includegraphics[width=0.20\textwidth, trim=2.39cm 1.2cm 2.39cm 1.2cm, clip]
           {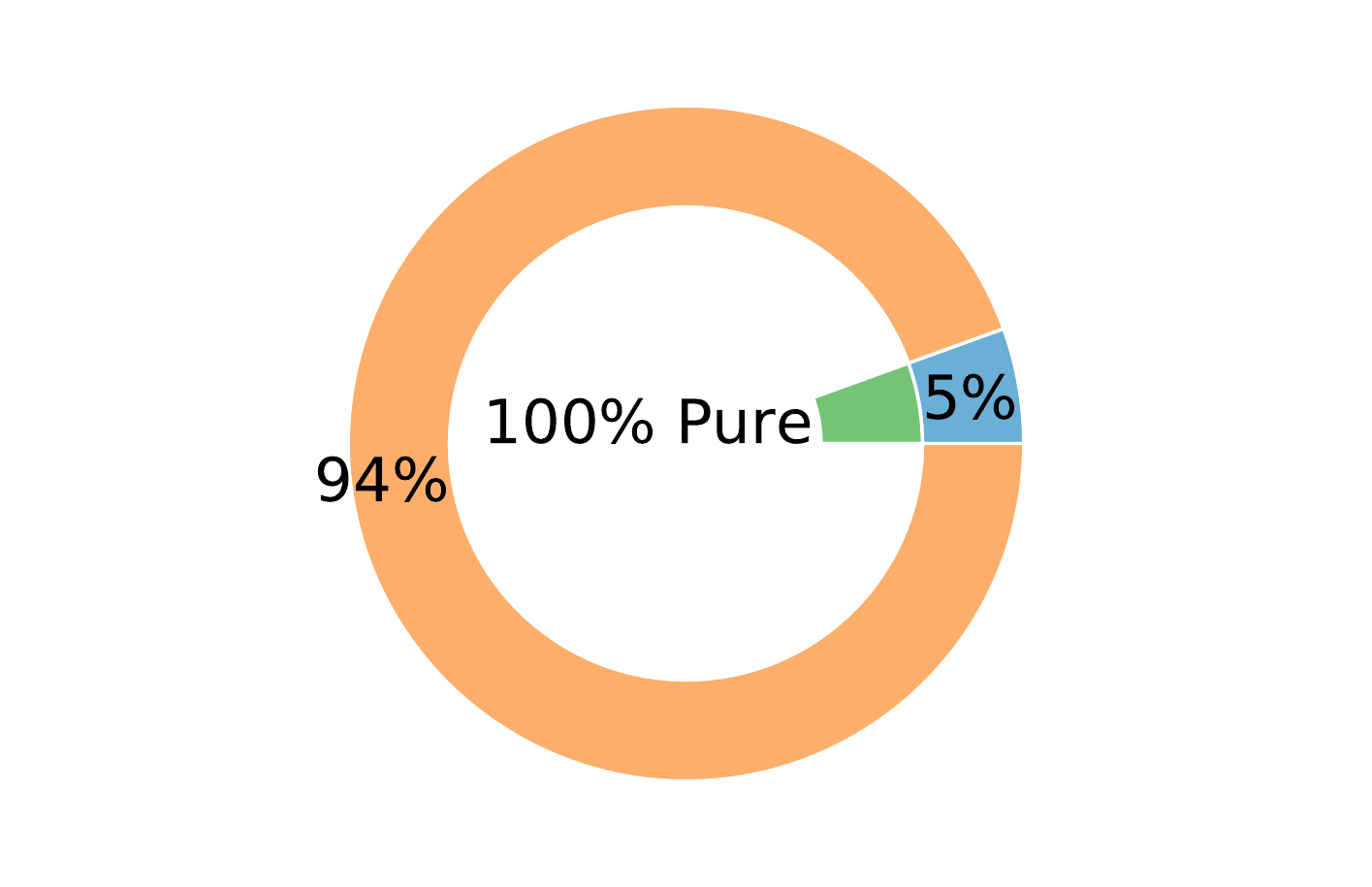}
        }
        \subfigure[\scriptsize Malware (Precision)]{%
            \label{fig:zoo_mixed_effective_malware_precision_pie}
           \includegraphics[width=0.20\textwidth, trim=2.39cm 1.2cm 2.39cm 1.2cm, clip]
            {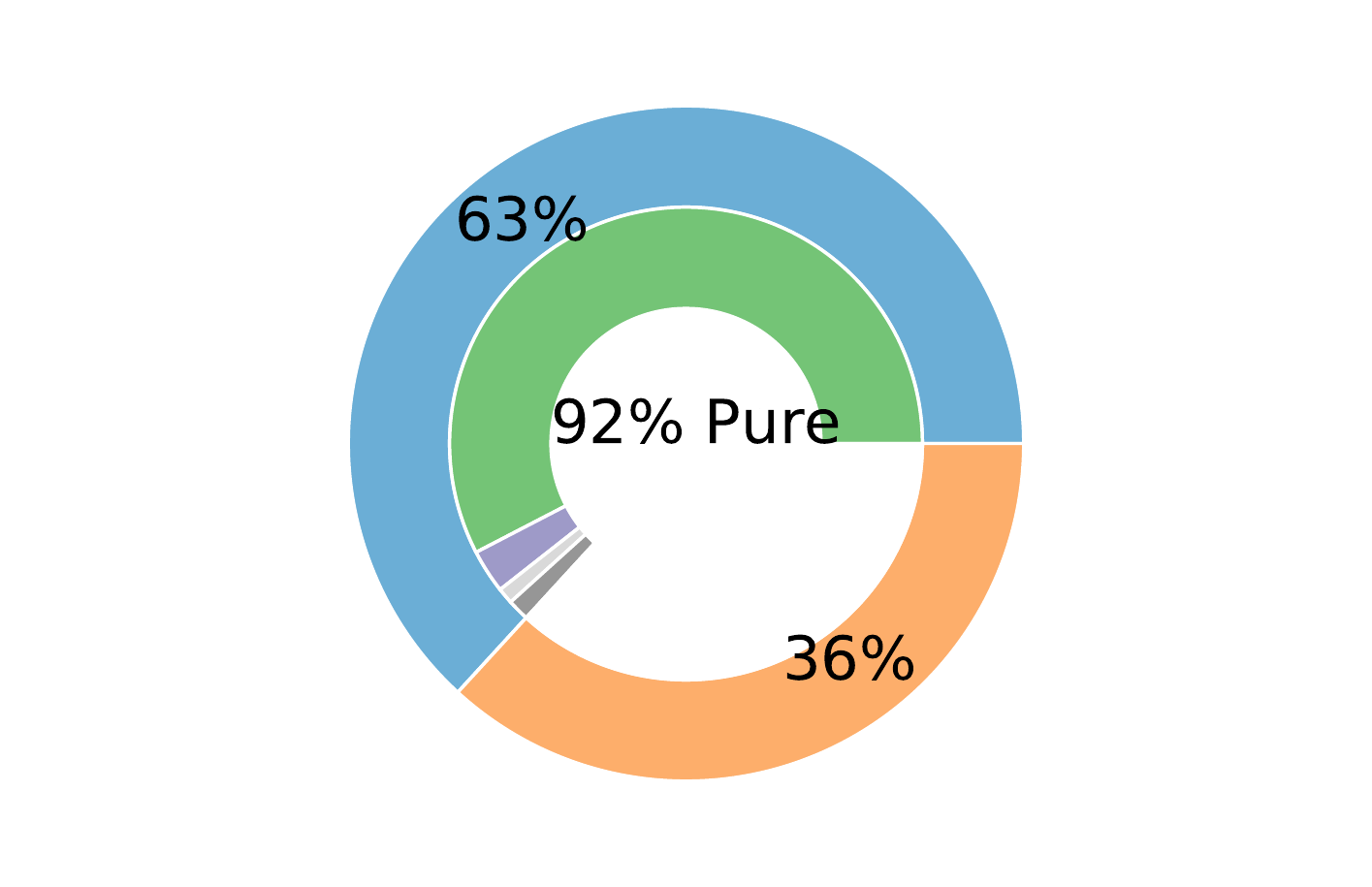}
        }
        \subfigure[\scriptsize Benign (Precision)] {%
           \label{fig:zoo_mixed_effective_benign_precision_pie}
           \includegraphics[width=0.20\textwidth, trim=2.39cm 1.2cm 2.39cm 1.2cm, clip]
           {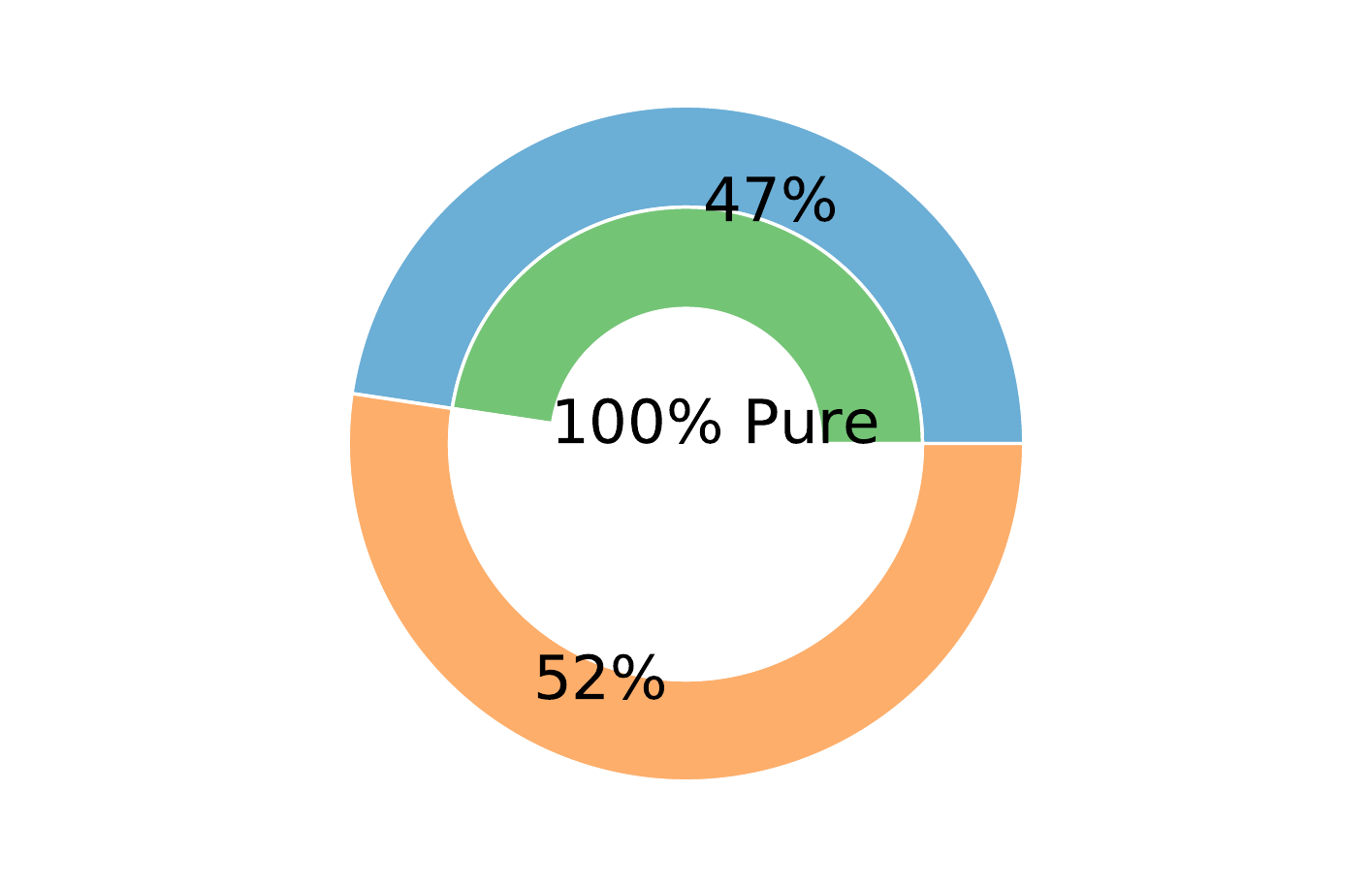}
        }                                                    
        \\
        {\small AndroZoo Dataset}
    \end{center}
   \caption{ \textsf{Cypider} Performance under Recall/Precision Settings on
   Mixed Scenario}
   \label{fig_mixedResultRPS}
\end{figure}
\end{scriptsize}

\begin{scriptsize}
\begin{figure}[ht!]
     \begin{center}        
        \subfigure[Recall Settings]{%
            \label{fig:malgenome_mixed_effective_recall_sim_network}
            \includegraphics[width=0.70\textwidth, trim=0.0cm 0.0cm 0.0cm 0.0cm, clip]
            {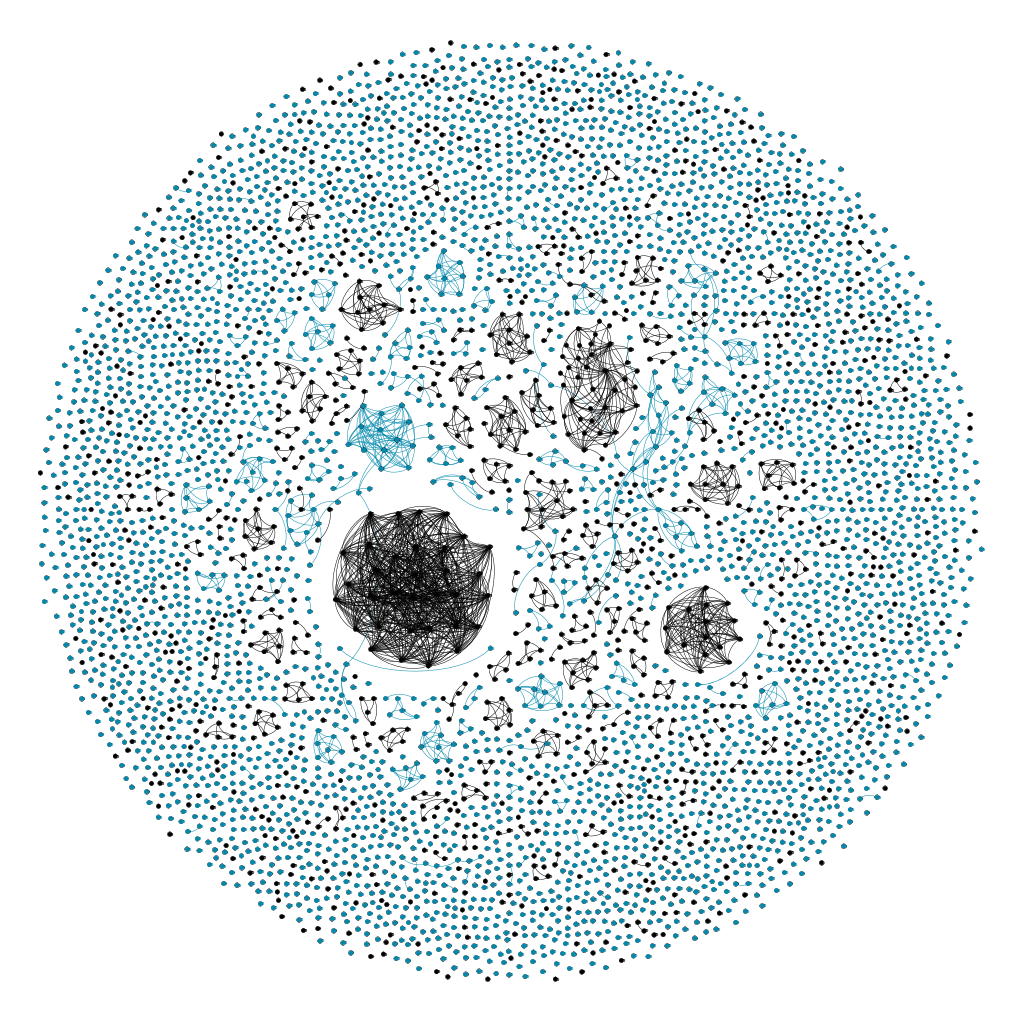}
        }\\
        \subfigure[Precision Settings] {%
           \label{fig:malgenome_mixed_effective_precision_sim_network}
           \includegraphics[width=0.70\textwidth, trim=0.0cm 0.0cm 0.0cm 0.0cm, clip]
           {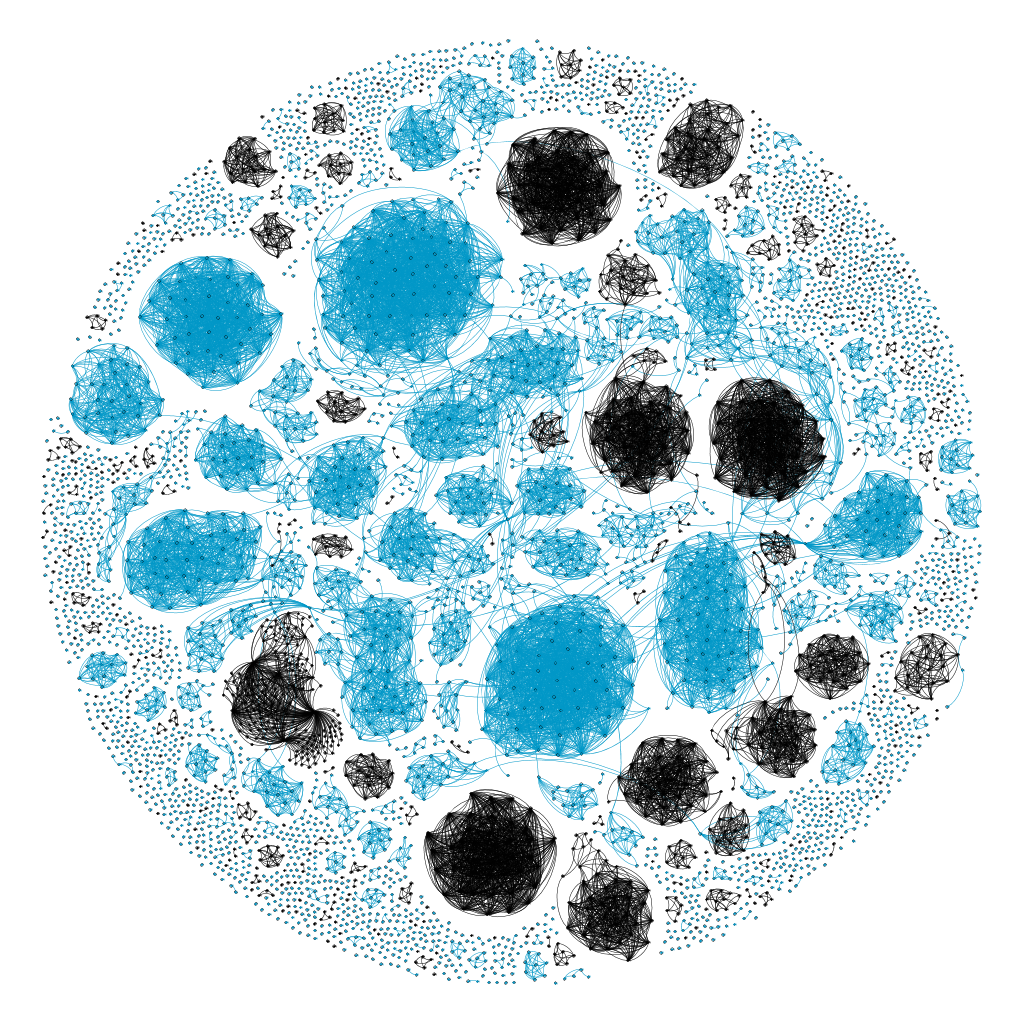}
        }
    \end{center}
    \caption{\textit{Malgenome} Mixed Dataset, Similarity Network of Recall/Precision Settings}
  \label{fig_mixGenomeNetRPS}
\end{figure}
\end{scriptsize}

The difference between recall and precision settings in the mixed scenario is more clear visually in the similarity networks. Figures \ref{fig_mixGenomeNetRPS} and \ref{fig_mixDrebinNetRPS} show the recall and precision similarity network of \textit{MalGenome} and \textit{Drebin} datasets respectively.  Darker color communities contain malware samples, and lighter ones contain benign samples. We notice a clear separation between the malicious communities and the benign ones. Also, more and bigger communities have been detected under the precision setting compared to the recall setting. Tables \ref{tab_numberCoveragePurityRecallPrecisionSettings} and \ref{tab_numberDetectedPureRecallPrecisionSettings} detail \textsf{Cypider} performance under the recall and the precision settings in terms of coverage/purity and number of detected/pure communities respectively.

\begin{scriptsize}
\begin{figure}[ht!]
     \begin{center}        
        \subfigure[Recall Settings]{%
            \label{fig:drebin_mixed_effective_recall_sim_network}
            \includegraphics[width=0.70\textwidth, trim=1.0cm 1.0cm 1.0cm 1.0cm, clip]
            {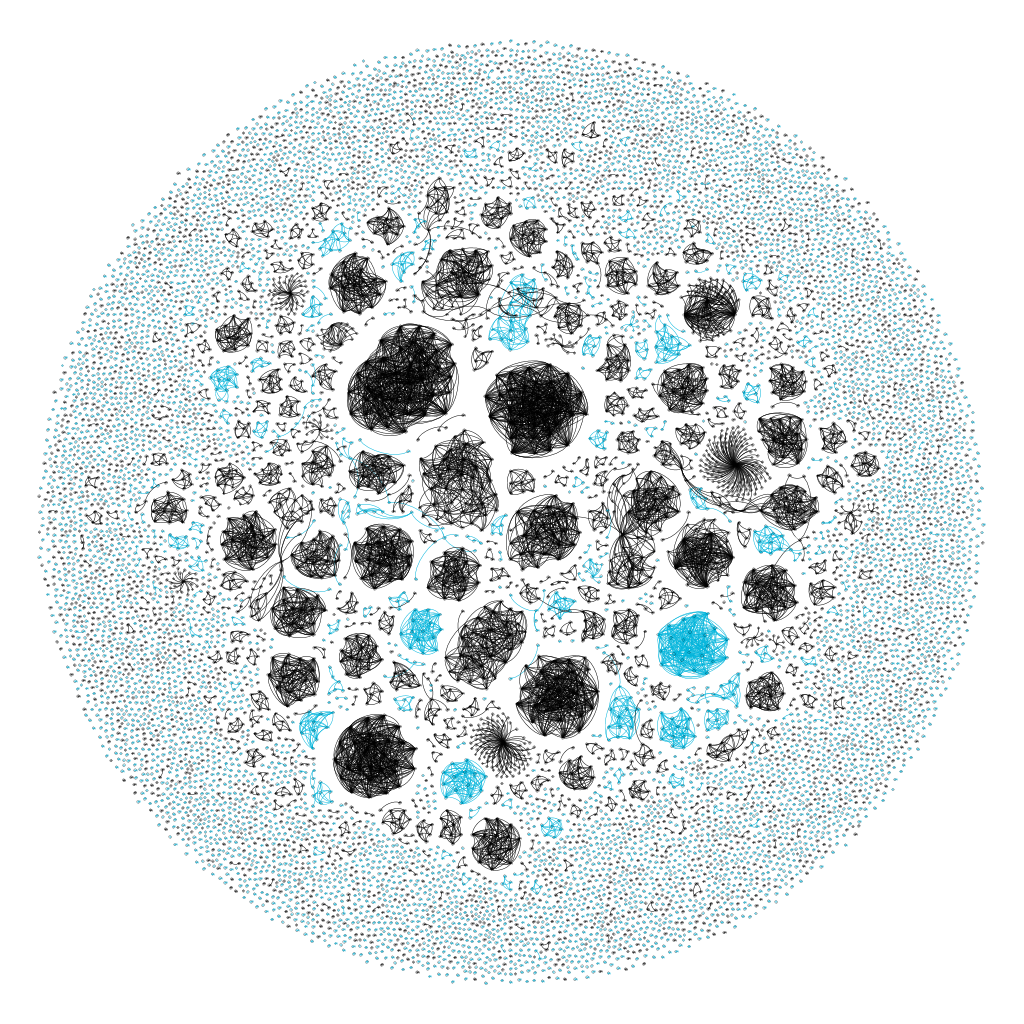}
        }
        \subfigure[Precision Settings] {%
           \label{fig:drebin_mixed_effective_precision_sim_network}
           \includegraphics[width=0.70\textwidth, trim=0.0cm 0.0cm 0.0cm 0.0cm, clip]
           {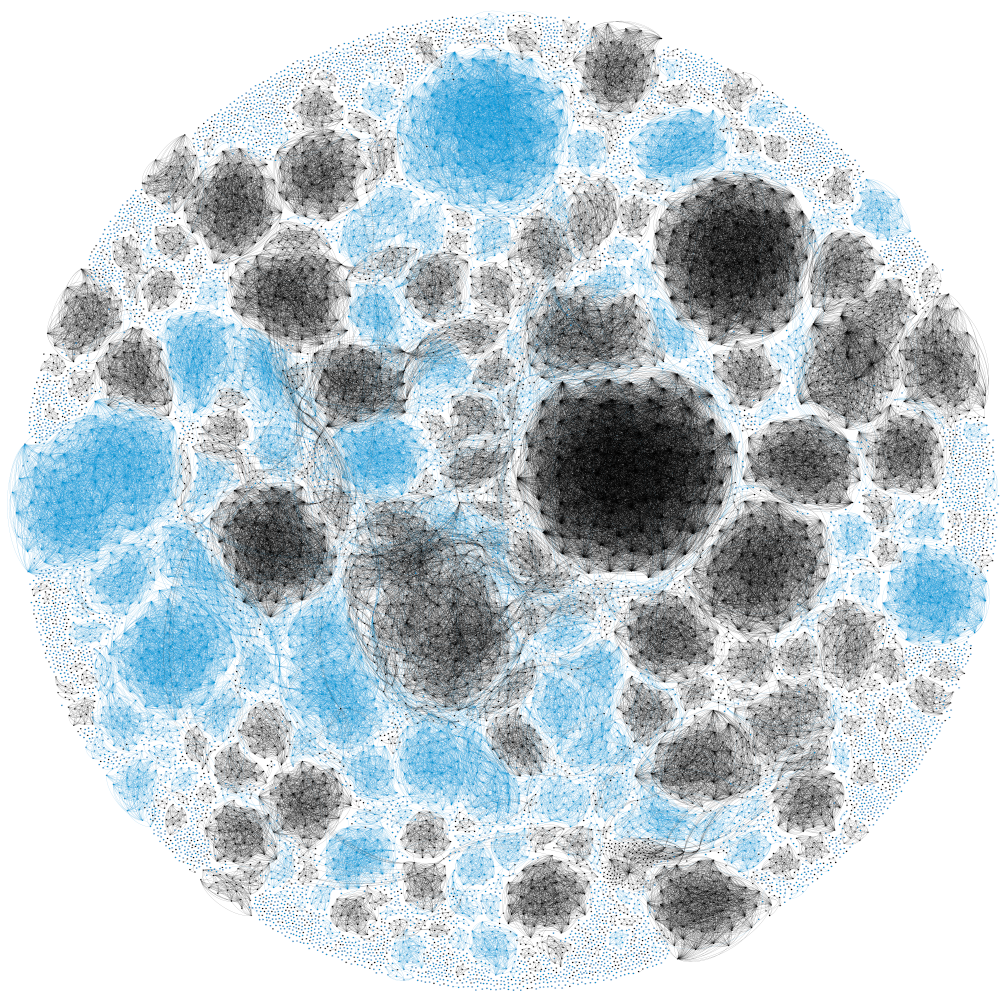}
        }
    \end{center}
    \caption{\textit{Drebin} Mixed Dataset Similarity, Network of Recall/Precision Settings}
    \label{fig_mixDrebinNetRPS}
\end{figure}
\end{scriptsize}

\begin{table}[ht]
\begin{scriptsize}
\centering
\resizebox{0.999\linewidth}{!}{
\begin{tabular}{lcc|ccc|ccc}
\hline
\multicolumn{3}{c|}{\textbf{Evaluation Setup}} &\multicolumn{3}{c|}{\textbf{Dataset Size}}  & \multicolumn{3}{c}{\textbf{Coverage/Purity \%}}            \\ 
\textbf{Dataset}                       & \textbf{Scenario}                 & \textbf{Settings}  & \textbf{\#Benign} & \textbf{\#Malware} & \textbf{\#Total} \\
\hline
\multirow{4}{*}{\textbf{Malgenome}}    & \multirow{2}{*}{\textbf{Malware}} & \textit{Recall}    & /      & 1.23k  & 1.23k    & /               & 15.56\%/99.0\%  & 15.56\%/99.0\%  \\
                                       &                                   & \textit{Precision} & /      & 1.23k  & 1.23k    & /               & 66.21\%/91.0\%  & 66.21\%/91.0\%  \\
                                       & \multirow{2}{*}{\textbf{Mixed}}   & \textit{Recall}    & 4.03k  & 1.23k  & 5.26k    & 3.7\%/100.0\%   & 15.48\%/99.0\%  & 6.46\%/100.0\%  \\
                                       &                                   & \textit{Precision} & 4.03k  & 1.23k  & 5.26k    & 39.66\%/100.0\% & 61.02\%/92.0\%  & 44.68\%/98.0\%  \\
\hline
\multirow{4}{*}{\textbf{Drebin}}       & \multirow{2}{*}{\textbf{Malware}} & \textit{Recall}    & /      & 5.55k  & 5.55k    & /               & 26.69\%/93.0\%  & 26.69\%/93.0\%  \\
                                       &                                   & \textit{Precision} & /      & 5.55k  & 5.55k    & /               & 73.08\%/84.0\%  & 73.08\%/84.0\%  \\
                                       & \multirow{2}{*}{\textbf{Mixed}}   & \textit{Recall}    & 7.64k  & 5.55k  & 13.19k   & 3.85\%/100.0\%  & 29.77\%/88.0\%  & 14.74\%/90.0\%  \\
                                       &                                   & \textit{Precision} & 7.64k  & 5.55k  & 13.19k   & 44.27\%/100.0\% & 72.0\%/83.0\%   & 55.93\%/91.0\%  \\
\hline
\multirow{4}{*}{\textbf{AndroZoo}}     & \multirow{2}{*}{\textbf{Malware}} & \textit{Recall}    & /      & 66.76k & 66.76k   & 0.0\%/0.0\%     & 18.48\%/100.0\% & 18.48\%/100.0\% \\
                                       &                                   & \textit{Precision} & /      & 66.76k & 66.76k   & 0.0\%/0.0\%     & 63.03\%/92.0\%  & 63.03\%/92.0\%  \\
                                       & \multirow{2}{*}{\textbf{Mixed}}   & \textit{Recall}    & 44.18k & 66.76k & 110.94k  & 5.52\%/100.0\%  & 18.7\%/100.0\%  & 13.45\%/100.0\% \\
                                       &                                   & \textit{Precision} & 44.18k & 66.76k & 110.94k  & 47.63\%/100.0\% & 63.19\%/91.0\%  & 56.99\%/94.0\%  \\
\hline

\end{tabular}
  }
\caption{Coverage and Purity Results of The Community Detection Framework Under
  Recall/Precision Settings} 
\label{tab_numberCoveragePurityRecallPrecisionSettings}
\end{scriptsize}
\end{table}

\begin{table}[ht]
\begin{scriptsize}
\centering
\resizebox{0.999\linewidth}{!}{
\begin{tabular}{lcc|ccc|ccc}
\hline
\multicolumn{3}{c|}{\textbf{Evaluation Setup}} &\multicolumn{3}{c|}{\textbf{Dataset Size}}  &  \multicolumn{3}{c}{\textbf{\#Communities/\#Pure }}\\ 
\textbf{Dataset} & \textbf{Scenario} & \textbf{Settings} &\textbf{\#Benign} & \textbf{\#Malware} & \textbf{\#Total} & \textbf{Bengin} & \textbf{Malware} & \textbf{Overall}\\
\hline
\multirow{4}{*}{\textbf{Malgenome}}   & \multirow{2}{*}{\textbf{Malware}} & \textit{Recall}    & /      & 1.23k  & 1.23k   &  /       & 15/15     & 15/15     \\
                                      &                                   & \textit{Precision} & /      & 1.23k  & 1.23k   &  /       & 35/31     & 35/31     \\
                                      & \multirow{2}{*}{\textbf{Mixed}}   & \textit{Recall}    & 4.03k  & 1.23k  & 5.26k   &  18/18   & 17/17     & 35/35     \\
                                      &                                   & \textit{Precision} & 4.03k  & 1.23k  & 5.26k   &  71/71   & 34/31     & 105/102   \\
\hline
\multirow{4}{*}{\textbf{Drebin}}      & \multirow{2}{*}{\textbf{Malware}} & \textit{Recall}    & /      & 5.55k  & 5.55k   &  /       & 95/89     & 95/89     \\
                                      &                                   & \textit{Precision} & /      & 5.55k  & 5.55k   &  /       & 155/136   & 155/136   \\
                                      & \multirow{2}{*}{\textbf{Mixed}}   & \textit{Recall}    & 7.64k  & 5.55k  & 13.19k  &  30/30   & 109/102   & 139/132   \\
                                      &                                   & \textit{Precision} & 7.64k  & 5.55k  & 13.19k  &  125/125 & 152/132   & 277/257   \\
\hline
\multirow{4}{*}{\textbf{AndroZoo}}    & \multirow{2}{*}{\textbf{Malware}} & \textit{Recall}    & /      & 66.76k & 66.76k  &  /       & 800/798   & 800/798   \\
                                      &                                   & \textit{Precision} & /      & 66.76k & 66.76k  &  /       & 1355/1291 & 1355/1291 \\
                                      & \multirow{2}{*}{\textbf{Mixed}}   & \textit{Recall}    & 44.18k & 66.76k & 110.94k &  176/176 & 828/826   & 1004/1002 \\
                                      &                                   & \textit{Precision} & 44.18k & 66.76k & 110.94k &  586/586 & 1321/1250 & 1907/1836 \\
\hline

\end{tabular}
  }

\caption{Number of Detected/Pure Communities of The Community Detection
  Framework Under Recall/Precision Settings}
\label{tab_numberDetectedPureRecallPrecisionSettings}
\end{scriptsize}
\end{table}

\section{Case Study: Obfuscation Resiliency Analyses}
\label{sec_obfuscationResliency}

In this section, we investigate the robustness of \textsf{Cypider} framework against common obfuscation techniques and code transformation in general.  We employ \textit{PRAGuard} obfuscated Android malware, which contains $11k$ samples, along with benign samples from \textit{AndroZoo} dataset. Table \ref{tab_obfuscationPG} details \textsf{Cypider} performance on the malware and the mixed scenarios. We compare \textsf{Cypider} performance before and after applying one or a combination of obfuscation techniques, as shown in Table \ref{tab_obfuscationPG}.

\begin{table}[h!]
\centering
\begin{scriptsize}
\resizebox{0.999\linewidth}{!}{
\begin{tabular}{cl|ccc|ccc}
\hline
\multicolumn{2}{c|}{\textbf{Evaluation Setup}} & \multicolumn{3}{|c|}{\textbf{Coverage/Purity \%}} & \multicolumn{3}{|c}{\textbf{\#Communities/\#Pure }}\\ 
 \textbf{Scenario} & \textbf{Obfuscation} & \textbf{Bengin} & \textbf{Malware} & \textbf{Overall} & \textbf{Bengin} & \textbf{Malware} & \textbf{Overall}\\
\hline
\multirow{8}{*}{\textbf{Malware}}     & \textit{Malgenome (Orignal)}   & /               & 66.2\%/92.3\%  & 66.2\%/92.3\%  & /     & 35/31 & 17/17   \\
                                      & \textit{(1) TRIVIAL}           & /               & 60.3\%/99.8\%  & 60.3\%/99.8\%  & /     & 34/34 & 35/31   \\
                                      & \textit{(2) STRING ENCRYPTION} & /               & 63.5\%/96.7\%  & 63.5\%/96.7\%  & /     & 34/32 & 35/31   \\
                                      & \textit{(3) REFLECTION}        & /               & 70.8\%/71.9\%  & 70.8\%/71.9\%  & /     & 30/27 & 35/31   \\
                                      & \textit{(4) CLASS ENCRYPTION}  & /               & 38.3\%/98.5\%  & 38.3\%/98.5\%  & /     & 27/26 & 35/31   \\
                                      & \textit{(1) + (2)}             & /               & 52.4\%/99.8\%  & 52.4\%/99.8\%  & /     & 42/42 & 35/31   \\
                                      & \textit{(1) + (2) + (3)}       & /               & 65.1\%/67.5\%  & 65.1\%/67.5\%  & /     & 38/34 & 35/31   \\
                                      & \textit{(1) + (2) + (3) + (4)} & /               & 36.3\%/99.7\%  & 36.3\%/99.7\%  & /     & 39/39 & 35/31   \\\hline
 \multirow{8}{*}{\textbf{Mixed}}      & \textit{Malgenome (Orignal)}   & 52.84\%/100.0\% & 45.44\%/95.0\% & 51.1\%/99.0\%  & 63/63 & 36/32 & 99/95   \\
                                      & \textit{(1) TRIVIAL}           & 50.98\%/100.0\% & 65.72\%/90.0\% & 54.3\%/97.0\%  & 66/66 & 38/32 & 104/98  \\
                                      & \textit{(2) STRING ENCRYPTION} & 52.47\%/100.0\% & 68.41\%/93.0\% & 56.06\%/98.0\% & 66/66 & 33/30 & 99/96   \\
                                      & \textit{(3) REFLECTION}        & 51.45\%/100.0\% & 77.23\%/63.0\% & 57.21\%/89.0\% & 65/65 & 29/21 & 94/86   \\
                                      & \textit{(4) CLASS ENCRYPTION}  & 52.84\%/100.0\% & 45.44\%/95.0\% & 51.10\%/99.0\% & 63/63 & 36/32 & 99/95   \\
                                      & \textit{(1) + (2)}             & 52.47\%/100.0\% & 68.41\%/93.0\% & 56.06\%/98.0\% & 66/66 & 33/30 & 99/96   \\
                                      & \textit{(1) + (2) + (3)}       & 53.66\%/100.0\% & 67.91\%/91.0\% & 57.01\%/97.0\% & 64/64 & 34/29 & 98/93   \\
                                      & \textit{(1) + (2) + (3) + (4)} & 49.04\%/100.0\% & 44.64\%/94.0\% & 48.01\%/99.0\% & 66/66 & 39/37 & 105/103 \\
\hline
\end{tabular}
}
\caption{Obfuscation Evaluation Using PRAGuard Dataset} 
\label{tab_obfuscationPG}
\end{scriptsize}
\end{table}

The evaluation results show that common obfuscation techniques have a limited effect on \textsf{Cypider} performance in general ($60\%-77\%$ coverage and $71\%-99\%$ purity). \textit{Class encryption} obfuscation decreased the coverage from $66\%$ in the non-obfuscated dataset to $36\%-38\%$. However, \textit{Class encryption} did not affect the purity performance. Similarly, \textit{Reflection} obfuscation technique dropped down the purity to $67\%-71\%$ compared to the original dataset but did not affect the coverage performance. To strengthen our findings (Table \ref{tab_obfuscationPG}) on \textit{PRAGuard} obfuscation dataset, we build our obfuscation dataset using \textit{Droid Chameleon} obfuscation tool. We obfuscated \textit{Drebin} malware dataset ($5k$ malware samples) and benign samples from \textit{AndroZoo} dataset ($5k$ malware samples). The result is $100k$ samples ($50k$ malware and $50k$ benign) from different obfuscation settings, as shown in Table \ref{tab_obfuscationCh}. Similar to \textit{PRAGuard} experiment, we compare \textsf{Cypider} performance before obfuscation (original \textit{Drebin} dataset) and after obfuscation. However, this experiment is different from the \textit{PRAGuard} one because both benign and malicious samples are obfuscated.

\begin{table}[ht!]
\centering
\begin{scriptsize}
\resizebox{0.999\linewidth}{!}{
\begin{tabular}{cl|ccc|ccc}
\hline
\multicolumn{2}{c|}{\textbf{Evaluation Setup}} & \multicolumn{3}{|c|}{\textbf{Coverage/Purity \%}}     & \multicolumn{3}{|c}{\textbf{\#Communities/\#Pure }}   \\
 \textbf{Scenario} & \textbf{Obfuscation}      & \textbf{Bengin} & \textbf{Malware} & \textbf{Overall} & \textbf{Bengin} & \textbf{Malware} & \textbf{Overall} \\
\hline
\multirow{10}{*}{\textbf{Mixed}}   & \textbf{Drebin (Original)           }  & 44.27\%/100.0\% & 72.0\%/83.0\%  & 55.93\%/91.0\% & 125/125 & 152/132 & 277/257 \\
                                   & \textit{Class Renaming              }  & 41.71\%/100.0\% & 72.61\%/83.0\% & 54.64\%/91.0\% & 129/129 & 156/135 & 285/264 \\
                                   & \textit{Method Renaming             }  & 42.02\%/100.0\% & 71.08\%/83.0\% & 54.19\%/91.0\% & 121/121 & 149/129 & 270/250 \\
                                   & \textit{Field Renaming              }  & 43.59\%/100.0\% & 72.01\%/83.0\% & 55.49\%/91.0\% & 128/128 & 148/127 & 276/255 \\
                                   & \textit{Code Reordering             }  & 43.43\%/100.0\% & 71.49\%/83.0\% & 55.19\%/91.0\% & 127/127 & 155/135 & 282/262 \\
                                   & \textit{Debug Information Removing  }  & 44.34\%/100.0\% & 72.09\%/83.0\% & 55.96\%/91.0\% & 117/117 & 151/130 & 268/247 \\
                                   & \textit{Junk Code Insertion         }  & 40.71\%/100.0\% & 71.81\%/83.0\% & 53.73\%/90.0\% & 124/124 & 153/132 & 277/256 \\
                                   & \textit{Instruction Insertion       }  & 42.65\%/100.0\% & 71.25\%/83.0\% & 54.63\%/91.0\% & 120/120 & 156/136 & 276/256 \\
                                   & \textit{String Encryption           }  & 43.2\%/100.0\%  & 72.16\%/83.0\% & 55.32\%/91.0\% & 133/133 & 147/127 & 280/260 \\
                                   & \textit{Array Encryption            }  & 43.55\%/100.0\% & 71.93\%/83.0\% & 55.42\%/91.0\% & 125/125 & 152/131 & 277/256 \\
\hline
\multirow{10}{*}{\textbf{Malware}} & \textbf{Drebin (Original)           }  & /               & 73.08\%/84.0\% & 73.08\%/84.0\% & /       & 155/136 & 155/136 \\
                                   & \textit{Class Renaming              }  & /               & 74.14\%/84.0\% & 74.14\%/84.0\% & /       & 160/137 & 160/137 \\
                                   & \textit{Method Renaming             }  & /               & 72.66\%/83.0\% & 72.66\%/83.0\% & /       & 159/136 & 159/136 \\
                                   & \textit{Field Renaming              }  & /               & 73.75\%/83.0\% & 73.75\%/83.0\% & /       & 155/132 & 155/132 \\
                                   & \textit{Code Reordering             }  & /               & 74.07\%/83.0\% & 74.07\%/83.0\% & /       & 158/135 & 158/135 \\
                                   & \textit{Debug Information Removing  }  & /               & 72.92\%/83.0\% & 72.92\%/83.0\% & /       & 155/132 & 155/132 \\
                                   & \textit{Junk Code Insertion         }  & /               & 73.86\%/83.0\% & 73.86\%/83.0\% & /       & 157/135 & 157/135 \\
                                   & \textit{Instruction Insertion       }  & /               & 73.96\%/85.0\% & 73.96\%/85.0\% & /       & 160/137 & 160/137 \\
                                   & \textit{String Encryption           }  & /               & 73.8\%/83.0\%  & 73.8\%/83.0\%  & /       & 155/132 & 155/132 \\
                                   & \textit{Array Encryption            }  & /               & 73.8\%/83.0\%  & 73.8\%/83.0\%  & /       & 155/133 & 155/133 \\

\hline

\end{tabular}
}
\caption{Obfuscation Evaluation Using DroidChameleon Tool on Drebin Dataset} 
\label{tab_obfuscationCh}
\end{scriptsize}
\end{table}

Table \ref{tab_obfuscationCh} details the result of \textsf{Cypider} framework on the different obfuscation techniques. The most noticeable result is that the obfuscation techniques provided by the \textit{DroidChameleon} tool have a limited effect on our clustering framework. All the performance metrics were kept stable on both non obfuscated and obfuscated samples under the malware and mixed scenarios. We argue that \textsf{Cypider} framework is robust to common obfuscation and code transformation techniques because our framework considers many APK contents for feature extraction. Therefore, the obfuscation techniques could affect one APK content, but \textsf{Cypider} could leverage other contents to fingerprint the malware sample and compute the similarity with other malware samples.


\section{Discussion} 
\label{sec:discussion}

\textsf{Cypider} framework achieves a very good result regarding the percentage of the detected malware and purity of the communities considering the difficulty of the task of malware clustering in general.  However, we believe that the detection performance of \textsf{Cypider} could be improved by considering more content vectors.  The more static coverage of the APK file is, the more broad and accurate the suspicious communities are. Fortunately, \textsf{Cypider} could easily add new vectors due to the \textit{majority-voting} mechanism to decide about the similarity. Using only static features leads to the right results. However, including dynamic features in \textsf{Cypider}'s detection process could boost detection since this covers Android malware that downloads and execute the payload at runtime.  \textsf{Cypider} with only static features could be a complementary solution to other detection approaches based on \textit{dynamic features}. As we have noticed in the evaluation results, there are more detected communities than the number of actual malware families. According to our analyses, multiple communities could be part of the same malware family. However, these communities could represent malware family variants. For example, \textbf{DroidKungFu} family is considered as one family in \textit{Drebin dataset}. However, there are many variants for this family in the \textit{Genome dataset} such as \textbf{DroidKungFu1}, \textbf{DroidKungFu2} and \textbf{DroidKungFu4}. Moreover, we notice that some malware instances could have various labels from the vendors. For example, certain instances of \textbf{FakeInst} malware have their community with no connection shared with the rest of the family. After a little investigation using the hashes of the instances, we identified a different label (\textbf{Adwo}), which represents a different malware family. Moreover, the evaluation shows the effectiveness of \textit{Cypider} in identifying Zero-day malicious apps since the detection was without any prior knowledge of the actual dataset. The unsupervised feature of \textit{Cypider} framework could make it handy for security practitioners.

The concept of malicious community, proposed by  \textit{Cypider}, could be similar to the malware family concept. However, there are some differences: i) One malware family could be detected in multiple communities of malicious apps. So, the community gives a more granular view of the malicious app similarity based on given statical features. ii) Security practitioners could decide about a given malicious app family using manual analyses. However, the community concept comes from a purely unsupervised automatic analysis. An important feature of 
\textsf{Cypider} that we can get from detected suspicious communities is the explainability, which facilitates defining what exactly is shared among the apps of a given community. The explainable results are due to the multiple content similarity links between the apps. For example, malware apps sharing the same IP addresses that are hard-coded in the binary are most likely to be part of a Command and Control (C\&C) infrastructure and are the sign of a botnet.

Obfuscation is considered a big issue in malware detection systems, including \textit{Cypider}, where the adversary uses an obfuscated content or transformation attacks. The latter could download the malicious code and execute it at runtime, is undetectable by \textsf{Cypider} unless the malware instances share other covered contents. However, \textit{Cypider} attempts to deal with obfuscated apps by considering a range of techniques. i)  \textit{Cypider} leverages multiple Android package contents (such as permissions), which allow the system to be more robust against obfuscation since it also uses other un-obfuscated contents. ii) In the case of Dex disassembly,  \textit{Cypider} uses instruction opcode instead of the whole instruction to compute the N-grams.  Therefore, the latter would be more resilient to obfuscation in the instruction operands. The previous techniques show their effectiveness in the evaluation since we used a real malware dataset in addition to apps from Google play (where the apps are supposedly obfuscated through ProGuard) and obfuscated malware datasets (PRAGuard and DroidChameleon). Besides, \textsf{Cypider} framework could be complementary to dynamic analysis solutions to achieve better results.  

\section{Related Work} 
\label{sec:related_work}

Previous work on Android analysis can be classified with respect to the analysis task as follows:

\begin{itemize}

\item \textit{Malware detection} The objective is to check if a unknown app is malicious or not. They mostly apply supervised learning approaches. The learning model is generated during a training phase and using labeled samples. In the testing phase, the learning model aims to identify if the analyzed sample is malware or benign.

\item \textit{Malware family attribution}: Given a set of known families, the objective of this task is to attribute an unknown sample to a known family.

\item \textit{Similarity-based detection}: The objective of this task is to find similarities among different apps without knowing any information about their types, i.e., benign or malware, or their families in case they are malware. 

\end{itemize} 

\subsection{Malware detection}

The Android malware detection methods focus on identifying whether the analyzed app is benign or malicious. They use three type of features: static, dynamic, and hybrid. The static methods \cite{arp2014drebin, karbab2016dna, feng2014apposcopy, yang2014apklancet, zhongyang2013droidalarm, sanz2014anomaly, kim2018multimodal, onwuzurike2019mamadroid, xu2018deeprefiner, xu2019droidevolver, allen2018improving, elish2015profiling, idrees2017pindroid, DBLP:journals/corr/abs-1712-08996, burnap2018malware, badhani2019cendroid}  depend on static features, which are extracted from the Apk file such as requested permissions, APIs, bytecodes, opcodes.  The dynamic methods \cite{canfora2016acquiring, karbab2017dysign, DBLP:journals/corr/KarbabDAM17, ali2016aspectdroid, zhang2013vetting, amos2013applying, wei2012android, huang2014asdroid, saracino2016madam, DBLP:journals/di/KarbabD19, DBLP:journals/corr/abs-1812-10327} use features that are derived from the app's execution. They are more resilient to obfuscations than static analysis methods. However, they incur additional cost in terms of processing and memory to run the app. The hybrid methods \cite{spreitzenbarth2013mobile, grace2012riskranker, lindorfer2014andrubis, bhandari2015draco, vidas2014a5, zhou2012hey, zhang2014semantics, martinelli2017bridemaid, jang2016andro, DBLP:conf/IEEEares/KarbabD18, DBLP:journals/corr/abs-1806-08893, ali2018toward}  use both static and dynamic features. In the case of Win32 malware detection, some static analysis \cite{sun2018deep, sharma2019malware}, dynamic analysis \cite{copty2018accurate, naval2015employing, wuchner2017leveraging, ding2018malware, rhode2018early, burnap2018malware,belaoued2019combined}, and hybrid methods \cite{han2019maldae} have been proposed.

\subsection{Malware family attribution}

Malware family attribution methods \cite{ali2015opseq, deshotels2014droidlegacy, lee2015screening, kim2015structural, suarez2014dendroid,lin2013identifying, faruki2015androsimilar, tian2017detection, fan2018android} focus on detecting variants of known families. In case of Win32 malware family attribution, the methods are classified as static \cite{ni2018malware}, dynamic \cite{mohaisen2015amal}, and hybrid \cite{chakraborty2017ec2}.  Other methods combine between malware detection and family attribution in case of Android OS \cite{kang2015detecting, canfora2016hmm, zhang2019scalable, karbab2018maldozer} and Win32 OS \cite{zhang2019feature, DBLP:conf/esorics/AlrabaeeK0D19}

\subsection{Similarity-based detection}

Similarity-based detection methods \cite{chen2015finding, sun2015droideagle, zhou2012detecting, crussell2012attack, zhou2013fast, crussell2013andarwin} focus on detecting repackaged malware. They adopt the unsupervised learning approach, i,e, they do not require training phase and they are directly applied on unlabeled samples. They assume that two or multiple apps that share the same code are likely to belong to the same family. Thus, they check if the apps are using the same malicious code (i.e., detection of malware families), or they are reusing the code of the same original app (i.e., code reuse detection).  Table \ref{tab_comp} summarizes the similarity-based detection methods with respect to (1) extracted feature, (2) similarity algorithm, (3) dataset size, (4) obtained findings from applying the method on the dataset, and (5) efficiency in terms of running time of the similarity algorithm.

DroidMOSS \cite{zhou2012detecting} aims to detect repackaged apps from the Android Market, which are redistributed to third-party marketplaces. It first applies a fuzzy hashing technique on the opcode of each app collected from the markets to generate the fingerprint of that app. Then, it measures the similarity between each app in the third-market and all the apps in the official Android market using the Edit distance. 

Zhou et al. \cite{zhou2013fast} proposed PiggyApp to detect piggybacked apps, which share the same primary module as the original app. It builds for each primary module its program dependency graph (PDG), and generates its semantic feature vector. To detect similar vectors, it uses a linearithmic search algorithm, which incurs time complexity  of $ O(n \log n)$, where $n$ is the number of apps. AnDarwin \cite{crussell2013andarwin,crussell2015andarwin} uses program dependency graph to extract a semantic vectors from each app.  AnDarwin finds similar code segments that are among all the semantic vectors of apps using Locality Sensitive Hashing (LSH).

DNADroid \cite{crussell2012attack} computes the similarity between two apps by comparing their program dependency graphs. It then applies subgraph isomorphism to find a mapping between nodes of two PDGs. MassVet \cite{chen2015finding} is designed for vetting apps at a massive scale, without knowing what malware looks like and how it behaves. It compares a submitted app with all apps already on the market by focusing on the difference between those sharing a similar GUI structure, which is known to be largely preserved during repackaging. Similarly, DroidEagle \cite{sun2015droideagle} uses the layout resources within an app to detect repackaged apps and phishing malware apps which are visually similar to the original app. It generates a layout tree from each app, and uses the Layout Edit Distance (LED) to measure the similarity among apps.  

Cypider aims to identify community of malware, which represent malware that are similar with respect to any set of features, which is different from the family concept that groups similar samples according to a predefined set of features. The fundamental idea of Cypider relies on building a similarity network between the input apps.  The app is represented as a vector of contents. If one content is shared between a pair of apps, an edge is established between the two apps. The result  is a similarity network, where the nodes are Android apps and the edges represent the high similarity with respect to one content between the apps.

\begin{table}[!ht]
\centering
\begin{scriptsize}
\resizebox{0.999\linewidth}{!}{
\begin{tabular}{lccccc}
\hline
{\bf Method}                                              & {\bf Feature} & {\bf Similarity algorithm}    & {\bf Dataset size} & {\bf Findings}                  & {\bf Efficiency} \\
\hline
DroidMOSS \cite{zhou2012detecting}                        & Opcode        & Edit distance                 & N/A                & 5\% to 13\% repackaged apps     & $ O(n^2)$        \\

PiggyApp \cite{zhou2013fast}                              & PDG           & linearithmic search algorithm & N/A                & 0.97\% to 2.7\% repackaged apps & $ O(n \log n)$   \\
\hline
Andarwin \cite{crussell2013andarwin,crussell2015andarwin} & PDG           & LSH                           & 265359             & 4295 cloned and 36106 rebranded &                  \\
\hline
DNADroid \cite{crussell2012attack}                        & PDG           & Subgraph isomorphism          & 75,000             & 141 cloned apps                 &                  \\
\hline
MassVet \cite{chen2015finding}                            & GUI           & DiffCom                       & 1.2 million        & 127429 malicious                & 10 sec/app       \\
\hline
DroidEagle \cite{sun2015droideagle}                       & GUI           & LED                           & 99626              & 1298                            & 3 hours          \\
\hline
Cypider                                                   & APK features  & similarity network            & 110k               & $60-80\%$ coverage              & about 60 sec     \\
                                                          &               &                               &                    & $85-99\%$ precision in the      &                  \\
                                                          &               &                               &                    & detected malicious communities  &                  \\
\hline
\end{tabular}
}
\end{scriptsize}
\caption{Comparison of similarity-based detection methods} \label{tab_comp}
\end{table}


\section{Conclusion}
\label{sec:conclusion} 

In this paper, we design and implement a novel, efficient, and scalable framework for Android malware detection, namely, \textsf{Cypider}. The detection mechanism relies on the community concept.  \textsf{Cypider} consists of a systematic framework that can generate a fingerprint for each community and identify known and unknown malicious communities. \textsf{Cypider} has been implemented and evaluated on different malicious and mixed data\-sets. Our findings show that \textsf{Cypider} is a valuable and promising approach for detecting application similarity and malicious communities in Android applications. \textsf{Cypider} needs only few seconds to build a network similarity of a large number of apps, and incurs about 60 seconds to perform community detection for $111k$ apps. The results of the community fingerprint are auspicious as $87\%$ of the detection is achieved.


\bibliographystyle{ieeetr}
\bibliography{references}

\end{document}